\documentclass[a4paper,11pt]{article}

\pdfoutput=1
\usepackage{jheppub}
\def\be{\begin{equation}}
\def\ee{\end{equation}}
\def\ba{\begin{eqnarray}}
\def\ea{\end{eqnarray}}
\def\beq{\begin{eqnarray}}
\def\eeq{\end{eqnarray}}
 
\def\mpl{M_{\rm Pl}}

\def\d{\mathrm{d}}

\def\({\left(}
\def\){\right)}
\def\[{\left[}
\def\]{\right]}

\def\nn{\nonumber}

\def\stu{St\"uckelberg }
\def\p{\partial}
\def\mupn{^\mu_{\ \nu}}
\def\<{\langle}
\def\>{\rangle}

\def\D{\mathcal{D}}

\def\GS{Gel`fand-Shilov }
\def\KL{K\"all\'en-Lehmann }
\newcommand{\tp}{\langle\tilde{\chi}(\tilde{x}_1)\tilde{\chi}(\tilde{x}_2)\rangle}

\usepackage{amssymb}
\usepackage{latexsym}
\usepackage{graphicx}
\usepackage{amsmath}
\usepackage{array}
\usepackage{graphicx} 
\usepackage[normalem]{ulem}

\allowdisplaybreaks[2]

\title{UV properties of Galileons: Spectral Densities}
\author{Luke Keltner,}
\author{Andrew J. Tolley}
\affiliation{CERCA/Department of Physics, Case Western Reserve University, 10900 Euclid Ave, Cleveland, OH 44106, USA}
\emailAdd{Lucas.Keltner@case.edu}
\emailAdd{Andrew.J.Tolley@case.edu}

\abstract{We propose a picture for the UV properties of Galileon field theories.  We conjecture that Galileons, and all theories incorporating the Vainshtein mechanism, fall into Jaffe's class of `non-localizable' field theories characterized by an exponential growth in their \KL  spectral densities. Similar properties have been argued to arise for Little String Theories and M-theory. For such theories, the notion of micro-causality and the time ordering used to define the S-matrix and correlation functions must be modified, and we give a Lorentz invariant prescription for how this can be achieved. In common with General Relativity (GR), the scattering amplitudes for Galileons are no longer expected to satisfy polynomial boundedness away from the forward scattering or fixed physical momentum transfer limits. This is a reflection of the fact that these theories are fundamentally gravitational and not local field theories. We attribute this to the existence of a locality bound for Galileons, analogous to the Giddings-Lippert locality bound for GR.  We utilize the recently developed Galileon duality to define a UV finite, Lorentz invariant, quantization of a specific Galileon theory for which the energy of all states are positive definite. We perform an explicit computation of the Wightman functions for this theory, and demonstrate the exponential growth associated with the locality bound.  In analogy with GR, the bound is correlated with the absence of Galileon Duality (i.e. Diffeomorphism) invariant local observables. We argue that these theories can be quantized in a manner which preserves Lorentz invariance and macro-causality and that the latter ensures that the superluminalities found in the low energy effective theory are absent in the full theory.
}

\begin{document}
\maketitle
\flushbottom

\newpage

\section{Introduction}

The discovery of cosmic acceleration, and the desire to develop theories of gravity in which the dynamics is distinct from general relativity in the infrared in order to tackle this problem, has led to the development of new classes of field theories incorporating screening mechanisms. A whole class of such theories incorporate a gravitational screening mechanism, known as the Vainshtein mechanism \cite{Vainshtein:1972sx}, whose existence is necessary for the observational viability of these theories. 
However, the status of these theories as consistent quantum field theories has remained uncertain. No explicit UV complete model which incorporates the Vainshtein mechanism is known, and there have been a number of arguments to suggest that such a completion may be impossible. The simplest model incorporating the Vainshtein mechanism is what is now known as the cubic Galileon model \cite{Luty:2003vm}. It arose as the effective description of the helicity-zero mode of the graviton in the Dvali-Gabadadze-Porrati (DGP) model \cite{Dvali:2000hr}. Specifically it was defined as the graviton mass $m \rightarrow 0 $ double scaling/decoupling limit of the DGP model in which the dynamics of the helicity-two modes decouple, $M_{\rm Planck} \rightarrow \infty$, whereas the low energy gravitational dynamics of the helicity-zero mode $\pi$ survive. In this limit an accidental non-linearly realized Galileon symmetry $\pi \rightarrow \pi + c + v_{\mu} x^{\mu}$ survives as a remnant of the higher dimensional diffeomorphism symmetry. Related to this, $\pi$ may be viewed as a Goldstone model for a spontaneously broken higher dimensional spacetime symmetry group (see Sec.~\ref{sec:coset}). \\
 
The authors of \cite{Nicolis:2008in} then generalized this structure to the most general scalar theory consistent with the same non-linearly realized Galileon symmetry which was intended as an effective description of some unknown infrared completion of gravity. It was subsequently realised that the entire structure of Galileon theories arise naturally as decoupling/scaling limits of Massive Gravity theories \cite{deRham:2010ik}, i.e. `hard mass' Massive Gravity theories (like `soft mass' DGP) are examples of IR completions to Galileons, and this structure was important for determining the unique ghost-free theories of massive gravity \cite{deRham:2010kj}. The Galileon structure has since been shown to arise in a whole host of infrared modifications of gravity which rely on giving a hard or soft mass to the graviton in a Lorentz invariant way (see \cite{deRham:2014zqa} for a review on some of these approaches). More recently, the Galileon has been shown to play a special role in soft limits of scattering amplitudes \cite{Cheung:2014dqa}, and the full tree-level scattering amplitudes of a special Galileon have been given in \cite{Cachazo:2014xea} using a novel `dimensional reduction' from $d+d$ dimensional Einstein gravity. This special Galileon has been shown to admit an additional symmetry (beyond the Galileon symmetry) in \cite{Hinterbichler:2015pqa}.
 \\

All of these developments have taken place at the level of the Low Energy Effective Field Theory (LEEFT) (or tree-level S-matrix \cite{Cachazo:2014xea}) with the understanding that some UV completion exists. However at the same time, by treating the Galileon decoupling limit as an ordinary field theory, doubt has been cast on the existence of any such UV completion \cite{Adams:2006sv}. In a standard field theory, the key properties are that it exhibits a UV completion whose S-matrix is Unitary, Analytic, Lorentz Invariant and Polynomially Bounded (in the form of the Froissart bound). Indeed any field theory satisfying the Wightman axioms will automatically satisfy polynomial boundedness and analyticity of the S-matrix, at least for fixed physical momentum transfer \cite{:1900qta,PhysRev.109.2178,Epstein:1969bg}.
It is easily shown that the cubic Galileon violates one (or more) of these properties \cite{Adams:2006sv} and that same argument extends to all Galileon models, and by extension Massive Gravity \cite{deRham:2010kj} and its Multi-gravity extensions \cite{Hassan:2011zd,Hinterbichler:2012cn}.  \\

We will argue here and elsewhere \cite{KeltnerTolley2} that it is the polynomial boundedness assumption that is not appropriate for Galileons\footnote{Polynomial boundedness is the assumption that the growth of the scattering amplitude as a function complex momenta $k_i$ is bounded by a polynomial, i.e. $A(\{ k_i \}) < C |\sum_i |k_i||^N$. Polynomial boundedness follows automatically for theories in which the Wightman functions are tempered distributions, however it also follows for all strictly localizable fields \cite{Epstein:1969bg} (at least for fixed physical momentum transfer) for which the Wightman functions grow slower than a linear exponential $W(\{ k_i \}) < C e^{A |\sum_i |k_i||}$.}. 
Our assumption will then be that the Wightman functions are no longer tempered distributions, which amounts to the UV description of Galileons admitting some degree of non-locality.
In fact it is generally believed that General Relativity and string theory violate polynomial boundedness, at least for fixed angle scattering or unphysical momentum transfer, in the region of small impact parameters for which black hole production can take place. This gravitational non-locality is intimately connected with the production of black holes controlling the high energy properties of scattering amplitudes and with the absence of off-shell local observables in a gravitational theory. We will argue that Galileon theories are in this sense fundamentally gravitational, with the Galileon duality playing the same role as diffeomorphism invariance in forbidding local off-shell observables.
\\

The idea that a theory whose infrared limit does not contain any dynamical metric and is a local effective field theory, should nevertheless be viewed as a gravitational theory, in the sense that it exhibits some high energy non-locality, is unusual, but is not new. Another well known example are Little String Theories (LSTs). Despite being field theories in the infrared limit, their UV description will not be a field theory with a UV fixed point. We will elucidate these points below. In \cite{Kapustin:1999ci} it is argued that if their exists a field theory description of LSTs, then they are Jaffe type field theories of quasi-local type \cite{Iofa:1969fj,Iofa:1969ex}. Here we will argue that analogous properties are true for Galileons, i.e. that the UV Galileon fields are Jaffe fields of non-localizable type \cite{Steinmann:1970cm}, with the only distinction from LSTs being the degree of non-localizability as defined by the order of growth of the \KL spectral density. A related example, which we argue has many properties similar to out conjectured UV description of Galileons, is the `asymptotic fragility' proposal for the description of infinitely long strings \cite{Dubovsky:2012wk,Cooper:2013ffa}.
\\

In short, we conjecture that the UV description of Galileons, and by extension all theories of Massive Gravity, Bi-gravity and Multi-gravity violates the condition of polynomial boundedness, but they respect Unitary, Lorentz Invariance and crucially Analyticity (the limitations imposed by the latter will be discussed elsewhere \cite{KeltnerTolley2}). The main arguments of this paper are that:

\begin{itemize}

\item We give evidence for the violation of polynomial boundedness by computing the exact \KL spectral density of a specific Galileon and demonstrating its exponential growth.  To do this we use the Galileon duality to define a specific quantization of a Galileon theory which is dual to a free theory. This fully quantum equivalence between non-localizable field theory and a localizable theory, which extends to the full S-matrix \cite{KeltnerTolley2} is an example of the Borchers equivalence, for which the S-matrix and asymptotic states are identical \cite{Taylor:1973nf}. For a Galileon in $d$ dimensions we find $\rho(E) \sim e^{(E/\Lambda)^{(d+2)/(d+1)}}$. 

\item The exponential growth of the spectral density is consistent with semi-classical arguments for which the classical action scales as $|S| \sim (E/\Lambda)^{(d+2)/(d+1)}$ demonstrating that much of the high energy behavior can be anticipated from semi-classical considerations \cite{Dvali:2010jz,Dvali:2010ns,Dvali:2011nj,Dvali:2011th}. It arises because the spectral density is dominated by intermediate states with a finite number of particles $N$ for which $ N \sim (E/\Lambda)^{(d+2)/(d+1)}$.

\item As a consequence of the exponential growth in the spectral density, we argue that Galileons fall into Jaffe's class of non-localizable field theories. We further conjecture that all field theories exhibiting the Vainshtein mechanism are non-localizable field theories.

\item We argue that Galileons exhibit a locality bound in space and time which is a property of all non-localizable field theories. This is connected with the fact that correlation functions are no longer tempered distributions and must be defined using a space of test functions which includes no functions of compact support. 

\item Although Galileons break strict micro-causality/locality, we argue that they nevertheless preserve macro-causality/locality. We give a prescription that defines macro-causality for the S-matrix and two-point correlation function out of vacuum, and argue that this implies the absence of superluminal propagation seen in the LEEFT in the UV theory.

\item It is known that non-localizable theories can exhibit standard properties, such as cluster decomposition, CPT, and an LSZ formalism \cite{Steinmann:1970cm}, but are non-Wilsonian in the sense that they exhibit no UV fixed point, and no local operator product expansion.

\item We give a Lorentz invariant prescription for defining time ordered correlation functions and generalized retarded products in terms of Wightman functions. This prescription makes use of entire analytic functions of momenta having the same order of growth at large complex momenta as the Wightman functions.

\item The absence of off-shell, local observables is consistent with the viewpoint that Galileon theories are fundamentally gravitational. This is consistent with their origin as decoupling limits of resonance (soft) or hard mass gravity theories, and the existence of the Galileon duality transformation which is simply a remnant of diffeomorphism symmetry that survives in the decoupling limit. 

\item Our picture of the UV properties of Galileons is consistent with the `asymptotic fragility' proposal  \cite{Dubovsky:2012wk,Cooper:2013ffa}, in the sense there is no UV fixed point, and the `classicalization' proposal, in that sense that the high energy scattering is expected to be semi-classical \cite{Dvali:2010jz,Dvali:2010ns,Dvali:2011nj,Dvali:2011th} and scattering should be dominated by the production of $N$-particle states with $N \sim (E/\Lambda)^{(d+2)/(d+1)}$.

\end{itemize}

This proposed picture for the UV properties of Galileons is radically different than standard low energy effective field theory expectations based on the idea that the Galileon strong coupling scale $\Lambda$ is the cutoff of the effective field theory (EFT). In particular, no new particle states need to arise at the scale $\Lambda$ in order to resolve unitarity. Rather, $\Lambda$ is a fundamental scale, intrinsic to the definition of the UV completion and the properties of the non-localizable fields\footnote{We shall continue to use the historical terminology `non-localizable' rather than `non-local' to distinguish these models from other distinct non-local field theories found in the literature. For instance a common approach in non-local field theories is to modify the Feynman propagator by an entire function without introducing an additional pole with the hope of improving Euclidean perturbation theory. From our perspective such a modification violates unitarity since the associated Wightman functions do not satisfy the positivity requirement of the \KL spectral representation, whereas all the Jaffe type non-localizable theories respect unitarity in the form of a positive \KL spectral representation. Furthermore we make no claims for improved perturbation theory, here the non-localizability only arises in the non-perturbative regime.
}, a role played analogously to the Planck scale in quantum gravity. There is no UV renormalization group fixed point, i.e. no conformal behavior at high energies. \\

Nevertheless, at energies $E \ll \Lambda$, the conjectured UV completion will be consistently described by the Galileon LEEFT. In a perturbative expansion in $E/\Lambda$, the exponential growth becomes polynomial, thus recovering local physics at low energies and large distances. Specifically as long as we look at separations $ |x| \gg r_*(E)$, where $r_*(E)= 1/\Lambda (E/\Lambda)^{1/(d+1)}$ is the Vainshtein/classicalization radius, and $E<\Lambda$ the LEEFT will be satisfied. As soon as $E>\Lambda$, we have $r_*(E) >1/\Lambda$ and there will be a region even at distances $>1/\Lambda$ where the LEEFT locality properties will be violated and there will be some manifestation of the non-localizability. These properties are analogous to the those expected for GR, as we discuss in Sec.~\ref{Gravity}.

\section{Galileons as Effective Field Theories and Beyond}

The traditional effective field theory (EFT) point of view on Galileons can be summarized by the usual EFT paradigm: {\it identify the low energy degrees of freedom and write down every local operator consistent with the low energy symmetries}. In the case of Galileons the symmetries are linearly realized Poincar\'e invariance and the non-linearly realized Galileon and shift symmetries $ \pi \rightarrow \pi + v_{\mu}x^{\mu} + c$. These symmetries may be formalized via the coset construction (see Sec.~\ref{sec:coset}) but the simplicity of the Galileon symmetry means that it is easy to write down the form of the most general interactions without this construction. \\

There are two classes of interactions that arise for the Galileon, those built out of the manifestly Galileon invariant combination $\Pi_{\mu \nu} = \partial_{\mu} \partial_{\nu} \pi/\Lambda^{\sigma}$ and a finite number of special Galileon interactions whose Lagrangian transforms as a total derivative under the Galileon transformation. Thus we may write the Wilsonian effective action for the Galileon as
\be
S_{\rm Galileon} = S_{\rm A} + S_{\rm B} 
\ee
where $S_{\rm A} $ are the lower derivative `Wess-Zumino' combinations (index contraction suppressed)
\be
S_{\rm A} = \int \d^d x  \Lambda^\sigma \pi \sum_{n=0}^d \alpha_n \epsilon \, \epsilon \, \Pi^n \eta^{d-n}
\ee
($\sigma = d/2+1$), $\epsilon$ is the Levi-Civita tensor, and $S_{\rm B} $ contains the infinite number of manifestly Galileon invariant combinations which we write schematically as 
\be
S_{\rm B} =  \int \d^d x \Lambda^{\sigma+1} \sum_{n,m} \beta_{n,m}(\Lambda_W/\Lambda) \frac{\partial^{2n}}{\Lambda^{2n}} \Pi^m \, ,
\ee
and a term of the form $\partial^{2n} \Pi^m$ includes all scalar contractions of all combinations of derivatives applied to different $\Pi$'s. There exists a non-renormalization theorem that $\pi$ loops to not renormalize $S_A$ \cite{2003JHEP...09..029L,2004JHEP...06..059N}, however the terms $S_B$ are renormalized. The non-renormalization theorem implies that the strong coupling scale $\Lambda$ and coefficients $\alpha_n$ do not flow, but it does not improve the regime of validity of the perturbative LEEFT. In a Wilsonian cutoff approach, all terms $\beta_n(\Lambda_W/\Lambda)$ must be included and will depend on the Wilsonian cutoff $\Lambda_W$.
\\

As is well known, the special finite number of contributions $S_{\rm A} $ lead to second order equations of motion. For this reason cosmological applications of Galileon theories usually focus on these finite number of terms. However, there is no requirement in an EFT that we should truncate to only those operators that have second order equations of motion. Rather all operators consistent with the symmetries should be included, and the would be Ostrogradski ghosts associated with the higher derivatives will have masses at what would then be the cutoff of the EFT $\Lambda_{\rm cutoff} \sim \Lambda$. Thus they lead to no unitarity violation in the regime of validity of the EFT and all higher derivative operators can be dealt with in a perturbative sense. \\

Assuming all the coefficients $\alpha_n$ and $\beta_{n,m}$ are of order unity, then the regime of validity of the EFT theory may be taken to be
\ba
\label{cond}
&& \partial \ll  \Lambda \\
&& \Pi \ll  1 \, .
\ea
We do not require $\pi \ll \Lambda$ because of the Galileon and shift symmetries. To be precise as long as the shift symmetry is exact or only approximately broken, a large vev for $\pi$ will not generate large quantum corrections as long as the above conditions (\ref{cond}) are met. \\
 
This regime may be qualitatively understood as the regime in which we expect the expansion of the effective action to converge. In reality, all EFT expansions are asymptotic and so have technically zero range of convergence, however this regime is the one in which we may reliably truncate the expansion to the first few terms.

\subsection{Pushing the boundary of the EFT}

Although the previous discussion accounts for the regime allowed by a traditional effective field theorist, it has been argued that we may be able to push beyond this scale. A simple counting of powers shows that all the terms in $S_{\rm B}$ have at least one more power of derivatives than $S_{\rm A }$ (see for example \cite{Endlich:2010zj}), i.e.
\be
S_{\rm B} \sim \frac{\partial}{\Lambda} S_{\rm A } \, .
\ee
Thus as long as $\partial \ll \Lambda$ it is conceivable that we may focus only on the  $S_{\rm A }$ part of the action, at least for determining semi-classical solutions. It may be plausible that it is even possible to have classically $\Pi \sim 1$ with the derivatives providing the expansion parameter necessary to make sense of the EFT. This is a rather modest improvement of the range, but does not lead to any significant changes in the phenomenology. \\

A more radical proposal goes under the umbrella of the {\bf Vainshtein mechanism} \cite{Vainshtein:1972sx}, and is central to the phenomenological viability of Galileon and massive gravity theories. Here the idea is that we may plausibly take backgrounds for which $\Pi \gg 1$ and still remain in the regime of validity of the EFT\footnote{For this it is essential to distinguish the strong coupling scale $\Lambda$ from the cutoff of the EFT.}. The argument for this is based on the following reasoning: First take the effective action to include only those terms from $S_{\rm A }$. Then determine a classical background solution which follows from the equations of motion for $S_{\rm A }$ which allows for a Vainshtein or strong coupling region for which $\Pi \gg 1$. This background generically renormalizes the kinetic term for fluctuations in such a way that the effective strong coupling scale becomes redressed to something which is parametrically larger than $\Lambda$
\be
\Lambda_{*} \sim Z^{\beta} \Lambda \, ,
\ee
where $\beta$ is a positive model dependent parameter and $Z\gg 1$ is the effective wave function renormalization \cite{Nicolis:2004qq,Brouzakis:2014bwa,deRham:2014wfa}. In this way we may argue that around a background with $\Pi \gg 1$ that the cutoff of the effective field theory is really $\Lambda_*$ and so we may now push into the new regime of validity
\be
\partial \ll \Lambda_* \,.
\ee
If $\delta \pi$ denotes the fluctuations around the background solution, then we may similarly argue that we can trust the reorganized EFT in the regime
\be
\partial \partial \delta \pi \ll \Lambda_*^\sigma \, . 
\ee

\subsection{Problems with the EFT description}

The Vainshtein proposal requires a very special behaviour of the UV completion of the Galileon theory. In a generic EFT, we would expect the scale $\Lambda$ to be associated with the mass of some new degree of freedom. If this new degree of freedom exists at this scale then we would expect that unitarity would require us to include it if we begin looking at energies up to the scale $\Lambda_* >\Lambda$. The Vainshtein mechanism implicitly assumes that this new degree of freedom does not need to be included which is generically false. \\

To see this more concretely we just need to do the previous calculation in a different order. Suppose we begin with the total action for the Galileon including $S_{\rm B}$. We then look for a background solution of this action which exhibits $\Pi \gg 1$ and perturb around it. Even after we have perturbed around this solution there will exist the following infinite set of contributions to the effective action which are schematically
\be
\delta_2 S_{\rm B} \sim  \int \d^d x \Lambda^{\sigma+1}  \sum_n \beta_{n,2} \frac{\partial^{2n}}{\Lambda^{2n}} \delta \Pi^2\ .
\ee
Even with the wave function renormalization, the expansion of these infinite set of terms will breakdown at the scale $\partial \sim \Lambda$. In fact it is precisely these terms which would be associated with a modification to the Galileon propagator due to the existence of other massive poles. The point is that wave function renormalization may change the scale of interactions but they do not change the position of poles in the propagator. A resolution to this is to imagine that the UV completion is such that $\beta_{n,2}$ all happen to vanish (or be significantly smaller in magnitude). This is conceivable, but implies an extremely nontrivial requirement on the UV theory for it to be consistent with the Vainshtein mechanism at low energies. \\

A perhaps more serious problem to the validity of the Galileon EFT description both at strong and weak coupling is the fact that it can be shown that no theory whose UV completion satisfies Lorentz invariance, unitarity, locality (polynomial boundedness) and causality (analyticity) can give rise to a Galileon EFT at low energies \cite{Adams:2006sv}.\footnote{The original arguments were made for the cubic Galileon model, but it may be easily show that these apply to the generic Galileon models.} Since every local quantum field theory satisfies these requirements there can be no local field theory for which Galileons are the low energy EFT, which means they are arguably not of interest as EFTs! It is conceivable of course that they admit a Lorentz violating UV completion (for suggestions along these lines for massive gravity see \cite{Blas:2014ira}), but then there was no reason to write down a Lorentz invariant theory in the first place. In fact the space of Lorentz violating Galileon theories is much larger due to the weakening of the symmetries.  As a result of these arguments many people have viewed Galileon theories, and by extension massive gravity theories, as sick from the get go. It is our intention here to show that such a conclusion would be too quick, and results from interpreting Galileon type theories as local field theories, rather than intrinsically gravitational theories.

\subsection{Beyond Effective Field Theories?}

Although not immediately apparent, the two previous problems are connected. It may be shown that in every theory known so far which exhibits classically the Vainshtein mechanism,  the S-matrix is such that it does not admit a standard `Wilsonian' or `local field theory' UV completion at the scale $\Lambda$. Thus the Vainshtein mechanism appears to be incompatible with a standard assumption that the scale $\Lambda$ is associated with new states in a {\bf local} field theory UV completion. This tells us that in these classes of theories we should view the UV completion in different terms \cite{Dvali:2012zc}. It will be our contention here that Galileons can be understood as a class of field theories considered in the 1960/1970's which are known as `non-localizable' field theories. These field theories are not Wilsonian in the sense that they are not defined via UV RG fixed points, and their operators do not satisfy a standard operator product expansion, they may nevertheless satisfy other standard properties such as cluster decomposition, analyticity, LSZ construction, crossing symmetry and, CPT symmetry \cite{Steinmann:1970cm}. Before getting to a discussion of this perspective we first clarify a number of arguments that have been made pertaining to the UV properties of these theories. \\

One resolution to the problems of the EFT of Galileons is to imagine that viewing Galileons as EFT theories, whilst not technically inconsistent in the EFT safe region $\partial \ll \Lambda$, $\Pi \ll 1$, is nevertheless fundamentally misleading since it assumes a standard UV completion at the scale $\Lambda$ which is known not to exist. Thus a more radical proposal, but one which is consistent with the Vainshtein mechanism, is that $\Lambda$ is not the cutoff of the Galileon theory, but a strong coupling scale above which the theory self-unitarizes without the introduction of new degrees of freedom\footnote{By self-unitarization we mean that there are no new poles in the propagator.}. This is the essence of the `UV completion via Classicalization' idea of \cite{Dvali:2010jz,Dvali:2010ns,Dvali:2011nj,Dvali:2011th} and we shall see that many aspects of this picture are consistent with the properties of non-localizable field theories. To understand these ideas it is worth commenting on what we mean by the breakdown of the EFT. \\

Let us for now focus on the finite Galileon terms $S_{\rm A }$ and ask the question, at what scale does unitarity breakdown? There is a well-known criterion for this, tree-level unitarity breaking, which proceeds as follows. It is usual to decompose the $S$-matrix in the form $\hat S = 1+ i \hat T$ and then compute the momentum conservation stripped transfer matrix $\langle f | \hat T | i \rangle = (2 \pi)^4 \delta^4(p_f-p_i) A(i \rightarrow f)$. Unitary is encoded in the optical theorem
\be
i (T^{\dagger}-T) = T^{\dagger} T
\ee
which between states amounts to the generalized optical theorem schematically given as
\be
i ( A(i \rightarrow f)^* - A(f \rightarrow i) ) = \sum_n (2 \pi)^4 \delta^4(p_n-p_i) A(i \rightarrow n) A(n \rightarrow f)  \, .
\ee
These equations impose bounds on the magnitude of the amplitude. Specifically, for $2 \rightarrow 2$ scattering, expressing the amplitude in partial waves
\be
A(s,t(\cos \theta)) = \sqrt{\frac{s}{s-4 m^2}} \sum_{l=0}^{\infty} (2 l+1) P_l(\cos \theta) \frac{1}{2 i} (a_l(s)-1) \, ,
\ee
then unitarity amounts to the condition that $|a_l(s)| \le 1$. \\

When computing the scattering amplitude $A(s,t,u)$ for $2 \rightarrow 2$ elastic scattering to tree-level for a Galileon, we would find in terms of the standard Mandelstam variables, schematically $A(s,t,u) \sim (s^3+t^3+u^3)/\Lambda^6$. Expressing this in terms of partial waves this may easily be shown to violate the optical theorem when $\sqrt{s} \sim \Lambda$ (assuming $\Lambda \gg m$)\footnote{We note that although the Galilon symmetry forbids a mass term $m$, since this is only a global symmetry we expect a small mass to be present. In particular in the context of massive gravity the Galileon arises as a massive particle. Including a small mass is also important for making use of analyticity arguments since it guarantees a small window in the Mandelstam plane where the amplitude is analytic around the real axis which is needed to prove Hermitian analyticity \cite{KeltnerTolley2}.}. For this reason it is usual to regard $\Lambda$ as the cutoff of the Galileon EFT, meaning that we cannot reliably trust the EFT at and above this scale since without new physics coming to the rescue it seems sure to violate unitary. \\

However this violation of the optical theorem is misleading for the following reason. We can equally well in perturbation theory express the $S$-matrix in terms of the reaction/reactance/Heitler matrix $\hat K$ as 
\be
\hat S = \frac{1- i \hat K/2}{1+i \hat K/2} \, .
\ee
Similarly $\hat K$ may be expressed in a momentum conservation stripped form 
\be
\langle f | \hat K | i \rangle = (2 \pi)^4 \delta^4(p_f-p_i) {\cal K}(i \rightarrow f) .
\ee
In this language, unitarity and the optical theorem is simply the statement that $\hat K$ is Hermitian which amounts to the linear condition 
\be
{\cal K}(i \rightarrow f)  = {\cal K}(f \rightarrow i)^* \, . 
\ee
If instead of computing perturbative corrections to $\hat T$ we compute them directly to $\hat K$, then at each order in perturbation theory we will find that $\hat K$ remains Hermitian\footnote{We may easily reorganize the standard perturbation theory expansion to directly compute corrections to $\hat K$. For instance, in the interaction picture, with $\hat H_I$ the interaction, explicitly to second order in perturbations $\hat K$ is given by
\be
\hat K =  \int_{-\infty}^{\infty} \d t \hat H_I(t) - \frac{i}{2}  \int_{-\infty}^{\infty} \d t \int_{-\infty}^{\infty} \d t'  \theta(t-t') [ \hat H_I(t), \hat H_I(t')] + \dots\, .
\ee
This is reminiscent of the {\it in-in} formalism in that non-time ordered correlators arise through the commutator.  This expression is manifestly Hermitian due to the properties of the commutator. The exact non-perturbative expression for $\hat K$ follows from the two equations \cite{Lippmann:1950zz}
\ba \hat K = \int_{-\infty}^{\infty} \d t \hat H_I(t) \hat V(t) \,  \, ,
 \,  \hat V(t) = 1 - \frac{i}{2} \int_{-\infty}^{\infty} \d t' \epsilon(t-t') \hat H_I(t') \hat V(t') \, ,
\ea
where $\epsilon(t-t') = \theta(t-t')-\theta(t'-t)$. Indeed the largely forgotten goal of the Lippmann-Schwinger paper \cite{Lippmann:1950zz} was to give a variational technique to compute $\hat K$ non-perturbatively.}. This follows automatically for any theory whose Hamiltonian is Hermitian. From this Hermitian $\hat K$ we may then reconstruct the unitary $\hat S$. Thus unitarity is never violated at any finite order in perturbation theory. From the perspective of the Feynman diagram expansion this may be viewed as a resummation of an infinite number of diagrams corresponding to the expansion 
\be
\hat S =({1-i \hat K/2}) \sum_{n=0}^\infty 2^{-n}(-i)^n \hat K^n \, . 
\ee
Truncating this expansion would imply a loss of unitary at the same scale, but there is no need to truncate, nor is there any need to phrase this in terms of the Feynman diagram expansion. This resummation may be performed explicitly at the level of the partial wave amplitudes and is straightforward to do in the elastic limit, or almost elastic limit in which only a finite number of intermediate particles contribute\footnote{We note in passing that precisely such a resummation was proposed as a means to deal with the perturbative expansion of nonrenormalizable field theories within the Efimov-Fradkin/superpropagator approach for which the real part of the perturbative amplitude grows exponentially \cite{Isham:1970rg}.}.  \\

An explicit example of this unitarization via resummation is the self-healing or self-unitarization mechanism of \cite{Aydemir:2012nz}. There it is shown that in QCD with arbitrary numbers of flavours $N_f$ and colours $N_c$, it is possible to create a hierarchy using $N_c$ and $N_f$ such that tree-level unitarity breaks down for the low energy Chiral Perturbation Theory (ChPT) EFT at scales below the actual cutoff of the ChPT EFT, i.e. the scale of new physics a.k.a. QCD. This may be resolved by computing $\hat K$ (in its partial wave representation), combining to give $\hat S$ and showing that unitarity is maintained. At a practical level this corresponds to resumming the partial wave amplitudes into a manifestly unitary form.
However there is physics in this resummation. In the $\hat K$ representation of the $S$-matrix it is clear that additional poles will generically arise already in perturbation theory due to the denominator. In the case of ChPT, the additional pole that arises is on the second Riemann sheet and so may be associated with a physical resonance. Thus while some new physics does arise at the scale of tree level unitarity violation, it is new physics described by the existing low energy EFT. 

Of course this is certainly not a universal cure, and if a pole arises on the physical sheet it can potentially lead to the emergence of a negative norm (ghost) state or failure of analyticity. The point is that {\bf the violation of tree level unitarity is not synonymous with the breakdown of the EFT or the need for new fundamental d.o.f.} as is often assumed\footnote{Although in many cases it is correct - e.g. the Higgs mechanism has now been confirmed as the correct solution for the unitarity violation of the Proca-Yang-Mills version of the standard model.}. Neither does it have to be the occurrence of a single resonance that cures the problem, but it could be composite resonant/bound multi-particle states.
\\

In summary then, we can never truly diagnose the breakdown of unitary using perturbation theory, we can only diagnose the breakdown of perturbation theory itself. If a theory has a Hermitian Hamiltonian, there is no logical impediment to the theory being unitary non-perturbatively despite the fact that it may violate tree-level unitarity at some scale. Some new physics does arise at the scale of tree-level unitarity violation, for instance a physical resonance, be it sharp or a composite metastable multi-particle state. However the description of this new physics may only require the existing EFT. If something like this happens for Galileon theories, then this resolves our apparent problem with incorporating the Vainshtein mechanism quantum mechanically. \\

\subsection{Absence of UV fixed point}

The idea that a theory can make sense of itself non-perturbatively despite being perturbatively non-renormalizable is part of the asymptotic safety program \cite{Weinberg:1980gg}\footnote{For discussions related to the present case see \cite{Kovner:2012yi,Codello:2012dx,Brouzakis:2013lla} and for different approaches \cite{Vikman:2012bx}}. In essence, one assumes the existence of a UV fixed point around which we can define the theory. For instance naively one may argue that for Galileon theories, at high energies we can suppose that the highest order Galileon operator in $S_A$ dominates. If one such operator dominates then the action is scale invariant, however not under the scaling associated with the engineering dimensions. \\

In $d$ dimensions, if an $n$-th order Galileon operator dominates at high energies and this defines the theory near a UV fixed point, then the conformal weight $\Sigma$ defined via 
\be
\langle \pi(x) \pi(y)  \rangle \sim \frac{1}{|x-y|^{2 \Sigma}} \, ,
\ee
will be $\Sigma = (d+2-2n)/n$. However unitarity, expressed through the need to have a well-defined \KL spectral representation, i.e. a unitary representation of the conformal group, requires for scalars $\Sigma \ge d/2-1$, i.e. the conformal weight must always be greater than that for a free scalar field in the appropriate dimension \cite{Mack:1975je,Grinstein:2008qk}. This inequality becomes 
\be
2 \ge n \, ,
\ee
which cannot be satisfied except for the trivial case $n=2$ which is just the usual free field.  In fact in the case of the $(d+1)$'th Galileon $n=d+1$ we shall perform an explicit computation of the Wightman function in Sec.~\ref{sec:Wightman} and we shall find an entirely different unitarity behaviour with no evidence of a UV fixed point. This is equivalent to the statement that the Wightman functions will not be tempered distributions, i.e. their Fourier transform will not be polynomially bounded.\\

We thus see that there is a fundamental conflict between the desire for a unitary theory and the existence of a UV fixed point. Our perspective will be quite different, if Galileons can be made sense of non-perturbatively without the introduction of additional degrees of freedom (by which we mean fundamental - resonance or bound state degrees of freedom are acceptable), then we still anticipate the length scale $\Lambda$ to control the high energy behaviour, i.e. there will be no UV fixed point and the Wightman functions will not be tempered. This is more typical of truly gravitational theories where, for instance, both the string scale and Planck scale determine the properties of the high energy theory and there is no sense in which these are UV fixed points. Analogous systems which are argued to be UV finite are Little String Theories  \cite{Kapustin:1999ci} where despite being `field theories' their UV behaviour retains remnants of their stringy behaviour and there is no UV fixed point. In fact, as we discuss later, our perspective is that Little String Theories are the most closely related to our conjectured behaviour of the UV properties of Galileons. More generally this idea of UV completion without a UV fixed point is referred to as `asymptotic fragility' in the scenario constructed in \cite{Dubovsky:2012wk,Cooper:2013ffa}, to distinguish it from asymptotic safety.

\subsection{Classicalization, Vainshtein and Galileons}

The `classicalization' proposal \cite{Dvali:2010jz,Dvali:2010ns,Dvali:2011nj,Dvali:2011th,Dvali:2014ila}  is another such proposal of how a theory may self-unitarize above its strong coupling scale $\Lambda$. Specifically it proposes that the scattering of particles at energies $E \gg \Lambda$ is dominated by the emergence of semi-classical configurations, `classicalons', which ultimately decay into many soft quanta \cite{Alberte:2012is}. The typical number of such quanta $N$ scales as $N \sim E \, r_*(E)$, where $r_*(E)$ is the Vainshtein/classicalization radius defined below. 
The distinction between this and the self-healing mechanism \cite{Aydemir:2012nz} is that the new states which emerge are not single resonances but quasi-classical bound multi-particle states. These are nevertheless not new degrees of freedom.
This picture is modeled after the `known' behaviour of gravity whereby high energy scattering will produce `messy' black holes which in turn go through the process of balding and ultimately evaporation. Galileon field theories satisfy precisely the conditions argued to lead to such a phenomena. That such an analogous behaviour can occur for a field theory such as Galileons is less surprising once we view the Galileons as the helicity-zero graviton in massive gravity. We shall go further and argue that any theory that classicalizes must have an inherent gravitational non-locality which is ultimately responsible for its different high energy behaviour. \\

We will find in subsequent sections arguments that support this picture using the Galileon duality. Specifically we will find that the high energy behaviour of Galileon fields is dominated by their semi-classical contributions to the path integral as manifested in the exponentially growing density of states $\rho \sim e^S \sim e^{E r_*(E)}$ for the Galileon operators. This behaviour is  consistent with the emergence of configurations with many soft-quanta whose `entropy' scales as $E r_*(E)$ where $r_*(E)$ is the Vainshtein/classicalization radius. Concretely the peak of the resonant part of the spectral density is from finite number of particle states for which $N \sim E r_*(E)$.
These properties rest on a nontrivial UV/IR mixing due to the Vainshtein mechanism which connects high energy scattering amplitudes with large semi-classical configurations. We attribute these properties to a fundamental gravitational non-locality that arises in these theories. \\

In Galileon theories, in common with the classicalization arguments \cite{Dvali:2010jz,Dvali:2010ns,Dvali:2011nj,Dvali:2011th}, associated with an incoming state of energy $E = \sqrt{s}$ we may define a radius $r_*$ which characterizes the region of strong coupling for scattering. This is none other than the Vainshtein radius for the system. To see this we note that for Galileons in four dimensions, the Vainshtein radius is usually given in terms of the mass of some external source which creates the background configuration as 
\be
r_V = \Lambda^{-1}\left( \frac{M}{\mpl} \right)^{1/3} \,. 
\ee
However if we compute the classical self-energy in the Galileon configuration associated with this source we find a finite contribution of the form 
\be
E \sim r_V^{-1} \left( \frac{M}{\mpl}\right)^2 \, .
\ee
That this self-energy is finite is the first clue to the better UV properties of these theories.
Rearranging these formula we find
\be
r_V(E) \sim \Lambda^{-1}\left(\frac{E}{\Lambda}\right)^{1/5} \, ,
\ee
which in the language of \cite{Dvali:2010jz}, is the classicalization radius $r_*(E)=r_V(E)$.  In a generic Vainshtein/classicalizing theory we may parameterize this as $r_*(E) = \Lambda^{-1}\left(\frac{E}{\Lambda}\right)^{2 \alpha-1} $ with $\alpha>0$ (here we use the notation $\alpha$ to denote the order of the spectral density to be defined later). In general dimensions $2 \alpha-1= 1/(d+1)$ for Galileons.
We shall continue to refer to this as the Vainshtein radius understanding that it is being expressed in terms of the energy of the Galileon field and not the energy of any potential source for the Galileon. It will become apparent that this is a more appropriate parameterization and it is independent of how we choose to couple to the Galileon to other fields.
\\

We know that classically the Vainshtein mechanism works in the sense that whenever an energy $E$ is localized in a region less than $r_*(E)$ the associated Galileon field will be highly non-linear, i.e. strongly coupled. In the context of massive gravity, the Vainshtein radius is for the helicity-zero mode of the graviton what the Schwarzschild radius is for the helicity-2 mode, namely the scale at which nonlinearities become important. This also means that in massive gravity, the effective size of the gravitational field around a massive source, e.g. a black hole, is set not by the Schwarzschild radius, but by the Vainshtein radius, at least from the point of view of scattering Galileons or particles coupled to the Galileon field. Particles which are not directly coupled to the Galileon will continue to see the Schwarzchild radius as the effective size.
\\

Although the Vainshtein radius is usually determined for spherically symmetric sources, it is easy to make a scaling argument that in any system, for which an energy $E$ is localized in a region of size $r_V$, then $r_V$ is the radius which determines the onset of strong coupling. Note that it is the localization of energy that is important for this argument. For instance for planar symmetric solutions of energy $E$ nothing special happens when scattering at these energy scales \cite{Akhoury:2011en}. However, this is a consequence of the fact that in these solutions the energy is not localized in all spatial directions. \\

In particular, this means that in scattering two high energy Galileons with incoming energy $\sqrt{s}$, when the impact parameter $b$ becomes comparable to $r_*({s})=r_V(s)$ we expect strong quantum non-linearities to kick in  \cite{Dvali:2010jz,Dvali:2010ns,Dvali:2011nj,Dvali:2011th}. A similar argument is made in the context of GR, when two particles are scattering at an impact parameter less than or equal to the Schwarzschild radius associated with the incoming state $r_*=R_S({s})$ then we expect a Black Hole to form. In the case of gravity, if the incoming particles have ultra-planckian energies $\sqrt{s} \gg \mpl$, then the scattering will become nonlinear at a length scale which is much bigger than the Planck length $R_S \gg \mpl^{-1}$. The larger the energy, the larger the distance at which scattering is nonlinear. However this implies a UV/IR mixing we would not expect to arise in any local field theory. In the case of GR this is attributed to the fact that it is a gravitational theory, and gravitational theories have a necessarily weaker notion of locality. Even away from the Black Hole production region, it is well known that ultra-planckian scattering in the eikonel regime is well described semi-classically using Einstein gravity regardless of the precise UV completion of gravity above the Planck scale \cite{'tHooft:1987rb}. 

\subsection{Loop counting parameter}

The emergence of semi-classical behaviour at high energies can be anticipated by a simple scaling argument.
Consider the leading order Galileon interactions in $d$ dimensions ($\sigma = d/2+1$)
\be
S_{\rm A} = \int \d^d x  \Lambda^\sigma \pi \sum_{n=0}^d \alpha_n \epsilon \,  \epsilon  \, \Pi^n \eta^{d-n} \, ,
\ee
and suppose we construct the S-matrix element between two coherent states 
\be
\langle \alpha | \hat S | \beta \rangle = \int \D \pi  \, e^{i S'_A } \, ,
\ee
where $S'_A = S_A +$ boundary terms associated with the coherent states. Associated with the incoming and outgoing coherent states we can define the energy by
\ba
&& E_{\alpha } = \int \d^{d-1} x \frac{1}{2 }\langle \alpha | \left( \dot \pi^2 + (\nabla \pi)^2 \right) | \alpha \rangle \, , \\
&& E_{\beta } = \int \d^{d-1} x \frac{1}{2 }\langle \beta | \left( \dot \pi^2 + (\nabla \pi)^2 \right) | \beta \rangle \, .
\ea
Let us suppose for simplicity that $E_{\beta} \sim E_{\alpha } \sim E$.
Next we define dimensionless space-time and field parameters,
\be
x=r_*\hat{x}\,,  \quad t = r_*\hat{t} \,,  \quad \pi=\sqrt{Z}\hat{\pi} \, .
\ee
The energy, $E$, dimensionally scales as
\begin{eqnarray}
E&\sim&\int d^{d-1}x\ \ \frac{1}{2}(\partial\pi)^2 \dots  = (r_*)^{d-3}Z  \int d^{d-1} \hat x\ \ \frac{1}{2}(\hat \partial \hat \pi)^2 \dots \,  ,
\end{eqnarray}
and so we may identify $E\sim (r_*)^{d-3}Z $.
All of the different terms in the Galileon action will become of the same order when $\partial \partial \pi \sim \Lambda^{\sigma}$. Reexpressing in terms of dimensionless variables this occurs when 
\be
\sqrt{Z} \frac{1}{r_*^2} \sim \Lambda^{\sigma} \, .
\ee
Solving for $Z$ ($Z= r_*^4 \Lambda^{2 \sigma}$) and substituting in the previous equation we find that the scale at which all interactions become comparable in fixed energy scattering is
\be
r_*(E) = \frac{1}{\Lambda} \left(\frac{E}{\Lambda} \right)^{\frac{1}{d+1}} \, ,
\ee
for any Galileon model.  As we have already emphasized this scale is precisely the Vainhstein radius expressed in terms of the energy of the Galileon field rather than the mass of the source which might create it. This is a more appropriate representation for the purposes of scattering processes. 

Finally, rewriting the action in terms of dimensionless variables we have
\begin{eqnarray}
S_A' &=&E \,  r_* \hat S_A' \, ,  \nn \\
&=& \left( \frac{E}{\Lambda}\right)^{\frac{d+2}{d+1}} \hat S_A'  \, .
\end{eqnarray}
$\hat S_A' $ is the dimensionless Galileon action with boundary terms (i.e. effectively with $\Lambda=1$ and $E=1$). This shows us that the proper loop counting parameter, i.e. the quantity playing the role of $\hbar$ is 
\be
L_{\text{lc}}=(\Lambda/E)^{\frac{d+2}{d+1}} \, ,
\ee
which decreases as energy increases.  In other words, in terms of dimensionless field and space-time variables, the path integral is
\be
\langle \alpha | \hat S | \beta \rangle = \int \D \hat \pi \, e^{i \left( \frac{E}{\Lambda}\right)^{\frac{d+2}{d+1}}  \hat S'_A } \, .
\ee
This suggests that quantum mechanics becomes less and less relevant as energy increases or more precisely that the semi-classical approximation to the path integral becomes better and better at high energies \cite{Dvali:2010jz,Dvali:2010ns,Dvali:2011nj,Dvali:2011th} (for similar observations from a different perspective see \cite{deRham:2014wfa}). This result is a manifestation of the UV/IR mixing implied by the Vainshtein mechanism. As it stands this argument is of course too simplistic since we have not demonstrated that however we resolve the UV divergences of the LEEFT will be consistent with this scaling. For instance, if we take the theory with a defined hard cutoff at $\Lambda_c$ then the rescaled theory will have a hard dimensionless cutoff at $r_*(E)\Lambda_c$. On the other hand perturbative unitarity continues to break down when the dimensionless energy is unity. If however the theory self-unitarizes by some mechanism, then this scaling will be correct.

\subsection{Including irrelevant operators, $S_{\rm B}$}

The previous argument was made including only those special Galileon terms in $S_A$ and one may expect it do be destroyed as soon as we include the infinite number of terms in $S_B$ which will generically be present in matching to a specific UV completion. However at this point the Galileon symmetry comes in useful! As we have already remarked, all the terms in $S_B$ have at least one additional derivative relative to those in $S_A$
\be
S_{\rm B} \sim \frac{\partial}{\Lambda} S_{\rm A } \, .
\ee
Naively one might think that then $S_{\rm B} \sim (E/\Lambda) S_{\rm A}$ and so at high energies $S_{\rm B}$ will dominate. However, in reality, in the semi-classical regime we should scale 
\be
\partial \sim \frac{1}{r_*(E)} \hat \partial
\ee
at least in determining the contributions to the semi-classical action which will determine the leading contribution to the phase shifts and absorption coefficients. In other words
\be
\frac{S_{\rm B}}{S_{\rm A}} \sim \frac{1}{\Lambda r_*(E)}  \rightarrow 0,  \quad \text{as } E/\Lambda \rightarrow \infty \, .
\ee
Thus not only do we expect that the high energy scattering to be semi-classical, we expect that the leading contributions to the scattering amplitudes to be determined entirely by the special Galileon operators $S_A$. This is closely analogous to the well-known behaviour of quantum gravity, that in certain limits, the leading contribution to the trans-Planckian scattering amplitude is determined entirely by the Einstein-Hilbert contribution to the action, regardless of the details of the UV completion \cite{'tHooft:1987rb,Banks:1999gd,Giddings:2007qq}. 

\subsection{Unitarization at High Energies and Non-polynomial boundedness}

\label{sec:unitarizationclassical}

How then do we expect the scattering amplitude to be consistent with unitarity at high energies given that it violates tree level unitarity? Consider the $2 \rightarrow 2$ scattering amplitude for a generic Galileon theory. The generic form of such a scattering amplitude that is consistent with unitarity is 
\be
A(s,t(\cos \theta)) = \sqrt{\frac{s}{s-4 m^2}} \sum_{l=0}^{\infty} (2 l+1) P_l(\cos \theta) \frac{1}{2i } \left( e^{2 i \delta_l(s)} e^{- 2\beta_l(s)}- 1\right) \, .
\ee
Here $\delta_l(s)$ is the scattering phase shift and $ e^{- 2\beta_l(s)}$ is the absorption coefficient reflecting the inelastic parts of the scattering amplitude. Unitarity is maintained provided that $\delta_l(s)$ is real and $\beta_l(s) \ge 0$. \\

If the conjecture is true, that at high energies the scattering is dominated by semi-classically configurations in the path integral, then the phase shift and absorption coefficients are simply determined by the real and imaginary parts of the classical action evaluated on classical configurations sourced by the scattered particles with the appropriate c.o.m energy and impact parameter $b = l /p$.
\ba
&& Re[S_{\rm classical}] = 2 \delta_l(s) \, ,\\
&& {\rm Im}[S_{\rm classical}] = 2 \beta_l(s) \, .
\ea
The classical configurations considered are in general complex solutions of the equations of motion which may tunnel through classically forbidden regions. It is these latter contributions that will give rise to a nonzero ${\rm Im}[S]$ as in standard instanton calculations.  
For example, in the case of scattering in gravity, this imaginary part arises because even in the absence of other interactions the two incoming particles could form a black hole. The likelihood for this to occur is determined by the Entropy for a black hole of energy $\sqrt{s}$ and angular momentum $l$, i.e. in this case we expect
\be
 {\rm Im}[S_{\rm classical}] = 2 \beta_l(s)  = S_{\rm entropy}(E,l) \, .
\ee
This is the `Black Hole ansatz' of \cite{Giddings:2007qq} (see \cite{Giddings:2009gj} for an updated review and \cite{Banks:1999gd} for earlier arguments). The remaining part of the ansatz is that $\delta_l(s)$ takes the form
\be
\delta_l(s) = \pi k(E,l) S_{\rm entropy}(E,l)/2 \, ,
\ee
where $k(E,l)$ is assumed to have a milder $E$ dependence. 
\\

We thus see that unitarity is guaranteed at high energies if the scattering becomes quasi-classical provided only that ${\rm Im}[S_{\rm classical}] >0$. This restriction is equivalent to requiring that the Euclidean action is positive definite. 
Since we are in the quasi-classical region we may expect to be able to replace the sum over $l$ by an integral, and then this integral may be performed by a saddle point/stationary phase approximation. In other words there will be some generically complex $l_*$ which minimizes the combined phase shift and exponential. We expect this either to be at the scale for which the impact parameter $b \sim |l_*|/\sqrt{s}$ is comparable to the Vainshtein radius, i.e. when $|l_*|\sim \sqrt{s} r_*(s)$ which becomes increasingly larger for high energies. For generic angles we utilize the large $l$ expansion for the Legendre polynomials
\be
P_l(\cos \theta) \sim - \frac{i}{\sqrt{2 \pi l \sin \theta}} \left(e^{i(l+1/2) \theta+ i \pi/4} -  e^{-i(l+1/2) \theta- i \pi/4}  \right) \, .
\ee
Then the amplitude will be
\be
A(s,t(\cos \theta)) \approx -\int_0^{\infty} \d l \frac{ \sqrt{l}}{\sqrt{2 \pi \sin \theta}}  \left(e^{2 i \delta_l -2  \beta_l + i(l+1/2) \theta+ i \pi/4} -  e^{2 i \delta_l -2  \beta_l -i(l+1/2) \theta- i \pi/4}  \right)
\ee
where primes denote derivatives with respect to $l$. 
In this case there are two saddle points for $l$ which will depend on the angles according to 
\be
\frac{\d \delta_l}{d l }(l_*)  +i\frac{\d \beta_l}{d l }(l_*) = \mp \frac{1}{2}\theta \, .
\ee
Assuming one or the other of the two phases contributions we get in the saddle-point approximation,
\be
A(s,t(\cos \theta)) \approx \mp \frac{ \sqrt{l_*}}{\sqrt{2  \sin \theta}}  \frac{1}{\sqrt{-i \delta_{l_*}'' + \beta_{l_*}'' }}e^{2 i \delta_{l_*} -2  \beta_{l_*} \pm i(l_*+1/2) \theta\pm i \pi/4}  \, .
\ee
Although explicitly constructing the classical configurations is extremely difficult, simple scaling arguments for Galileons in four dimensions predict the large $s$ behaviour
\ba
&& \delta_{l_*}(s) \sim  \sqrt{s} r_*(s) c_0(\theta)= \Lambda^{-6/5} s^{3/5} c_0(\theta) \, ,
\ea
and similarly for $\beta_{l_*}(s)$. To see this it is sufficient to consider a semi-classical configuration of energy $\sqrt{s}$ and compute the action integrated over the light crossing time for the strong coupling region $r_*(s)$.  \\

Thus, rather than blowing up, we may expect the scattering amplitude for fixed angle $\theta$ to oscillate or decay exponentially at large $s$ as 
\be
A(s,t(\cos \theta)) \sim e^{-c(\theta) s^{3/5}/\Lambda^{6/5}} \, .
\ee
For generic Vainshtein/classicalizing theories we would be led to 
 \be
 A(s,t(\cos \theta))  \sim e^{-c(\theta)  \sqrt{s} r_*(s)} \sim e^{-c(\theta) s^\alpha/\Lambda^{2 \alpha}} \, ,
 \ee 
 which is consistent since for Galileons in four dimensions $r_*(s)=r_V(s) \sim \Lambda^{-1 } (\Lambda^{-2} s)^{1/10}$  \cite{Dvali:2010jz,Dvali:2010ns,Dvali:2011nj,Dvali:2011th}. 
 The limit we have just described is the standard quasi-classical limit in which the classical scattering cross section is recovered and so we may expect the total cross section to grow as the square of the impact parameter which we identify with $r_*(s)$
\be
\sigma_{\rm total}(s) \sim  \pi \frac{l_*^2}{s} \sim  \pi r_*^2(s) \, .
\ee
In the above, the growth parameter $\alpha$ in $r_*(s) \sim \Lambda^{-1} \left( \sqrt{s}/\Lambda\right)^{2 \alpha-1}$ will be correlated with the growth properties of the Wightman functions. \\

The emergence of the generically non-integer power for $s$, and hence $\Lambda$ in the exponent, at first sight seems peculiar from the perspective of perturbation theory, but it can arise from the growth of certain entire functions which can be obtained by resumming the perturbative expansion. We shall see this behaviour explicitly in the computation of the \KL spectral densities. This is the characteristic behaviour of a non-localizable field theory whose momentum space Wightman functions for Heisenberg picture operators grow exponentially as $W(k) \sim e^{(k/\Lambda)^{2 \alpha}}$.\\

This conjectured high energy behavior can be entirely consistent with unitarity, but what it does violate is polynomial boundedness. The power $\alpha=3/5$ ensures that there will be some direction in the complex $s$ plane for fixed angles in which the amplitude grows exponentially. This follows simply from the analytic properties of functions in the complex plane.
More precisely, the conjectured behaviour $A(s,t(\cos \theta)) \sim e^{-c(\theta) s^{3/5}} $ violates the Cerulus-Martin bound \cite{cerulus1964lower} whose derivation assumes polynomial boundedness. Specifically the Cerulus-Martin bound states that the fixed angle scattering amplitude cannot fall off faster than
\be
\label{amplitudeguess}
A(s,t(\cos \theta)) |> C \,  e^{- f(\theta) s^{1/2} \ln (s/s_0)} \, .
\ee
A simple minded argument for this bound comes from noting that if the amplitude falls of as $e^{-a s^{\alpha}}$ then on continuing $s$ into the complex plane $s \rightarrow e^{i \theta} s$ we find $e^{- a |s|^{\alpha} (\cos(\alpha \theta) + i \sin (\alpha \theta)}$ which for $\alpha>1/2$ will inevitable grow exponentially in the upper half complex $s$ plane thus violating polynomial boundedness.
Again this is not in contradiction with unitarity since this occurs for unphysical momenta, but it does imply an inherent non-locality.  \\

The semi-classical ansatz for the high energy cross-section grows faster than allowed by the usual Froissart bound. For massless Galileons, or for Galileons whose intermediate scattering states include massless particles, this is acceptable since the Froissart bound does not apply. For the massive case, i.e. for which there is a mass gap for all states, this behaviour would be forbidden for a local field theory, but is consistent with the bounds for non-localizable theories which we discuss in more detail in \cite{KeltnerTolley2}. 
The form (\ref{amplitudeguess}) is allowed for non-localizable theories where the bounds on the growth of the amplitude in the whole complex plane are much weaker $|A(s,t)| \le D \, e^{c |s|^{\alpha}}$ \cite{Fainberg:1971ia} and so the Cerulus-Martin bound does not apply. These amplitudes may then violate the Froissart bound \cite{Fainberg:1971ia} which makes the explicit use of polynomial boundedness. It is important to note though that while this behavior violates polynomial boundedness at fixed angles, if $\alpha \le 1$ we nevertheless expect that the fixed momentum transfer $t\le 0$ amplitude, including in particular the forward scattering limit, to be polynomial bounded. This follows from the fact that ${\rm Im}(A(s,t)) < {\rm Im}(A(s,0))$ for real $t \le 0$ and use of the Phragm\'en-Lindel\"of theorem. We will discuss these issues in more detail elsewhere \cite{KeltnerTolley2}. \\

Thus if the classicalization proposal for unitarization at high energies is correct also for Galleons with a mass (regardless of how small), then these theories are necessarily non-localizable. The non-localizability appears to be consistent with the arguments of \cite{Dvali:2010jz,Dvali:2010ns,Dvali:2011nj,Dvali:2011th,Dvali:2014ila} where the exponential decay of the elastic scattering amplitude is interpreted in terms of the product of intermediate quasi-classical multiple particle states, `classicalons', of size $r_*(s)$. As in the Black Hole example, the production of such states from scattering implies an inherent non-locality over the scale $r_*(s)$. Even in the massless case we expect this non-locality to show up in the violation of polynomial boundedness, at least away from forward scattering and fixed physical momentum transfer limits \cite{KeltnerTolley2}. \\

Thus, we see that there is, in principle, no obstruction to UV completing the theory in a Lorentz invariant, unitary and even analytic way. However we must give up polynomial boundedness \cite{KeltnerTolley2}. This is the signature of an inherent non-locality which we regard as signalling that these theories are quintessentially gravitational as we discuss in the next section. 

\section{Gravity and Locality}

\label{Gravity}

The fundamental distinction between gravitational theories and ordinary gauge theories is the following theorem: {\it `There are no {\bf local} diffeomorphism invariant observables in gravity'}. The theorem is simple to prove since even a scalar field transforms under a diffeomorphism $\delta_{\xi} \phi(x) = \xi^{\mu}(x) \partial_{\mu}\phi(x)$, but its implications to attempts to quantize gravity are profound, as was noted early on by Wigner and Salecker \cite{Wigner:1957ep,Salecker:1957be}. In particular, it is no longer clear how to apply the notion of locality and  causality in the strong quantum gravity regime since the usual definition of the vanishing of the commutator of fields $\phi(x)$ outside the lightcone is neither gauge invariant nor independent of the background metric used to define the field. The normal approach to deal with this is to focus on a gauge invariant quantity such as the S-matrix, and find a substitute for locality and causality, namely polynomial boundedness and analyticity. It has been argued, however, that graviton scattering amplitudes do not satisfy polynomial boundedness, at least at fixed angles \cite{Giddings:2007qq,Giddings:2009gj}, and that any UV completion of gravity includes an inherent non-locality \cite{Giddings:1999jq,Giddings:2005id,Giddings:2006vu} that could even potentially explain aspects of the information loss paradox \cite{Giddings:2006sj,Giddings:2007pj}. We briefly review the arguments in favor of an inherent gravitational non-locality

\subsection{Giddings-Lippert locality bound and UV/IR mixing}

The existence of a locality bound in gravity is also connected with the existence and properties of black holes. In scattering two high energy particles at high energies, it is expected that when the impact parameter becomes less than the Schwarzschild radius associated with the centre of mass energy, 
\be
b \le R_s( \sqrt{s}) \, ,
\ee
a black hole will form. As the energy increases the black hole increases in size, indicating a UV/IR mixing. This UV/IR mixing in itself runs counter to the expectations of a local field theory for which UV physics decouples from IR and is indicative of some, albeit mild, degree of non-locality. In particular, Giddings and Lippert have proposed a fundamental bound on non-locality in gravitational theories \cite{Giddings:2001pt,Giddings:2004ud} related to this. To state this bound, consider a smeared field $\hat \phi_{x,p} = \int \d^4 x' f_{x,p}(x') \hat \phi(x')$ which effectively creates a wavepacket at position $x$ with momentum $p$. For example this may be achieved with a Gaussian wave packet
\be
\hat \phi_{x,p} = \int {\d^3 x'} \frac{1}{(\sqrt{2 \pi} L)^3}\hat  \phi(\vec x' ,t) e^{i \vec p.(\vec x- \vec x')- \frac{1}{2 L^2 }(\vec x-\vec x')^2} \, .
\ee 
In a local field theory there is no problem constructing a dressed two-particle state of the form $| \psi \rangle = \hat \phi_{x_1,p_1} \hat \phi_{x_2,p_2} | 0 \rangle$ where $| 0 \rangle$ is the gravitationally dressed vacuum. Furthermore, by the Wightman axioms, if the test functions $f_{x_1,p_1}(x)$ and $f_{x_2,p_2}(x)$ have compact support (which is allowed if the correlation functions are tempered distributions) and the support of one region lies entirely at spacelike separations for the support of the other then we would have for the equal time commutation relation
\be
[\hat \phi_{x_1,p_1}, \dot{\hat \phi}_{x_2,p_2} ] = 0 \, . 
\ee
If there is overlap as in the case of the Gaussian wave packets then we would expect a weaker statement to the effect of 
\be
\label{commutator}
| [\hat \phi_{x_1,p_1}, \dot{\hat \phi}_{x_2,p_2} ] | \le \frac{1}{(\sqrt{2 \pi} L)^3} e^{- | \vec x- \vec x'|^2/(2 L^2)}
\ee

However in a gravitational theory, once the following condition is violated
\be
| \vec x_1- \vec x_2 | \ge R_S( | \vec p_1 + \vec p_2|) \, ,
\ee
gravity becomes sufficiently strong that a black hole forms, or at least that the fluctuations in the geometry are so strong that it is no longer possible to talk about locality in the usual sense.  Concretely this would imply a violation of the locality of the Wightman axioms through a violation of the condition (\ref{commutator}). \\

In practice what does this violation of locality mean? From the perspective of the S-matrix for particle scattering, which by the LSZ construction is related to the time ordered Wightman functions, we expect a violation of locality to imply a violation of polynomial boundedness since the latter is derived on the basis of locality \cite{GellMann:1954db,:1900qta,PhysRev.109.2178,Epstein:1969bg}. This violation may only show up in some unphysical directions in the complexified momentum space since unitarity typically constrains the growth in the physical directions. This is indeed observed in the eikonel behaviour of scattering amplitudes, although it is conjectured not to arise in the forward scattering limit $t=0$ and fixed momenta transfer $t<0$ \cite{Giddings:2007qq,Giddings:2009gj}.  \\

From the perspective of the Wightman functions, we will argue that the failure of locality shows up in the failure of one of the central assumptions of the Wightman axioms, namely that the correlation functions are tempered distributions which is a sufficient condition to imply polynomial boundedness of the S-matrix \cite{GellMann:1954db,:1900qta,PhysRev.109.2178,Epstein:1969bg}. All tempered distributions have a Fourier transform. Our conjecture for the properties of the Wightman functions implies that they exist in momentum space, but do not admit a Fourier transform back to real space, because of the locality bound. More precisely the Wightman functions will be argued to be \GS distributions.

\subsection{Quantum Mechanics of M-theory}

A closely related argument which emphasizes a different aspect of the role of black holes was given by Aharony and Banks \cite{Aharony:1998tt}. Consider quantum gravity (e.g. M-theory?) with a fixed asymptotic form for the geometries, e.g. Minkowski or AdS. There will exist some asymptotic symmetry group which we assume to contain at least time-translations. Quantum mechanics then guarantees the existence of a generator for translations, i.e. a Hamiltonian, $\hat H$ which can be used to define Heisenberg operators
\be
\hat O(t) = e^{i \hat H t} \hat O e^{- i \hat H t} \, .
\ee
We have been intentionally agnostic about what the operator is or even its space dependence to be as general as possible. As in field theory, it is natural to define Wightman functions and their associated \KL spectral representation
\be
W(t) = \langle 0 | \hat O^{\dagger}(t) \hat O(0) | 0 \rangle  = \int_0^{\infty} \d E e^{-i E t} \rho_O(E) \, ,
\ee
where the spectral density is manifestly positive
\be
 \rho_O(E) = \sum_n \delta(E-E_n) | \langle 0 | \hat O | n \rangle|^2 \, ,
\ee
$|n \rangle$ are a complete set of energy eigenstates (including bound states) and the only assumption is that all these states have positive energy (technically bounded below is sufficient since a constant part of $\hat H$ drops out of $\hat O(t)$.)  \\

In order for the Wightman function to exist it is necessary that the spectral density for the chosen operator $\rho_O(E)$ to grow no faster than an exponential. A typical operator will have a spectral density which scales with the density of states. For a field theory whose UV behavior is determined by a UV fixed point we expect the density of states in $d$ dimensions in finite volume $V$ to behave as $\rho(E)\sim e^{c V^{1/d} E^{(d-1)/d}}$. To see this we note that for gas of massless particles at temperature $T$ the energy in $d$ spacetime dimensions will be $E \sim V T^d$ and so the entropy will scale as $S_{\rm entropy } \sim E/ T \sim E /(E/V)^{1/d}$, hence the density of states is 
\be
\rho(E) \sim e^{\rm S_{\rm entropy}}\sim e^{c' V^{1/d} E^{(d-1)/d}} \, .
\ee
This growth is sufficiently slow, i.e. slower than a linear exponential, in all dimensions that the Wightman function exist. There will also exist special operators whose growth is at most polynomial. These are the operators usually used to build the theory and its interactions. 
 \\

By contrast for a gravitational theory, we expect the density of states at high energies to be determined by the density of states for black holes which scales as 
\be
\rho(E) \sim e^{S_{\text{BH entropy}}} = e^{c (E/\mpl)^{\frac{d-2}{d-3}}} \, .
\ee
In other words, as in the previous S-matrix argument, we expect the physics at high energies to be dominated by the production of black holes which are known to have exponentially large density of states. 
Contrary to the field theory case, this grows faster than a linear exponential in any dimension $d>3$. This stronger growth is related to the negative specific heat capacity of Black Holes in asymptotically Minkowski spacetime. Thus, operators whose spectral density scales as the density of states $\rho_O(E) \sim \rho(E)$ can no longer give rise to well defined Wightman functions, i.e. they are no longer associated with well defined localized Heisenberg fields.  \\

The resolution is to look at operators for which the high energy behavior is sufficiently cutoff, meaning in practice that the spectral densities are cutoff above the Planck scale. This can easily be achieved by working with modified operators whose inner product between the vacuum and energy eigenstates decays exponentially at high energies, e.g. $\hat O'=e^{-L^2 \hat H^2/2 } \hat O e^{- L^2 \hat H^2/2 }$, however this resulting operator will not be local.
Inevitably this implies that there is no precise notion of time locality, i.e. no tempered distribution $W(t)$ for the original operators $\hat O(t)$. Alternatively this high energy cutoff may be achieved by smearing the original operator over a length scale $L$ comparable to the Planck scale $\mpl^{-1}$. 
\be
\hat O(t) \rightarrow \hat {\cal O}(t)=\int_{-\infty}^{\infty} \d t' \, \frac{1}{\sqrt{2 \pi} L}e^{-\frac{1}{2L^2 }(t-t')^2} \hat O(t') \, .
\ee
The Wightman function of the smeared operator is well defined
\be
\tilde W(t) =  \langle 0 | \hat {\cal O}^{\dagger}(t) \hat {\cal O}(t) | 0 \rangle  = \int_0^{\infty} \d E \, e^{-\frac{L^2}{2}E^2}e^{-i E t} \rho_O(E) \, ,
\ee
provided that $L^2\ge 2 c \mpl^{-2}$. There is no contradiction with the validity of the EFT at low energies since as long as we simultaneously make observations at energies below $\mpl$ and distances above $\mpl^{-1}$ the Wightman functions can be replaced by their smeared values which satisfy the usual requirements of a local field theory. 
\\

\subsection{Giddings-Lippert bound for time non-locality}

At a mathematical level, the claim of \cite{Aharony:1998tt} is that the Wightman functions are no longer tempered distributions, however they do exist as distributions but must be defined using test functions from an appropriate \GS space \cite{Gelfand:112483} $S_{\alpha}$, which contain no functions of compact support. If $f(t)$ and $g(t)$ are drawn from such a space then
\be
W(f,g) = \langle 0 | \hat O^{\dagger}(f) \hat O(g) | 0 \rangle  \, ,
\ee
exists and is finite. The fact that we must use test functions which do not have compact support to define the Wightman functions already indicates that the operator is non-local. \\

We may use this conjectured behaviour of the Wightman function spectral densities to derive an obvious generalization of the Giddings-Lippert bound. Suppose that $\hat O(x)$ is now a space-time dependent field. Again we express the vacuum expectation value using the \KL spectral representation\be
\langle 0 | \hat O^{\dagger}(x) \hat O(x') | 0 \rangle = \int_0^{\infty} \d \mu \, \rho_O(\mu) \int_{k^2=-\mu^2} \d \tilde k \,  e^{ik.(x-x')}  \, .
\ee
where
\be
 \int_{k^2=-\mu^2} \d \tilde k =  \int \frac{\d^{d} k}{(2 \pi)^d} \, \theta(k^0) 2 \pi \delta(k^2 + \mu) \, .
\ee
The conjecture of \cite{Aharony:1998tt} similarly implies $\rho_O(\mu) \sim e^{c (\mu/\mpl)} $ at large $\mu$ for $d=4$. 
Now Giddings and Lippert \cite{Giddings:2004ud} consider the equal time commutation relation, and smear these operators with wave packets which give a spread in space. For instance, defining the smeared operator 
\be
\hat O_{x,p} = \int {\d^{3} y} \frac{1}{(\sqrt{2 \pi} L)^{3}}\hat  O(\vec y ,t) e^{i \vec p.(\vec x- \vec y)- \frac{1}{2 L^2 }(\vec x-\vec y)^2} \, .
\ee 
then the vacuum expectation value of two equal time product of two such operators is given by
\ba
\langle 0 | \hat O^{\dagger}_{x,p} \hat O_{x',p'}  | 0 \rangle &=& \int_0^{\infty} \d \mu \, \rho_O(\mu) \int_{k^2=-\mu^2} \d \tilde k \,  e^{i \vec k.(\vec x-\vec x')} e^{-\frac{(\vec p-\vec k)^2}{2 L^2} -\frac{(\vec p'-\vec k)^2}{2 L^2} }  \, , \nn  \\
&=& \int \frac{\d^{4} k}{(2 \pi)^4} \, \theta(k^0) \theta(-k^2) 2 \pi \rho_O(-k^2 )  e^{i \vec k.(\vec x-\vec x')} e^{-\frac{(\vec p-\vec k)^2}{2 L^2} -\frac{(\vec p'-\vec k)^2}{2 L^2} }  \, .
\ea
It is transparent that despite the spatial smearing the $k^0$ integral does not converge. Thus the non-locality implied by the pure spatial form of the Giddings-Lippert bound applied at equal times is not sufficient. We must extend the bound to include an inherent non-locality in time
\be
| t_1- t_2 | \ge R_S( | \vec p_1 + \vec p_2|+ |p_1^0 +p^0_2| ) \, .
\ee
Of course in a Lorentz invariant theory it is not surprising that non-locality in space implies non-locality in time, nevertheless we see that the finiteness of the Wightman functions has more to do with smearing them in time. To demonstrate this consider now the time smeared operators 
\be
\hat O_{x,E} = \int_{-\infty}^{\infty} {\d \tau} \frac{1}{\sqrt{2 \pi} L}\hat  O(\vec x ,\tau) e^{-i  E(t-\tau) - \frac{1}{2 L^2 }(t-\tau)^2} \, .
\ee 
Then we have
\ba
\hspace{-15pt} \langle 0 | \hat O^{\dagger}_{x,E} \hat O_{x',E'}  | 0 \rangle &=& \int_0^{\infty} \d \mu \, \rho_O(\mu) \int_{k^2=-\mu^2} \d \tilde k \,  e^{i \vec k.(\vec x-\vec x')-i k^0(t-t')} e^{-\frac{(E-k^0)^2L^2}{2} -\frac{(E'-{k^0})^2L^2}{2} }  
\ea
Now convergence of all the integrals is guaranteed by the factor $e^{-{k^0}^2 L^2 } = e^{-{\mu}L^2 }e^{-{\vec k}^2 L^2 }$. Reorganizing we have
\ba
\langle 0 | \hat O^{\dagger}_{x,E} \hat O_{x',E'}  | 0 \rangle &=& e^{-\frac{(E^2+{E'}^2) L^2}{2 }}\int_0^{\infty} \d \mu \, \rho_O(\mu) e^{- \mu L^2} \int_{k^2=-\mu^2} \d \tilde k \,  e^{i \vec k.(\vec x-\vec x')-i k^0(t-t' -i (E+E') L^2)}  e^{-\vec k^2 L^2} \, .\nn
\ea
In order for the $\mu$ integral to converge we need $L^2 \ge  c/\mpl^2$. We see from the form of the integral that there is a spread in the time dependence of the order
\be
\Delta(t-t') \sim (E+E') L^2 \ge (E+E') c/\mpl^2 \ge R_S(E+E') \, .
\ee
This is the time non-locality version of the Giddings-Lippert bound. 

\subsection{Effective Field Theory point of view}

These argument are sufficiently simple that they are meant  to apply to any UV completion of gravity in all dimensions greater than $3$. As we discuss in the next section, a field whose spectral density grows slower than a linear exponential are known as strictly localizable fields, and those whose spectral density grows faster than a linear exponential are known as non-localizable fields. Hence the claim is that one of the key distinctions between a local field theory and a gravitational theory is: \\

{\bf In a gravitational theory, the spectral densities of generic operators grow faster than linear exponentials, i.e. operator valued fields in quantum gravity are fundamentally non-localizable}. \\

All of this is consistent with the original observation that no gauge invariant local observables exist in gravitational theories. How can we see this from the point of view of the EFT description of gravity? Consider again quantum gravity, treated now in an EFT sense, for spacetimes whose asymptotic symmetry group includes time translations. As long as we are at low energies we may express the metric as $g_{\mu\nu}(x)= \bar g_{\mu \nu}(x) + h_{\mu\nu}(x)$, where $\bar g$ is the background solution with the desired asymptotics. Then follow some gauge fixing formalism for which the constraints may be solved locally in time, for instance the BSSNOK formalism. By construction this formalism is designed so that the solution of the constraints only requires inverting elliptic spatial operators. This reduces the system to the physical degrees of freedom in a way which remains local in time. We may then solve the Heisenberg equations of motion for the remaining physical degrees of freedom perturbatively using the Yang-Feldman formalism \cite{Yang:1950vi} or by one of the many other approaches. The Yang-Feldman approach has the virtue for example of making causality and locality manifest at each order in perturbation theory. Even if solving the constraints involves inverting $\nabla^2$, the Yang-Feldman equations for the propagating degrees of freedom always take the schematic form
\be
\hat A \hat h_P(x) = - \frac{2}{\mpl^2} \hat T_P(x)
\ee
where $\hat A$ is some hyperbolic operator of the schematic form $\hat A \sim \Box$ and $\hat T_P(x)$ is the effective stress tensor source including non-linearities in $\hat h_P$. This may then be solved in a local/causal way using the retarded propagator
\be
\hat h_P(x) = \hat h_P^0(x) - \int \d^4 y \, G_R(x-y)  \frac{2}{\mpl^2} \hat T_P(y) \, .
\ee
where $G_R$ is the retarded propagator for $\hat A$, and $ \hat h_P^0(x)$ is the quantized free physical field . \\

We can now compute the spectral densities of some time-local operators (but not necessarily space-local) $\hat O(t) = F(\hat h_P(x))$ perturbatively as an expansion in $E/\mpl$. At any finite order in perturbations there are a finite number of counterterms that need to be added to remove divergence, nevertheless matching against some specific UV completion will determine specific values for these counterterms. The spectral density will now be expressed as a perturbative expansion (with possible logarithmic terms from loops)
\be
\rho_{\hat O}( E) = \sum c_{n,m} \frac{E^{2n}}{\mpl^{2n}} \( \ln \frac{E}{\mpl} \)^m \(\ln^2 \dots \)+ \dots
\ee
Since the EFT is local, and the operators defined are local in time, at any finite order in perturbations the Wightman functions will exist as tempered distributions. Similarly at any finite order in perturbations the S-matrices will be polynomially bounded. In this language passing to the UV completion simply corresponds to summing the series to an infinite order. If this sum converges absolutely and $\rho_{\hat O}( E)$ grows slower than a linear exponential then the Wightman function would remain well defined. But this would imply that it was possible to define gauge invariant observables which are local in time (even if they were non-local in space due to the solution of the constraint equations)  which we know is impossible from time diffeomorphism invariance alone. Even if this series were asymptotic then assuming it can be made sense of by Borel summation or some similar procedure, as long as it grows slower than a linear exponential we would be left with the same contradiction. Thus the absence of local gauge invariant observables can only be consistent if $\rho_{\hat O}( E)$ grows as fast as or faster than a linear exponential. This may be easily achieved even if the coefficients $c_n$ are on average positive and fall of at a particular rate. For instance if $c_n \sim c^n/n!$ in $d=4$  we would recover the prediction $\rho_{\hat O}( E)  \sim e^{c (E/\mpl)^2}$. Of course we cannot from an EFT point of view determine the $c_n$ precisely because of the need to add counterterms, they can only be determined by matching against a specific UV completion, nevertheless we can at least see how the locality of the EFT of gravity can be consistent with the non-locality of the UV completion based on the convergence properties of the perturbative expansion. The effective field theorist that truncates the expansion at small finite order will see no evidence of non-locality, but the intrepid field theorist that computes to large orders, and can perform the EFT matching, will begin to see the non-locality by the convergence properties.

\subsection{The case of Little String Theory}

Above we have reviewed suggestive arguments that imply that in gravitational theories, the spectral densities of certain gauge invariant observables grow exponentially fast, consistent with the requirement that such operators cannot be local. Operators associated with these observables are known as non-localizable fields and we shall review the definition of these in Sec.~\ref{sec:nonlocalizabledefn}. Another class of theories in which this behaviour has been argued to occur are Little String Theories  (LST's) \cite{Berkooz:1997cq,Seiberg:1997zk}. LSTs describe M-theory compactified on a $T^5$ and are thus six dimensional theories. They are obtained as decoupling limits of N coincident fivebranes in string theory in which $M_{\rm Planck} \rightarrow \infty$, $g_s \rightarrow 0$ keeping the string scale $M_{\rm string}$ fixed. Since $M_{\rm Planck} \rightarrow \infty$ they are no longer gravitational in the usual sense, and so one might be tempted to think that they are field theories. Indeed, the infrared limit of LSTs are normal six dimensional superconformal field theories (SCFTs), but the full LST cannot be a local field theory (i.e. there is no UV fixed point) because it includes non-trivial string effects since $M_{\rm string}$ is kept finite. As a result we expect this theory to maintain certain non field theory features of string theory - hence the name `Little String Theory'. In particular it is expected that it preserves T-duality. This alone implies that LSTs cannot be local field theories since the T-duality transformation is itself non-local. In \cite{Kapustin:1999ci} it was proposed that LSTs are quasi-local theories, i.e. non-localizable field theories of the Jaffe type with $\alpha = 1/2$ (see Sec.~\ref{sec:nonlocalizabledefn} for the definition). \\

The salient feature which establishes this interpretation is that the Wightman functions of LSTs can be shown to have the exponential growth characterstic of the Hagedorn density of states \cite{Aharony:1998tt,Peet:1998wn,Minwalla:1999xi}. In particular the \KL spectral density for the two-point function is expected to grow as
\be
\rho(\mu) \sim e^{\frac{cM_{\rm string} \sqrt{\mu}}{\sqrt{N}}}
\ee
for an order unity constant $c$, where $N$ is the number of five-branes. A similar growth is expected to arise for all correlators, and this growth cannot be removed with a simple wavefunction renormalization. Our perspective will be that LSTs are the closest known analogue theories to Galileons in the sense they are local field theories in the infrared limit, but they are fundamentally non-localizable in the UV, having no UV fixed point, and an exponentially growing density of states. Both LSTs and Galileons are obtained as double scaling limits of gravitationally theories. Both theories exhibit a duality symmetry which is a remnant of a symmetry of the original theory they are defined as decoupling limits of. This analogy is summarised in the table below. 

\begin{center}
    \begin{tabular}{| l | l | l | l |}
    \hline
    Theory & Little String Theory & Galileon  \\ \hline \hline
   Parent Gravitational Theory & M-theory compactified on $T^5$ & Massive Gravity/DGP etc.  \\ \hline
   Decoupling Limit  Theory & LST & UV Galileon  \\ \hline
   Infrared Limit & Six Dimensional SCFT & Galileon LEEFT  \\ \hline
    Decoupling/Scaling Limit & $g_s \rightarrow 0$, $M_{\rm Planck } \rightarrow \infty$ & $m \rightarrow 0$, $M_{\rm Planck } \rightarrow \infty$  \\ \hline
    Strong Coupling Scale  & $\Lambda= M_{\rm string}/\sqrt{N}$ & $\Lambda =  m^\frac{4}{(d+2)} M_{\rm Planck}^{\frac{d-2}{d+2}}$  \\ \hline
  Infrared Region  & $E \ll M_{\rm string}/\sqrt{N}$ & $E \ll \Lambda$  \\ \hline
    Symmetry of Parent Theory & M-theory T-duality & Diffeomorphism Invariance \\ \hline
    Duality of D.L. Theory & T-duality & Galileon Duality \\ \hline
   Growth of Spectral Density & $\ln \rho(\mu) \sim {cM_{\rm string} \sqrt{\mu}}/{\sqrt{N}}$ & $ \ln \rho(\mu) \sim \frac{\mu^{(d+2)/(2 (d+1))}}{\Lambda^{(d+2)/(d+1)}}$ \\ \hline
   `Vainshtein' Radius & $r_*(E) = \sqrt{N} M_{\rm string}^{-1}$  & $r_*(E) =  \frac{1}{\Lambda} \left( \frac{E}{\Lambda}\right)^{\frac{1}{(d+1)}}$  \\ \hline
    Jaffe Class & Quasi-local $\alpha =1/2$  & Non-localizable $\alpha>1/2$ \\ \hline
    \end{tabular}
\end{center}  

Since LSTs are argued to be quasi-local, the analogue of the Vainhstein/classicalization radius is independent of energy scale. In \cite{Kapustin:1999ci} it is argued that LSTs should be understood as Jaffe type quasi-local field theories and we shall follow an analogous description in the case of Galileons in what follows.
\subsection{Galileon Duality and the Locality Bound}

\label{sec:localitybound}

We have previously reviewed the arguments to suggest that gravitational theories, and certain decoupling limits of M-theory/string theory, exhibit some degree of `gravitational non-locality'. This non-locality will lead to a violation of the usual Wightman axioms, specifically the temperedness assumption. This in turn leads to a violation of the assumptions used to derive the polynomial boundedness of the S-matrix, about which we shall discuss more elsewhere \cite{KeltnerTolley2}. \\

In the following we conjecture that these same properties hold for Galileon theories, in other words Galileon theories are intrinsically gravitational. Our main concrete evidence for this proposal is that we are able to give an explicit off-shell UV complete quantization of a specific Galileon model, and demonstrate that its Wightman functions grow exponentially as
\be
\rho(E) \sim e^{E \, r_*(E)}
\ee
where $r_*(E) = \Lambda^{-1} (E/\Lambda)^{\frac{1}{d+1}}$ is the Vainshtein/classicalization radius. This exponential growth means that there do not exist local off-shell observables for Galileons, although there do exist precisely defined non-localizable off-shell fields. This exponential growth means that we can only talk about locality for separations satisfying the analogue of the Giddings-Lippert locality bound
\be
|x| \gg r_*(E) = \frac{1}{\Lambda} \( \frac{E}{\Lambda} \)^{\frac{1}{d+1}} \, .
\ee
Mathematically, this corresponds to stating that Galileon fields are no longer `operator valued tempered distributions' but rather `operator valued \GS distributions' \cite{Gelfand:112483}. \\

Although we only demonstrate this explicitly for one specific example of a Galileon model whose S-matrix is trivial, we conjecture that the same qualitative exponential growth of the off-shell Wightman functions is true for all interacting Galileon models. This is reasonable since all Galileon models have the same scaling of $r_*(E)$ regardless of the choice of coefficients, and the phenomenology of the Vainshtein mechanism is qualitatively the same in all such models. Furthermore, a semi-classical argument reproduces the same conjectured behaviour for all Galileon models regardless of choice of the coefficients. \\

This exponential growth of the correlation functions will feed into an exponential growth of the scattering matrix in some directions of the complex plane (albeit constrained by the requirements of analyticity and unitarity). Thus, we conjecture that the interacting Galileon models necessarily violate polynomial boundedness of their scattering amplitude \cite{KeltnerTolley2}. \\

Another viewpoint on the origin of the locality bound is provided by the existence of the duality between various Galileon models which we review in Sec.~\ref{sec:DualityMap}. At the level of the field theory Lagrangian, the duality is an equivalence map, enacted by a non-local field redefinition, between two naively distinct Galileon theories. However, as we review in Sec.~\ref{sec:DualityMap} from the point of view of massive gravity where the duality was derived in \cite{Fasiello:2012rw,deRham:2013hsa} (see also \cite{Curtright:2012gx}), or from the coset construction considered in \cite{Creminelli:2014zxa,Kampf:2014rka}, the duality is simply a choice of diffeomorphism gauge. In other words, by viewing the Galileon models as a gauge fixed version of a diffeomorphism invariant theory, which is the natural interpretation both in the massive gravity and coset perspectives (see Sec.~\ref{sec:DualityMap}), the Galileon duality becomes a symmetry, just as T-duality of LST is a symmetry of M-theory. Assuming that the quantization respects this symmetry, which seems desirable, then following the usual arguments we would led to conclude that there are no local off-shell gauge invariant operators in Galileon theories. Of course we may choose to fix unitary gauge to go back to the original Galileon Lagrangian, but then we would be left with the statement that there are no local off-shell operators invariant under the Galileon duality transformation, which is the remnant of the previous statement. This is analogous to the statement that in LST, there are no local off-shell operators invariant under T-duality, which as we have argued is one of the clues that LSTs are not local field theories. \\

Thus in our proposed picture, the Galileon duality transformation, far from being an accident, is viewed as a central property of the Galileon theories which enforces the locality bound. Furthermore the duality must hold at the quantum level, since it is simply a remnant of a local symmetry of the UV completion (assuming that the latter is not anomalous). We shall in fact show that the same exponential growth of the Wightman functions occurs in two different duality frames, albeit for two different operators, consistent with the duality map.

\section{Strictly Localizable versus Non-localizable field theories}

\label{sec:nonlocalizabledefn}

The interest in field theories with a weaker notion of locality than standard perturbatively renormalizable field theories dates back to the 1960's, largely to two programs. On the one hand there were attempts within the axiomatic approach to quantum field theory to determine the minimal requirement for incorporating locality into the axioms of field theory. A precise condition was given by Jaffe \cite{Jaffe:1967nb,Jaffe:1966an} and Meimam \cite{meiman1964causality} that distinguished (in Jaffe's language) so-called {\bf strictly localizable} field theories, and {\bf non-localizable} field theories. On the other hand there was an interest, sparked by Guttinger \cite{guttinger1967dynamics,guttinger1966unrenormalizable,guttinger1959non}, Efimov \cite{efimov1965formulation,Alebastrov:1973np,Alebastrov:1973vw}, Fradkin \cite{fradkin1963application}, Volkov \cite{volkov1968method,volkov1968quantum,Volkov:1969br}, and others \cite{schroer1964concept,Pfaffelhuber:1971qz,Isham:1973by,Isham:1970rg,delbourgo1969infinities,Delbourgo:1970au,Albeverio:1973ck,Albeverio:1973cm} to give resummation techniques to field theories that were non-renormalizable from the standard point of view, but could potentially be partially resummed into a renormalizable field theory. This approach was sometimes referred to as the `super-propagator' \cite{Biswas:1973yr,Flume:1972iu,Lehmann:1971gq,Atakishiev:1971ec,Blomer:1971rx,Krause:1972kw} or `Efimov-Fradkin' \cite{Zumino:1969tt,Lee:1969ni,Houard:1973yj} method and at that time the most plausible application was to non-renormalizable theories of the Weak nuclear force. The resummed operators generically fall into the class of fields with an exponentially growing spectral density. It was even argued that this approach may be applicable to gravity \cite{Delbourgo:1969ha,Isham:1970aw,Isham:1972pf} and potentially tame the na\"ive non-renormalizability of gravity. This original program failed and was ultimately replaced by the now standard field theory tools of renormalization group and effective field theories. In particular the modern perspective is that all non-renormalizable non-gravitational field theories should be viewed as EFTs. It will be our contention here however, that Galileon theories are intrinsically gravitational and are explicit realizations of such non-localizable field theories.

\subsection{Definition of Strictly Localizable, Quasi-local and Non-Localizable fields}

The physical meaning of the definition of a strictly localizable versus a non-localizable field is easy to understand by considering its associated Wightman functions written in the form of a \KL spectral representation. Consider some operator $\hat O(x)$, its position space Wightman function may be defined as 
\be
W(x,y) =\langle 0 | \hat O^{\dagger}(x) \hat O(y) | 0 \rangle =  \int_0^{\infty} \d \mu \, \rho_{O}(\mu) \, W_{\mu}(x,y)  \, ,
\ee
where $\rho_{O}(\mu) $ is the \KL spectral density. Unitarity is encoded in the statement that $\rho_{O}(\mu)\ge 0 $ for all $\mu\ge 0$ and is real. Stability is encoded in the statement that $\rho_{O}(\mu) $ only has support for $\mu \ge 0$. The function $W_{\mu}(x,y)$ is the Wightman function of a free scalar field of mass $\sqrt{\mu}$.  \\

A {\bf strictly localizable field} is one for which the spectral density integral $\int \d \mu$ converges so that $W(x,y)$ can be given meaning for $x \neq y$ without smearing (i.e. using test functions of compact support). In the case of four dimensions $W_{\mu}(x,y)$ has the asymptotic properties 
\ba
&&W_{\mu}(x,y) \approx \frac{(2 \sqrt{\mu})^{1/2}}{(4 \pi |x-y|)^{3/2}} e^{-\sqrt{\mu}|x-y|} \,, \quad \, (x-y)^2 >0 \,, \quad  \mu |x-y|^2 \gg 1 \, ,\\
&& W_{\mu}(x,y) \approx -i e^{-i \pi/4}\frac{(2 \sqrt{\mu})^{1/2}}{(4 \pi |x-y|)^{3/2}} e^{-i \sqrt{\mu}|x-y|} \,, \quad \, (x-y)^2 < 0 \,, \quad  \mu |x-y|^2 \gg 1 \, ,
\ea
and a similar exponential behavior occurs in general dimensions. It is clear that the convergence of the spectral density integral requires that the spectral density grows no faster than a linear exponential in $\sqrt{\mu}$, 
i.e. that
\be
\lim_{\mu \rightarrow \infty} \sqrt{\mu} \frac{d}{d \mu} \ln \rho_{O}(\mu) =0
\ee
or equivalently that
\be
\int_1^{\infty} \d \mu \frac{\ln \rho_{O}(\mu)}{\mu^{3/2}} < \infty \, .
\ee
There are obvious generalizations of this condition to the $n$-point Wightman functions. If $W(\{ k_i \})$ denotes the Fourier transform of the $n$-point Wightman function as a function of complex momenta then we require the bound
\be
W(\{ k_i \}) < C e^{A |\sum_i |k_i||}
\ee
Whenever these conditions are satisfied for {\it all} $n$-point Wightman functions, the operator is said to be strictly localizable. \\

\label{defnonlocal}

A {\bf non-localizable} field is then any field for which 
\be
\lim_{\mu \rightarrow \infty} \sqrt{\mu} \frac{d}{d \mu} \ln \rho_{O}(\mu) =\infty
\ee
or
\be
\int_1^{\infty} \d \mu \frac{\ln \rho_{O}(\mu)}{\mu^{3/2}} = \infty \, .
\ee
When this condition is true the position space Wightman functions do not exist, i.e. we may not use test functions of compact support. However, the Wightman functions do exist as we shall see when defined using an alternative class of test functions, namely those from the appropriate \GS space \cite{Gelfand:112483}. In this case the Wightman functions are still distributions but not tempered ones. The description of such distributions is well developed \cite{Gelfand:112483}. A non-rigorous but practical way to understand this is to note that the momentum space Wightman functions are well defined, and so provided we choose test functions which are sufficiently delocalized in position space, then their Fourier transforms will fall off sufficiently rapidly at large $k$ to compensate the exponential growth of the spectral density. Explicitly we can define the smeared Wightman function by
\be
W(f,g) = \int_0^{\infty} \d \mu \int \frac{\d^d k}{(2 \pi)^d} f^*(k) g(k) 2 \pi \rho_{O}(-k^2) \, .
\ee
Motivated by the theory of entire functions, we define the `order' of the spectral density $\alpha$ by the requirement that
\be
\lim_{\mu \rightarrow \infty} \frac{ \ln ( \ln \rho_{O}(\mu))}{\ln \sqrt{\mu}} = 2 \alpha  >0 
\ee
and the `type' $\sigma$ by
\be
\lim_{\mu \rightarrow \infty} \frac{ \ln \rho_{O}(\mu)}{\mu^{\alpha}} = \sigma > 0 \, .
\ee
In other words, the high $\mu$ behaviour of the spectral density is taken to be
\be
\rho_{O}(\mu) \sim e^{\sigma \mu^{\alpha}} \times \text{subdominant terms} \, .
\ee
It is then clear that a suitable choice of momentum space test functions are the set of infinitely differentiable functions for which $f(k),g(k) < C e^{- \frac{1}{2}\sigma |k|^{2 \alpha}}$ as $|k| \rightarrow \infty$ for some finite constant $C$. This gives a (non-rigorous) definition of the appropriate \GS space \cite{Gelfand:112483} which is usually denoted as $S_{\alpha}$. When the test functions are drawn from such a space, the smeared Wightman function is well-defined. The Fourier transform of such test functions define the dual space $S^{\alpha}$. Their properties are well understood \cite{Gelfand:112483} and, roughly speaking, correspond to functions which are no more localized than a function of the form 
\be
f(x) \sim D \, e^{- \beta |x-a|^{2\alpha/(2\alpha-1)}} \, .
\ee
Test functions of compact support are thus excluded. An obvious simple example is the case $\alpha = 1$ for which the set of momentum space test functions fall of as Gaussians and hence by the usual Hardy uncertainty principle argument that position space functions cannot be localized more than a Gaussian.
\\

This discussion should be contrasted with the `traditional' case, {\bf tempered localizable} fields, which are operator valued tempered distributions for which the spectral density grows at most as a polynomial. For tempered fields, only a finite number of subtractions need to be performed to give the position space Feynman propagators meaning. In the case of non-localizable fields an infinite number of subtractions, i.e. an infinite number of renormalizations should be performed. This connects with the notion that non-localizable field theories have to do with perturbatively non-renormalizable field theories whereas localizable fields have to do with renormalizable field theories \cite{schroer1964concept}. Precisely one of the historical reasons for interest in non-localizable fields was to give a potential renormalizable description of perturbatively non-renormalizable field theories. In particular we note that the usual notion of operator product expansions is only applicable in the case of tempered localizable fields for which the short distance behaviour of a product of two, operators $\hat A(x)$ and $\hat B(y)$ diverges by no more than an inverse power of $|x-y|$. This behaviour is only applicable if $\langle \psi | \hat A(x) \hat B(y) | \psi \rangle $ is a tempered distribution in $(x-y)$ and if an analogue were to exist in the case of non-localizable field it would have to be substantially different \cite{Isham:1973by}. 
\\

\begin{figure}  
\begin{center}  
\includegraphics[height=2.7in,width=4.5in,angle=0]{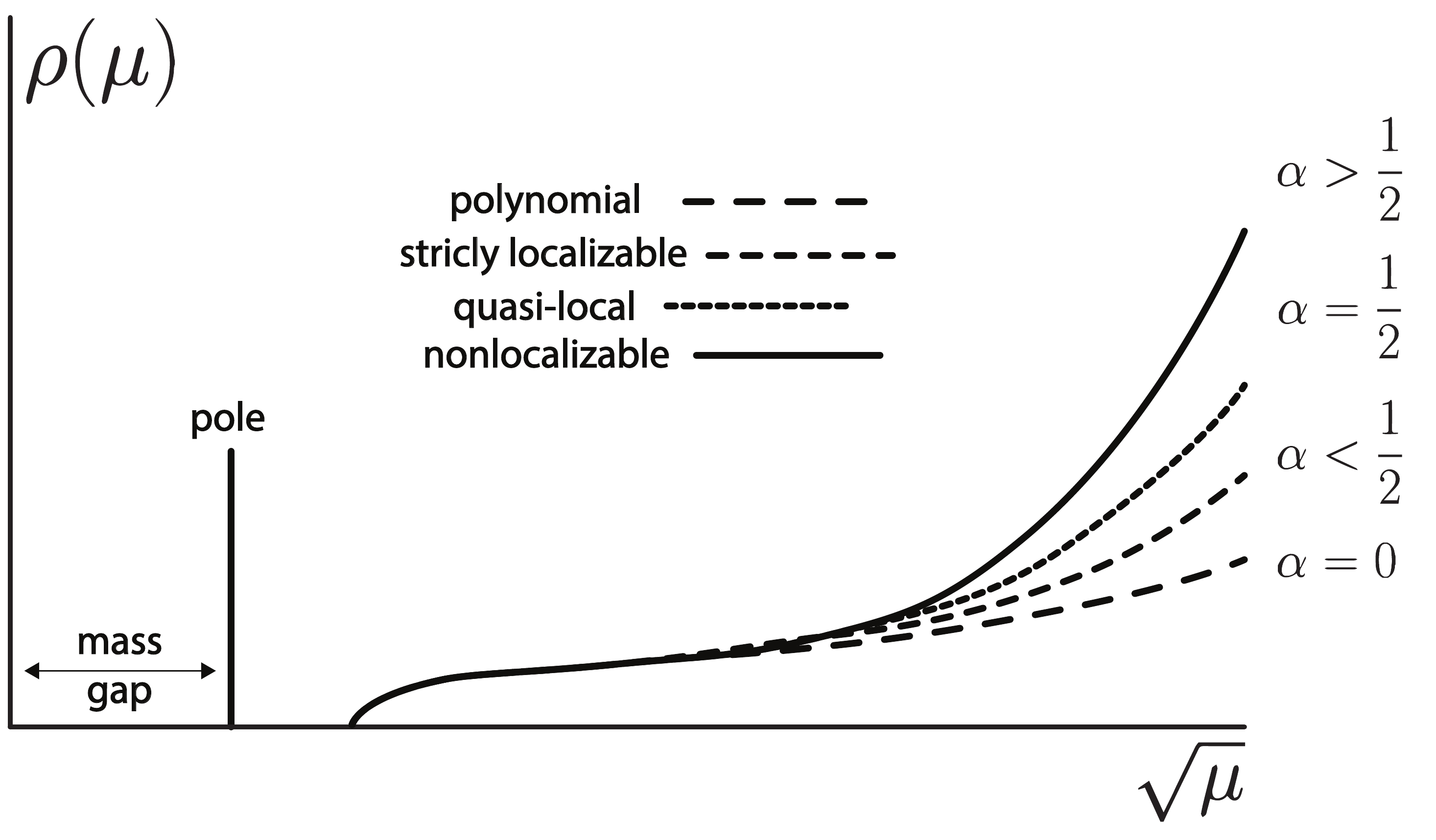}  
\caption{\small \sl Schematic form of the spectral density for a theory with a mass gap in each of the possible cases. In addition, bound states may exist between the pole and the beginning of the continuum. In the massless limit, applicable for the examples computed in the text, the continuum begins at the position of the pole located at $\mu =0$. \label{fig:contour}}  
\end{center}  
\end{figure}  

Finally, as a special case we can consider a {\bf quasi-local} field for which
\be
\lim_{\mu \rightarrow \infty} \sqrt{\mu} \frac{d}{d \mu} \ln \rho_{O}(\mu) = {\rm finite} \neq 0 \, .
\ee
is a limit of a non-localizable field with $\alpha \rightarrow 1/2$. This case is clearly special since the position space Wightman function does exist at sufficiently large space-time separations. As a consequence, such fields retain some degree of locality even though they violate strict micro-locality/micro-causality. Interestingly, it is precisely this case that is expected to be applicable to Little String Theories \cite{Kapustin:1999ci} where the exponential growth of the spectral density is associated with the Hagedorn behaviour for a string. \\

In summary the nature of the field is detemined by the order $\alpha$ of the spectral density. If $\alpha>1/2$ the field is non-localizable, $\alpha=1/2$ it is quasi-local, $0<\alpha<1/2$ strictly localizable and if $\alpha=0$ it is a tempered localizable field (see Fig.~1). Textbook renormalizable field theories correspond exclusively to the latter case $\alpha=0$.

\subsection{Functions of a free field}

As an illustrative example of a strictly localizable field whose Wightman function is not tempered, consider a free field operator $\hat \phi(x)$ which for simplicity of calculation we take to be a massless field. Define the operator $\hat O(x)$ to be the normal ordered $::$ exponential \cite{jaffe1965entire,Rieckers:1971sw,Lehmann:1971gq}
\be
\hat O(x) = \sum_{n=1}^{\infty} \frac{g^n}{n!} : \hat \phi(x)^n : = : e^{g \hat \phi(x)} -1: \, .
\ee
An elementary application of Wick's theorem tells us that the two-point Wightman function for $\hat O(x)$ is given by
\ba
\langle 0 | \hat O(x) \hat O(y) | 0 \rangle  &=& \sum_{n,m=1}^{\infty} \frac{g^{n+m}}{n! m!} \langle 0 | : \hat \phi(x)^n: : \hat \phi(y)^m: | 0 \rangle  \\
&=& \sum_{n=1}^{\infty} \frac{g^{2n}}{n!} \left( \langle 0 |  \hat \phi(x) \hat \phi(y) | 0 \rangle \right)^n \\
\label{OO}
&=& e^{g^2 \langle 0 |  \hat \phi(x) \hat \phi(y) | 0 \rangle }-1  \, .
\ea
For a massless field in four dimensions the Wightman function is given by
\be
 W(x,y)=\langle 0 |  \hat \phi(x) \hat \phi(y) | 0 \rangle = \frac{1}{4 \pi^2} \frac{1}{(\vec x-\vec y)^2 - ((x^0-y^0)-i \epsilon)^2} \, .
\ee
The Wightman function is a `generalized function' or `distribution' \cite{Gelfand:112483} so even products of it must be interpreted with great care. The problem with the eq.~(\ref{OO}) is that in effect it includes the exponential of a delta function from the support of the Wightman function on the light cone, which is meaningless as it stands. To deal with this more rigorously we must smooth the operators with test functions to give meaning to the operator product. However, to proceed it is simplest to directly compute the spectral density via the perturbative expansion. Concretely we have
\ba
\theta(k^0) 2 \pi \rho_{O}(-k^2)  &=& \sum_{n=1}^{\infty} \frac{g^{2n}}{n! ( 4 \pi^2)^n}\int \d^4 x e^{- i k.x} \frac{1}{(\vec x^2 - (x^0-i \epsilon)^2)^n}  \\
&=& \theta(k^0) 2 \pi \delta(-k^2) + \sum_{n=2}^{\infty} \frac{g^{2n}}{n! } \theta(k^0) \,  \Omega_n(-k^2)
\ea
where $\Omega_n(-k^2)$ is the $n$-particle phase space density 
\be
\Omega_n(-k^2) = \left[ \prod_{i=1}^n \int \d \tilde k_i \right] \, (2 \pi)^4 \delta^{(4)}(k - \sum_{i=1}^n k_i) \, .
\ee
This is given by (see Appendix~\ref{app1})
\be
\Omega_n(\mu) = \frac{1}{ (16 \pi^2)^{n-1} (n-2)! (n-1)!} \mu^{n-2} \, .
\ee
The perturbative summation converges to give a pole contribution plus an entire function
\ba
\label{rho0}
2 \pi \rho_{O}(\mu) &=&  2 \pi \delta(\mu) + \sum_{n=2}^{\infty} \frac{2^4 \pi^2 g^{2n}}{ (16 \pi^2)^n n! (n-2)! (n-1)!} \mu^{n-2} \\
&=& 2 \pi \delta(\mu) + \frac{g^4}{32 \pi^2} {}_PF_{Q} \(;2,3,\frac{g^2 \mu}{16 \pi^2} \) \, ,
\ea
which behaves asymptotically as $\mu \rightarrow \infty$ as
\be
\rho_{O}(\mu) \sim e^{\frac{3 g^{2/3} \mu^{1/3}}{2^{4/3} \pi^{2/3}}} \frac{1}{\sqrt{3}} \left( \frac{2 g^4}{\mu^4 \pi}\right)^{1/3}
\ee
We thus see that $\hat O(x)$ corresponds to a strictly localizable field with $\alpha = 1/3$ and has faster than polynomial growth. Unitarity is preserved since $\rho(\mu)>0$ for all $\mu$ as is evident since each coefficient in the entire series is positive. Note that since the continuum contribution to the spectral density is an entire function, there is no ambiguity with regards to its non-perturbative definition, i.e. there is no need for Borel summation.  \\

An example with a stronger high energy growth is given by the operator 
\be
\hat O(x) =: \frac{\hat \phi(x)}{1-g \hat \phi(x)}: = \sum_{n=0}^{\infty} g^n : \hat \phi(x)^{n+1} : \,.
\ee
The position space Wightman function is naively
\be
\langle 0 |\hat O(x) \hat O(y) | 0 \rangle = \sum_{n=0}^{\infty} \frac{g^{2 n} (n+1)!  }{(4 \pi^2)^{n+1}} \frac{1}{((\vec x-\vec y)^2 - ((x^0-y^0)-i \epsilon)^2)^{n+1}} \, . 
\ee
This series is asymptotic and so ill-defined without further specification. A natural guess would be to try to Borel resum the series, however there is no guarantee that such an approach preserves unitarity.  

The `resolution' which ensures unitarity is to directly compute the spectral density as before
\ba
2 \pi \rho_{O}(\mu) &=& 2 \pi \delta(\mu) + \sum_{n=1}^{\infty} g^{2 n} (n+1)! \, \Omega_{n+1}(\mu) \, ,\\
&=&  2 \pi \delta(\mu) + \sum_{n=1}^{\infty} g^{2 n} \, \frac{n+1}{ (16 \pi^2)^{n} (n-1)! } \mu^{n-1} \, , \\
&=& 2 \pi \delta(\mu) + e^{\frac{g^2 \mu}{16 \pi^2}} \frac{g^2 (g^2 \mu+32 \pi^2)}{256 \pi^4} \, .
\ea
Unlike the position space correlation function, the spectral density is convergent (it is in effect a double Borel transform of the position space correlator - see Appendix~\ref{doubleBorel}) for all $g$. In other words this series gives an unambiguous non-perturbative definition of the Wightman function. 
The asymptotic form of the spectral density is
\be
2 \pi \rho_{O}(\mu)  \sim e^{\frac{g^2 \mu}{16 \pi^2}} \frac{g^4 \mu}{256 \pi^4}    \, ,
\ee
hence this is an example of a non-localizable field with $\alpha=1$. We see that the failure of the position space expansion to converge is now consistent with the fact that the field was not strictly localizable.  One may be concerned that $\hat O(x)$ is ill-defined due to the lack of convergence at $\hat \phi(x) = 1/g$, however we can resolve this by making $g$ complex and taking the limit to real $g$. Since the spectral density is entire, it is automatically the boundary value of its analytic continuation. There is no non-perturbative ambiguity in determining the spectral density. \\

Remarkably, this exponential behaviour is possible to anticipate based on very simple semi-classical considerations which will anticipate what we find using the Galileon duality.  The strong coupling region for the operator $\hat O(x)$ is the region $\hat \phi(x) \sim 1/g$. Consider a semi-classical field configuration which is strongly coupled inside a region of radius $r_*$, i.e. the Vainshtein/classicalization  radius. The energy of the field in this region will scale as
\be
E \sim \int_{|x| < r_*} \d^3 x  \, {{\hat \phi}'}{}^2 \sim r_* \frac{1}{g^2} \, .
\ee
Turning this around, the Vainshtein/classicalization radius for this operator is
\be
r_*(E) \sim E g^2  \sim \sqrt{\mu} g^2 \, .
\ee
Finite contributions to the action (e.g. from complex instantons) will then scale as $S(\mu) \sim E r_*(E) \sim \mu g^2$. The spectral density thus grows as
\be
\rho(\mu) \sim  e^{S(\mu)} \sim  e^{g^2 \mu}\, .
\ee
This example nicely anticipates our general theme, that the Jaffe order of growth, $\alpha$, is determined semi-classically.
 
\subsection{Higher $n$-point functions}

The construction of higher $n$-point functions proceeds analogously by using the appropriate spectral representation form for the $n$-point function \cite{Kallen:1958ifa,Kallen:1961wza,Kallen:1959kza}.
For example, for the exponential function $\hat O(x) = \sum_{n=1}^{\infty} \frac{g^n}{n!} : \hat \phi(x)^n : = : e^{g \hat \phi(x)} -1: $ we have the 3-point function
\be
\langle 0 | \hat O(x_1)\hat O(x_2)\hat O(x_3) | 0 \rangle = \sum_{n=1}^{\infty} \sum_{m=1}^{\infty} \sum_{p=1}^{\infty} \frac{g^{2(n+m+p)}}{n!m!p!} W_{12}^nW_{13}^m W_{23}^p \, ,
\ee
where 
\be
W_{ij} = \frac{1}{4 \pi^2} \frac{1}{(\vec x_i-\vec x_j)^2 - ((x_i^0-x_j^0)-i \epsilon)^2} \, .
\ee
We then express this in Fourier space as
\be
\langle 0 | \hat O(x_1)\hat O(x_2)\hat O(x_3) | 0 \rangle = \int \frac{\d^4 k_{12}}{(2 \pi)^4} \int \frac{\d^4 k_{13}}{(2 \pi)^4} \int \frac{\d^4 k_{23}}{(2 \pi)^4}   \,  e^{i \sum k_{ij}.(x_i-x_j) }\rho(k_{12},k_{13},k_{23}) \, ,
\ee
where the 3-point function spectral density is
\be
\rho(k_{12},k_{13},k_{23}) = \sum_{n=1}^{\infty} \sum_{m=1}^{\infty}  \sum_{p=1}^{\infty} \frac{g^{2(n+m+p)}}{n!m!p!} \Omega_n(-k_{12}^2)\Omega_m(-k_{13}^2)\Omega_p(-k_{23}^2) 
\ee
and it is understood that the integration is over the range $k_{ij}^2<0$ and $k_{ij}^0 >0$.

As a consequence of the exponential form of the interaction we have in this case
\be
\rho(k_{12},k_{13},k_{23}) =(2 \pi)^3 \rho_0(-k_{12}^2) \rho_0(-k_{13}^2) \rho_0(-k_{23}^2)  \, ,
\ee
where $\rho_0(-k^2)$ is the 2-point function spectral density (\ref{rho0}).  We thus see the spectral densities are still entire functions and that the high energy behaviour of the 3-point function in each $-k_{ij}^2$ is the same as that for the 2-point function. 
This implies 
\be
\langle 0 | \hat O(x_1)\hat O(x_2)\hat O(x_3) | 0 \rangle  = \langle 0 | \hat O(x_1)\hat O(x_2) | 0 \rangle \langle 0 | \hat O(x_1)\hat O(x_3) | 0 \rangle \langle 0 | \hat O(x_2)\hat O(x_3) | 0 \rangle \, . 
\ee
which was to be expected given the exponential form of $\hat O(x)$, however we should remember that this expression only has meaning after smearing with appropriate test functions.

\subsection{Canonical Quantization of Vainshtein theories and non-localizablilty}

\label{canonical}

Using the duality map, we will argue below that Galileon theories correspond to non-localizable quantum field theories. It is tempting to conjecture that this property is true of all theories exhibiting the Vainshtein/classicalization mechanism. This connection is based on the fact that in these theories, the Vainshtein/classicalization radius $r_*(E)$ grows with energy which means that the (Euclidean) action on associated semi-classical configurations grows with energy faster than a linear growth $S \sim E r_*(E)$. This simultaneously means that the semi-classical approximation is better at high energies and then we may anticipate that the spectral density of a typical operator is determined semi-classically as $\rho \sim e^S$. The order of growth of the spectral density is then
\be
\alpha  = \frac{1}{2} + \frac{1}{2} \frac{d \ln r_*}{d \ln E} >\frac{1}{2} \, ,
\ee
implying that the associated operators are non-localizable. Turning this around, any UV embedding of a theory which exhibits the Vainshtein mechanism for which the UV theory is a local field theory will inevitably destroy the Vainshtein mechanism since locality would enforce $r_*$ to decrease at high energies so that $\rho$ grows slower than an a linear exponential. This is consistent with known attempts \cite{Kaloper:2014vqa}. \\

We will now see that the non-localizability of Vainshtein type theories can be anticipated from a naive application of the rules of canonical quantization in the interacting representation. The S-matrix may be formally defined in terms of the interaction Hamiltonian density ${\cal H}_{\rm int}(x)$ as
\be
\hat S = T e^{-i \int_{-\infty}^{\infty} \d^4 x \, \hat {\cal H}_{\rm int}(x)} \, .
\ee
For the S-matrix to be Lorentz invariant, we require that this notion of time ordering is Lorentz invariant. This requires that 
\be
\left[ \hat {\cal H}_{\rm int}(x) ,  \hat {\cal H}_{\rm int}(x') \right]  =0 \,,  \quad (x-y)^2>0 \, .
\ee
In order for this to be the case the field $ \hat {\cal H}_{\rm int}(x) $ must be a strictly localizable field otherwise the product of operators in position space has no meaning. \\

In the interacting representation $\hat {\cal H}_{\rm int}(x) $ is built out of the free fields which themselves do commute outside the light cone, and so a sufficient condition for $\hat {\cal H}_{\rm int}(x) $ to be strictly localizable is that it is built out of a finite number of powers of fields and their derivatives. However, as we have seen we can also have strict localizability even with an infinite number of powers of the field, as exemplified by the exponential example $\hat {\cal H}_{\rm int}(x) = :e^{g\hat \phi(x)} : -1 $.
The characteristic feature of theories exhibiting the Vainshtein mechanism is that their Hamiltonian is built out of an infinite number of powers of fields in such a way that the Hamiltonian density is not a strictly localizable field. The origin of this is that the Vainshtein mechanism requires a modification of the kinetic term. A finite number of local corrections to the kinetic term in the Lagrangian corresponds to an infinite number of terms in the Hamiltonian as a result of the Legendre transform between them. 

\subsubsection{Anti-DBI}

\label{antiDBI}

As a simple example of this phenomena consider the DBI Lagrangian with a possible switch for the signs in the kinetic term
\be
{\cal L} =\epsilon \Lambda^4  - \epsilon \Lambda^4 \sqrt{1+ \epsilon (\partial \phi)^2/\Lambda^4 } \, .
\ee
The `right sign' DBI theories is $\epsilon = +1$ and the `wrong sign' anti-DBI theory $\epsilon=-1$. It is know that for $\epsilon =+1$ the low energy S-matrix is consistent with the possibility of a UV completion with a local theory \cite{Adams:2006sv} and in fact it is easy to find local completions using a two field system (e.g. \cite{Tolley:2009fg}) or as a brane moving in higher dimensions. On the other hand for $\epsilon =-1$ there is no local UV completion\footnote{If we construct the two field model as in \cite{Tolley:2009fg} the second field would be a ghost and so there is no unitary UV completion. If we construct the brane EFT one of the extra dimensions would be time-like and this would lead to violations of causality and unitarity.}. \\

Let us nevertheless proceed naively and construct the interacting Hamiltonian in order to canonically quantize this theory
\be
{\cal H}_{\rm int} = -\epsilon \Lambda^4 + \epsilon \Lambda^4 : \sqrt{1 + \epsilon P^2/\Lambda^4} \sqrt{1+ \epsilon (\nabla \phi)^2 } : - \frac{1}{2} : P^2 : - \frac{1}{2} : (\nabla \phi)^2 :\, ,
\ee
where $P$ is the momentum density conjugate to $\phi$. Due to the square root structure this Hamiltonian contains an infinite number of terms and so we should check whether ${\cal H}_{\rm int}(x)$ is a localizable field or not. In fact it is non-localizable which may be easily seen by analyzing each square root separately. Defining 
\be
\hat O(x) = : \epsilon \sqrt{1 + \epsilon P^2(x)/\Lambda^4} : = \sum_{n=0}^{\infty} \frac{\Gamma[3/2]}{\Gamma[3/2-n] \Gamma[n+1] }  \frac{\epsilon^{n+1} }{\Lambda^{4n}} : P^{2n}(x) :
\ee 
then we have the Wightman function
\be
\langle 0 | \hat O(x) \hat O(y) | 0 \rangle  = \sum_{n=0}^{\infty} \frac{(\Gamma[3/2])^2 (2 n)!}{(\Gamma[3/2-n] \Gamma[n+1])^2  } \frac{1}{\Lambda^{8n}} (\langle 0 | \hat P(x) \hat P(y) | 0 \rangle)^{2n} \, . 
\ee
The spectral density is given by
\be
\theta(k^0) 2 \pi \rho_{O}(k) =   \theta(k^0) 2 \pi \delta(k^2) + \sum_{n=1}^{\infty} \frac{(\Gamma[3/2])^2 (2 n)!}{(\Gamma[3/2-n] \Gamma[n+1])^2  } \frac{1}{\Lambda^{8n}}D_{2 n}(k)
\ee
where 
\ba
D_n(k) &=& \int \d^4 x \, e^{-ik.x} \left( -\frac{3 {x^0}^2+\vec x^2}{2 \pi^2 (\vec x^2- (x^0-i \epsilon)^2)^3} \right)^n \, ,\\ 
&=& \sum_{r=0}^n \frac{n!}{r! (n-r)!}   \frac{(-1)^r 4^{n-r}}{(2 \pi^2)^n} \int \d^4 x \, e^{-ik.x} \frac{1}{ (\vec x^2- (x^0-i \epsilon)^2)^{3 n-r}} (- x_0^2 )^{ (n-r)} \, , \\
&=& \sum_{r=0}^n \frac{n!}{r! (n-r)!}   \frac{(-1)^r 4^{n-r}}{(2 \pi^2)^n}   \frac{\partial^{ 2 (n-r)}}{\partial k_0^{ 2 (n-r)}}  \int \d^4 x \, e^{-ik.x} \frac{1}{ (\vec x^2- (x^0-i \epsilon)^2)^{3 n-r}} \, , \\
&=& \sum_{r=0}^n \frac{n!}{r! (n-r)!}   \frac{(-1)^r 4^{n-r}}{(2 \pi^2)^n}  (4 \pi^2)^{3n-r} \frac{\partial^{ 2 (n-r)}}{\partial k_0^{ 2 (n-r)}}  \Omega_{3n-r}(k_0^2-\vec k^2) \, .
 \ea
The asymptotics of the spectral density at large $k^0 \gg |\vec k|$ are then
\ba
D_n(k) &=& \sum_{r=0}^n \frac{n!}{r! (n-r)!}   \frac{(-1)^r 4^{n-r}}{(2 \pi^2)^n}  (4 \pi^2)^{3n-r} \frac{\partial^{ 2 (n-r)}}{\partial k_0^{ 2 (n-r)}}  \Omega_{3n-r}(k_0^2) \, , \\
D_n(k) &=& k_0^{4n-4} \sum_{r=0}^n \frac{n! (-1)^r 4^{n-r} (4 \pi^2)^{3n-r} (6n-2r-4)!}{r! (n-r)! (2 \pi^2)^{n} (16 \pi^2)^{3n-r-1} (3n-r-2)! (3n-r-1)! (4n-4)!}  \nn , \\
D_n(k) &=& k_0^{4n-4} \frac{2^n \pi^{3/2-2n} (3n-5/2)!}{(4n-4)! (3n-1)!} {}_{2}F_{1}\(1-3n,-n,5/2-3n,1/4\) \, . 
\ea
At large $n$ this may be approximated as 
\be
D_n(k) = a k_0^{4n-4} \( \frac{3}{4} \)^{n}
\ee
where $a \approx 0.85$. With this approximation we have
\ba
&&  2 \pi \rho_{O}(k) =   2 \pi \delta(k^2) + \\
&& \frac{21k_0^4 a}{8192 \pi^2 \Lambda^8} {}_PF_{Q}\( \frac{1}{2}, \frac{1}{2} , \frac{11}{12} ,\frac{13}{12} , \frac{17}{12},\frac{19}{12} ; \frac{5}{8} , \frac{7}{8} , 1,\frac{9}{8} , \frac{7}{6} , \frac{4}{3} , \frac{11}{8} , \frac{3}{2} , \frac{5}{3} , \frac{11}{6} ,2 ; \frac{9 k_0^8}{16777216 \pi^4 \Lambda^8}\) \nn
\ea
which at large $k_0$ behaves as
\be
2 \pi \rho_{O}(k) \approx   2 \pi \delta(k^2) +\frac{128 \sqrt{2} \pi}{9 k_0^4} e^{\frac{3^{4/3}}{8 \pi^{2/3}} \frac{k_0^{4/3}}{\Lambda^{4/3}}} \, .
\ee
This exponential growth is precisely what is expected from semi-classical arguments. The Vainshtein radius for this system is $r_* = \Lambda^{-1} \left( E/\Lambda \right)^{1/3}$ and so the semi-classical action scales as $ S \sim \left( E/\Lambda \right)^{4/3}$. An analogous calculation may be performed for the full interaction Hamiltonian with a similar result. \\

From an EFT point of view we need not be concerned about this behaviour since we should treat the interaction perturbatively where at each finite order of perturbations ${\cal H}_{\rm int}$ remains local. More precisely, in an EFT field theory we can add to the original Lagrangian an infinite number of increasingly irrelevant operators consistent with the symmetry which will contribute to increasing powers of $k/\Lambda$ in the spectral density computation. Thus we cannot take this exponential growth of the leading order operators too seriously since it is possible that the additional irrelevent operators cancel the high energy growth. In the case of the DBI model, $\epsilon=+1$ we expect this to be the case since we know we can UV complete this model in a local way, e.g. via a two-field system or by viewing it as the LEEFT of a brane moduli in extra dimensions. For the anti-DBI case $\epsilon=-1$ there will be no local UV completion, and this suggests that in this case there is no way to resolve the non-localizability of the interaction, and that this means that this theory is fundamentally non-localizable. However, it is also known that it is only in the case of $\epsilon=-1$ that there is an active Vainshtein mechanism when this field is coupled to sources. This suggests that the Vainshtein mechanism and the non-localizability go hand in hand. For different reasons similar conclusions were reached in \cite{Dvali:2012zc} with regards to the classicalization proposal.

\subsubsection{Goldstone model}

A similar model which is simpler at the level of the Lagrangian but more complicated at the level of the Hamiltonian is the Goldstone model where the Lagrangian is just
\be
{\cal L} = -\frac{1}{2} (\partial \phi)^2 + \frac{\epsilon}{4\Lambda^4} (\partial \phi)^4 \, .
\ee
As in the DBI case the Vainshtein radius for this system is $r_* = \Lambda^{-1} \left( E/\Lambda \right)^{1/3}$ and so we anticipate that the spectral density scales as $e^{\left( E/\Lambda \right)^{4/3}}$. Unlike the DBI case there are only a finite number of terms in the Lagrangian, nevertheless there are an infinite number of terms in the interaction Hamiltonian which takes the inelegant form
\be
{\cal H}_{\rm} = \frac{1}{2} : P^2:  + \frac{1}{2} : (\nabla \phi)^2: + \frac{\epsilon}{4\Lambda^4} :(3 v^2 + (\nabla \phi)^2) (v^2 - (\nabla \phi)^2) : \, ,
\ee
where $v$ is the solution of the cubic equation
\be
P = v+ \frac{\epsilon}{\Lambda^4} v(v^2 - (\nabla \phi)^2) \, .
\ee
As before, we may compute the spectral density of this interaction and we find, due to the cube root structure of the solution for $v$, the predicted exponential growth in the spectral density. Once more in the `right sign' case, $\epsilon =+1$, we would regard this as harmless and resolvable by adding irrelevant operators to the Lagrangian. In the wrong sign $\epsilon=-1$ case, which is the case which exhibits the Vainshtein mechanism, we expect to be stuck with the non-localizability. 

\subsubsection{Cubic Galileon}

Let us now consider how this argument applies to the relevant case of the cubic Galileon model. Here the Lagrangian is very simple
\be
{\cal L} = -\frac{1}{2} (\partial \pi)^2 \left( 1 + \frac{1}{\Lambda^3} \Box \pi \right) \, .
\ee
and the Hamiltonian interaction takes the form
\be
{\cal H}_{\rm int} = \frac{1}{2}:\frac{1}{1+ \nabla^2 \pi/\Lambda^3}P^2 :+\frac{1}{2\Lambda^3}: (\nabla \pi)^2 \nabla^2 \pi  : \, . 
\ee
The second term is manifestly local and harmless. The `dangerous' part of the Hamiltonian comes from the operator $\hat O(x) = \frac{1}{1+ \nabla^2 \pi/\Lambda^3}$. Following the same procedure if we compute the spectral density of this operator we find that it grows exponentially as $ \rho_O \sim e^{\left( E/\Lambda \right)^{6/5}}$ which is the expected behavior given that the Vainshtein radius is $r_* = \Lambda^{-1} \left( E/\Lambda \right)^{1/5}$. \\

Unlike in the previous two examples there is no sign issue since the sign of the interaction is degenerate with the field redefinition $\pi \rightarrow - \pi$. Both signs are bad in that it is known that the cubic Galileon does not admit a local Lorentz invariant UV completion \cite{Adams:2006sv}. Consequently, and consistent with the picture developed from the duality described below, we anticipate that the cubic, and indeed all Galileon models, are non-localizable field theories. \\

One interesting difference with the Galileon model is that even if we attempt to view it as an EFT, it is known that because of the Galileon symmetry all the additional operators $S_{\rm B}$ that we can write down that are Galileon invariant have additional derivatives. This means that the contributions to the spectral density from these additional operators always have a different form than those from the leading order operators. In the semi-classical regime the additional suppression amounts to a suppression by
\be
\frac{\partial }{\Lambda} \sim \frac{1}{\Lambda r_*} = \left( \frac{\Lambda}{E} \right)^{1/5} \, .
\ee
If we assume that the semi-classical approximation to the spectral density is good at high energies then we find that it will be dominated by its prediction from the leading operators $S_A$ with negligible corrections from $S_B$. In this sense, the Galileon symmetry prevents any possible additional irrelevant operators from resolving the non-localizability of the leading order ones which is consistent with the fact that the Galileon theory does not admit a local UV completion. This is of course not a rigorous argument because it could be that the semi-classical approximation is not good (despite naive expectations based on the path integral scaling), nevertheless it sits well with the known properties of Galileons.

\subsubsection{{\rm Im}plications of a non-localizable Hamiltonian}

If, as we have argued, it is the case that all Vainshtein type theories have a non-localizable interaction Hamiltonian then we are led to the following consequences:
\begin{itemize}

\item There can be no quantization of a Vainshtein type theory for which the fields are localizable. If there were it should be equivalent to the canonical procedure which requires only locality in time.

\item We can maintain the canonical quantization, i.e. causal and unitary evolution in time and micro-causality, at the price of breaking Lorentz invariance. 

\item If we insist on maintaining Lorentz invariance, the quantization will not be canonical non-perturbatively (although at any finite order in perturbations it will be canonical) since there will be no local evolution in time. 

\end{itemize}

To reiterate the last point in a different language, and to set ourselves up for our future perspective: we can potentially quantize Vainshtein type theories, e.g. Galileon models, in a Lorentz invariant unitary way, but we cannot maintain micro-causality (at least with its usual definition) since that would require a well-defined operator that generates time translations which we have excluded (we cannot maintain Lorentz invariance without that operator being localizable). At least for one Galileon model, albeit one with a trivial S-matrix, we shall be able to provide precisely such a quantization procedure using the Galileon duality transformation described in the next section.

\section{Classical Duality Map}

\label{sec:DualityMap}

In this section we review the essential details of the Galileon duality discovered in \cite{Fasiello:2012rw,deRham:2013hsa,Curtright:2012gx} (see also \cite{Creminelli:2013fxa}). We will make use of this duality in the next sections to provide a non-perturbative definition the Galileon theory which is dual to a free field. However, here we shall only be concerned with the classical properties of the map. 

The origin and role of the duality can be viewed in a number of equivalent complementary ways, as a field dependent diffeomeorphism, a field redefinition, a Legendre transformation, a change of representation of the Galileon coset, or equivalently as a choice of diffeomorphism gauge in the coset construction. We briefly review these different ideas, and also discuss how the duality extends to coupling to matter as was first observed in \cite{deRham:2014lqa}.

\subsection{The Duality as a coordinate transformation}

The Galileon duality transformations are a one parameter family of invertible field redefinitions. Given a field $\pi(x)$ we can define the dual field $\tilde \pi(\tilde x)$ via the implicit relations\footnote{We choose a slightly different sign convention as in \cite{deRham:2013hsa} so that the fields are equivalent when $s=0$, $\tilde \pi(\tilde x)=\pi(x)+\mathcal{O}(s)$.} \cite{deRham:2013hsa}
\ba
\label{D1}
\D_s: \left\{\begin{array}{rcl}
x^\mu & \longrightarrow & \tilde x^{\mu} = x^{\mu} + \frac{s}{\Lambda^{\sigma}}\partial^{\mu} \pi(x) \,  ,\\[5pt]
\varphi_\mu(x)=\p_\mu \pi(x)& \longrightarrow & \tilde \varphi_\mu(\tilde x)=\tilde \p_\mu \tilde \pi(\tilde x)=\varphi_\mu(x)
\end{array}\right.\,.
\ea
Here $\Lambda$ is a fixed energy scale and $s$ is the parameter of the group transformation and $\sigma=d/2+1$ where $d$ is the number of spacetime dimensions. 
Below we shall give explicit formula that are more practical to use than these implicit relations for actually determining the dual fields. The implicit relations are however practical to determine the transformation properties of the action.  The inverse of the transformation is $\D_s^{-1}=\D_{-s}$,
\ba
\D_{-s}: \left\{\begin{array}{rcl}
\tilde x^\mu & \longrightarrow & x^{\mu} = \tilde x^{\mu} - \frac{s}{\Lambda^{\sigma}}\tilde{\partial}^{\mu} \tilde \pi(\tilde x) \,  ,\\[5pt]
\tilde \p_\mu \tilde \pi(\tilde x) & \longrightarrow & \p_\mu \pi(x)=\tilde \p_\mu \tilde \pi(\tilde x)
\end{array}\right.\,.
\ea
These implicit relations can equivalently be written as
\ba
\label{D1 pi}
\D_s:  \pi(x)&\longrightarrow & \tilde \pi(\tilde x) = \pi( x) + \frac{s}{2 \Lambda^{\sigma}} (\partial \pi(x))^2 \, ,\\
\label{D2 pi}
\D_{-s}:  \tilde \pi(\tilde x)&\longrightarrow & \pi(x) = \tilde \pi(\tilde x) -\frac{s}{2 \Lambda^{\sigma}} (\tilde \partial \tilde \pi(\tilde x))^2 \,.
\ea
The infinitesimal form of the duality transformation is
\be
\delta \pi(x) = \tilde \pi(x) - \pi(x)= - \frac{s}{2 \Lambda^\sigma}  (\partial \pi(x))^2+ {\cal O}(s^2) \, .
\ee
We see that the infinitesimal transformation is a local field redefinition.

\subsection{Duality Group}
The duality map forms a continuous group which can be extended to the full group $GL(2,R)$ considered in \cite{Kampf:2014rka}. We shall however focus on the only truly non-trivial part of the group determined by the above transformation laws. Performing a second duality transformation with parameter $s'$ starting from $\tilde \pi(\tilde x)$ and denoting the new dual field as $\hat \pi(\hat x)$ we have
\ba
\D_{s'}: \left\{\begin{array}{rcl}
\tilde x^\mu & \to&  \hat x^\mu = \tilde x^\mu +\frac{s'}{\Lambda^\sigma}\tilde \p^\mu\tilde \pi (x)= x^\mu+\frac{s+s'}{\Lambda^\sigma}\p^\mu\pi (x)\\
\tilde \pi(\tilde x) &\to& \hat \pi (\hat x)= \tilde \pi(\tilde x)+\frac{s'}{\Lambda^\sigma}\(\tilde \p \tilde \pi(\tilde x)\)^2
=\pi(x)+\frac{s+s'}{\Lambda^\sigma}\(\p \pi(x)\)^2
\end{array}
\right. \,,
\ea
where we used the relations \eqref{D1} and \eqref{D1 pi}. This leads to  a combined transformation
\ba
\D_{s'}\circ\D_s = \D_{s+s'}\,.
\ea
In other words the duality map forms an abelian group with transformation law $s''=s+s'$. 
Again we note that this group transformation leaves invariant the derivatives of the Galileon fields
\ba
\partial_{\mu} \pi(x) = \tilde \partial_{\mu} \tilde \pi (\tilde x) = \hat \partial_{\mu} \hat \pi (\hat x)\, .
\ea
Defining the Galileon invariant combination
 \ba
\Pi\mupn(x)=\frac{1}{\Lambda^{\sigma}}\eta^{\mu\alpha}\partial_\alpha\partial_{\nu} \pi(x)\,,
\ea
and similarly $\tilde \Pi\mupn(x)=\tilde \partial^{\mu}\tilde \partial_{\nu} \tilde \pi(\tilde x)/\Lambda^{\sigma}$ and with the hat variables then we have in matrix notation the relations
\ba
\label{Pi 4}
&& \tilde \Pi^{-1} = \Pi^{-1}+s\  \mathbb{I}  \, ,\\
\label{Pi 5}
 && \hat \Pi^{-1} = \tilde \Pi^{-1}+ s'\   \mathbb{I}\, , \\
\label{Pi 6}
 && \hat \Pi^{-1} = \Pi^{-1}+ (s+s')\  \mathbb{I} \, .
\ea
A particularly useful result is that
\be
{\rm det}[1+s \Pi] {\rm det}[1-s \tilde \Pi]  =1 \, .
\ee

\subsection{Covariant Coset construction}

\label{sec:coset}

Here we review the coset construction for the Galileon duality \cite{Creminelli:2014zxa,Kampf:2014rka} emphasizing the duality as a choice of diffeomorphism gauge. 
Our starting point is the Galileon group Gal(d,1) which has been shown to be the natural starting point for the coset construction of the Galileon models \cite{Goon:2012dy}. The full set of generators are momenta $P_A$, Lorentz transformations $M_{AB}$, Galileon boosts $K_A$ and Galileon translation (shift symmetry) $C$. The Galileon algebra is (we use physics convention which includes the factor of $i$ )
\ba
&& [M_{AB},M_{CD}] =i  \left( \eta_{AC} M_{BD} - \eta_{AD} M_{BC} -\eta_{BC} M_{AD} + \eta_{BD} M_{AC}  \right) \\
&& [M_{AB},P_C] = i\eta_{AC} P_B-i \eta_{BC} P_A \\
&& [M_{AB},K_C] = i \eta_{AC} K_B-i \eta_{BC} K_A \\
&& [P_A,K_B] = i \eta_{AB} C \\
&& [C,P_A] = [C,K_A]=0
\ea
It is transparent that in this algebra $P_A $ and $K_A$ play the same role. Furthermore we may take linear combinations of $P_A$ and $K_A$ to define the dual Galileon boosts $\bar K_A = K_A -s  P_A$ with the algebra remaining unchanged under the replacement $K_A \rightarrow \bar K_A$ \cite{Creminelli:2014zxa}. This ambiguity is the origin of the duality transformation. However rather viewing it as an automorphism of the algebra, we shall reinterpret as a change in coordinates, i.e. diffeomorphism gauge. This point of view was emphasized in \cite{Kampf:2014rka} and more clearly connects with the interpretation of the duality in massive gravity. \\

When dealing with the breaking of spacetime symmetries, in defining a coset it is usual to split the generators into the broken and unbroken generators giving special role to the unbroken translations (for a recent discussion and references see \cite{Delacretaz:2014oxa}). For instance for the current breaking pattern $Gal(d,1)/ISO(d-1,1)$, Galileon boosts $K_A$ and shifts $C$ are broken, and rotations $M_{AB}$ unbroken. Translations generated by $P_A$ are assumed unbroken.  The coset element is then parameterized in an asymmetric way in which we include the broken generators and the unbroken translations
\be
\Omega = e^{i P_{A} x^A} e^{i \pi(x) C} e^{i \phi^A(x) K_A} \, ,
\ee
where $\pi(x)$ and $\phi^A(x)$ are the Goldstone fields from the broken Galileon shifts and boosts.
However in the specification of any space, a given point $x$ spontaneously breaks translations. Thus it is more symmetric to view that all translations are always broken, in other words the coset is $G/H = Gal(d,1)/SO(d-1,1)$ so that $H$ is the isotropy (not isometry) group. This is precisely what we mean when we specify Minkowski spacetime as the coset Minkoswki= $ISO(d-1,1)/SO(d-1,1)$.\\

Since all translations are being viewed as broken, we express the coset element in terms of only broken generators with associated Goldstone fields
\be
\Omega = e^{i P_{A} Y^A(x)} e^{i \phi^A(x) K_A} e^{i \pi(x) C}  \, .
\ee
In this picture $Y^A(x)$ are the Goldstone fields for the translations which are viewed as broken by the need to define a point $x^A$. In other words these are the coset analogues of the \stu fields in the massive gravity/bi-gravity construction of the duality \cite{Fasiello:2012rw,deRham:2013hsa,deRham:2014lqa}.
\\

Since $\Omega $ is a one-form, it is manifestly invariant under diffeomorphisms even though we have not chosen to gauge translations. This is as it should be since the coset formalism, as is well known, separates the concept of gauged translations from diffeomorphisms. Thus even though there is no `gravity', we must demand that the actions we construct respect diffeomorphism invariance and this gives any theory of a nonlinearly realized spacetime symmetry a gravitational flavor. \\

The Maurer-Cartan one-form is given by
\ba
\label{MC1}
-i \Omega^{-1} d \Omega &=&  dY^A P_A + d \phi^A K_A + ( \phi_A dY^A+ d \pi) C  \nonumber \\
&=& \omega_P^A P_A + \omega_K^A K_A + \omega_C C
\ea
where we have introduced the one-forms $\omega_P^A, \omega_K^A,\omega_C$ for future notational convenience. In this form $P_A$ and $K_A$, and hence $Y^A$ and $\phi^A$  are treated symmetrically. Indeed defining $\hat \pi  = \pi - Y^A \phi_A$ we have equivalently 
\be
-i \Omega^{-1} d \Omega =  dY^A P_A + d \phi^A K_A + (-Y_A d\phi^A+ d \hat \pi) C  
\ee
which is equivalent to (\ref{MC1}) under the interchange $Y,P \leftrightarrow -\phi,-K$.

\subsection*{Legendre transformation}

The next step in the Galileon coset construction is to impose the inverse Higgs constraint \cite{Goon:2012dy}
\be
\omega_C =  \phi_A dY^A+ d \pi = -Y_A d\phi^A+ d \hat \pi = 0 \, .
\ee
These equations can be formally solved as $\phi^A = - \partial \pi/\partial Y_A$ and $Y^A = - \partial \hat \pi/\partial \phi^A$ so that
\ba
&& \hat \pi = \pi - Y^A \partial \pi/\partial Y_A  \\
&& \pi = \hat \pi - \phi^A \partial \hat \pi /\partial \phi^A \, ,
\ea
which are just a Legendre transform and its inverse. This confirms that the coset construction naturally reproduces the Legendre transform picture for the duality described in \cite{Curtright:2012gx}.

\subsection*{Galileon Wess-Zumino actions}

The one-forms $\omega_P^A, \omega_K^A,\omega_C$ are the natural building blocks from which the Galileon Lagrangians are constructed. The resulting action should be diffeomorphism invariant and Lorentz invariant. Since we are imposing the inverse Higgs constraint $\omega_C=0$ then naively we can only construct the Lagrangian directly from $\omega_P^A$ and $\omega_K^A$. However, it transpires that all wedge product terms are total derivatives and all other terms will lead to higher order equations of motion. The resolution is to construct the Galileon Lagrangians as Wess-Zumino terms \cite{Goon:2012dy} by introducing an auxiliary (d+1)-th dimension and defining the Lorentz-invariant (d+1) co-cycles
\be
A_n = \epsilon \, \omega_C (\wedge \, \omega_K)^{n-1} (\wedge \, \omega_P)^{d+1-n} 
\ee
for $n=1 \dots d+1$ and we have used a shorthand notation where it is understood that all the Lorentz indices are contracted with the Levi-Civita symbols so for example in $d=4$ $A_2 = \epsilon_{ABCD}  \, \omega_C \wedge \omega_P^A\wedge \omega_P^B\wedge \omega_P^C\wedge \omega_K^D $. The $A_n$ are all exact
\be
A_n = d \beta_n \, ,
\ee 
where
\be
\beta_n=\epsilon \pi (\wedge \, \omega_K)^{n-1}  (\wedge \, \omega_P)^{d+1-n}  - \frac{(n-1)}{2 (d+2-n)}\phi^A \phi_A  \,  \epsilon (\wedge \, \omega_K)^{n-2} (\wedge \, \omega_P)^{d+2-n} \, .
\ee
To see this it is enough to use $d \omega_P = d \omega_K=0$ and so
\be
d \beta_n = \epsilon d \pi (\wedge \, \omega_K)^{n-1}  (\wedge \, \omega_P)^{d+1-n} - \frac{(n-1)}{(d+2-n)}\phi^A d \phi_A  \,  \epsilon (\wedge \, \omega_K)^{n-2} (\wedge \, \omega_P)^{d+2-n} \, ,
\ee
and noting that after a little rearrangement  \cite{Goon:2012dy}
\be
(n-1) \phi^A d \phi_A  \,  \epsilon (\wedge \, \omega_K)^{n-2} (\wedge \, \omega_P)^{d+2-n} =- (d+2-n)  \phi^A d Y_A  \,  \epsilon (\wedge \, \omega_K)^{n-1} (\wedge \, \omega_P)^{d+1-n} .
\ee
whence $d \beta_n = \epsilon \, ( \pi + \phi^A dY_A) (\wedge \, \omega_K)^{n-1} (\wedge \, \omega_P)^{d+1-n} = A_n$.
Since the $A_n$ are exact, the Wess-Zumino terms can be written as
\be
S_n=\int_{M^{d+1}} A_n = \int_{M^d} \beta_n  \, .
\ee
These define the terms that arise in the action $S_A$.

\subsection*{Diffeomorphism gauge choice}

Let us reiterate that all of the above formula are manifestly diffeomorphism invariant and we have made no choice between $P_A$ or $K_A$ about which generators represent translations and which Galileon boosts. To write the action in standard field theory language we must choose a gauge. Since both $Y^A$ and $\phi^A$ are Lorentz vectors, it is natural to choice a one-parameter ($s$) family of gauges in the form
\be
Y^A(x) + s \, \phi^A(x) = x^A \, ,
\ee
where $x^A$ are the coordinates on the manifold. Implicit in this gauge choice is the identification of $\bar K_A = K_A - s P^A$ as the generator of Galileon boosts since $P_A Y^A+ K_A \phi^A = P_A x^A + \bar K_A \phi^A$. However it is important to stress that by phrasing this as a diffeomorphism gauge choice, we have disassociated ourselves with the identification of who is the correct generator of Galileon boosts or translations. It is clear that any choice for $s$ is equivalent and since this is just a gauge choice they must lead to equivalent representations of the physics. This is exactly how the duality is viewed in the context of massive gravity and bigravity \cite{Fasiello:2013woa,deRham:2013hsa,deRham:2014lqa}, where it amounts to a one-parameter choice of gauges for the St\"uckelberg fields $\phi_{\rm Stuck}^A = x^A + s \partial^A \pi$. Furthermore, as shown in \cite{deRham:2014lqa}, the duality easily extends to the coupling to any matter fields simply by writing the standard diffeomorphism invariant Lagrangian for matter coupled to $\phi^A$ and $Y^A$ and their derivatives in a diffeomorphism invariant way.
\\

In this gauge the inverse Higgs constraint implies
\be
d\pi +\phi_A(dx^A - s d \phi^A) =0 \, .
\ee
Defining $\tilde \pi  = \pi  - \frac{1}{2} s \phi^A \phi_A$ this amounts to  $d \tilde \pi  + \phi_A  dx^A =0$ which is easily solved as
\be
\phi_A(x)  = - \partial_{\mu} \tilde \pi(x) \, ,
\ee
$\tilde \pi$ is the dual Galileon field to $\pi$ under a finite $s$ transformation defined through the duality map $\tilde \pi = \pi -\frac{1}{2} s \partial_{\mu} \tilde \pi \partial^{\mu} \tilde \pi$.
We note in particular that
\be
\(\frac{\partial  \pi}{\partial y}  + \phi_{A} \frac{\partial Y^A}{\partial y}\) = \(\frac{\partial  \pi}{\partial y}  -s  \phi_{A} \frac{\partial \phi^A}{\partial y}\)=\frac{\partial  \tilde \pi}{\partial y} \, ,
\ee
where $y$ is the auxiliary (d+1)-th dimension, so that the Galileon terms simply become (up to a constant)
\be
S_n = \int_{M_4 }\,  \tilde \pi  \, \epsilon   (\wedge \, \omega_P)^{5-n} (\wedge \, \omega_K)^{n-1} \, .
\ee
It thus remains only to express $\omega_P$ and $\omega_K$ in terms of $\tilde \pi$
\ba
&& \omega_P^A = dx^A +s d \partial^A \tilde \pi(x) = (\delta^A_B +s \partial^A \partial_B \tilde \pi(x)) dx^B \, , \\
&& \omega_K^A = -  d \partial^A \tilde \pi(x) = ( - \partial^A \partial_B \tilde \pi(x)) dx^B \, .
\ea
Defining $\omega_{\bar P}^A = \omega_P^A+s \omega_K^A$ we have $\omega_{\bar P}^A = dx^A$ and so 
\ba
S_n &=& \int_{M_d }\,  \tilde \pi  \, \epsilon   (\wedge \, (\omega_{\bar P}^A-s \omega_K^A)^{d+1-n} (\wedge \, \omega_K)^{n-1} \, , \\
S_n &=& \sum_{r=0}^{d+1-n} \frac{(d+1-n)!}{(d+1-n-r)! r!} (-s)^r \tilde S_{n+r} \, ,
\ea
where $\tilde S_n$ denotes the usual Galileon terms with the simple replacement $\pi \rightarrow \tilde \pi$. This is precisely the duality map identified in \cite{deRham:2013hsa,Curtright:2012gx}. This is consistent with \cite{Creminelli:2014zxa,Kampf:2014rka}.  \\

Let us reiterate that the value of $s$ is purely a gauge choice. Provided there is no diffeomorphism anomaly, it is impossible for any physics to depend on the precise value of $s$. In this sense all representations of the coset, i.e. choices of $s$, are equivalent. 

\subsection{Duality in the presence of matter}

The duality transformation easily extends to the case of Galileons coupled arbitrarily to matter \cite{deRham:2014lqa}. This result is clear once we recognize the duality as a diffeomorphism, be it in the coset language or the massive gravity/bigravity language.  
The matter Lagrangian and the coupling to matter remain local after the duality transformation. 

In this article we will be mostly concerned with two specific duality frames. For this reason we shall set the group parameter $s=1$ from now on. The first frame will consist of a $(d+1)$'th order Galileon field $\pi$ which is dual to a free field coupled arbitrarily to a scalar matter field $\chi$.  This frame will naturally be called the $(\pi,\chi)$ frame.  The second frame will consist of the free field $\rho$ (dual to $\pi$) and its associated scalar matter field $\tilde \chi$ (dual to $\chi$) which we will call the $(\rho,\tilde \chi)$ frame.  Below we review how to transform from one frame to another.

\subsubsection*{$(\pi$, $\chi)$ Duality frame}

The Galileon action which is dual to a free field can be conveniently written as
\be\label{gwm}
S[\pi,\chi]=\int \d^dx \left[  -\frac{1}{2}\det(1+\Pi(x))(\partial\pi(x))^2+{\cal L}_m(\pi(x),\chi(x)) \right] \ .
\ee
where ${\cal L}_m(\pi,\chi)$ represents any arbitrary coupling the Galileon $\pi$ may have with a matter field $\chi$.  The above action is the system we wish to quantize and we will do so by motivating a definition of the quantized version using the Galileon duality.

\subsubsection*{$(\rho$, $\tilde \chi)$ Duality frame}

To transform $\eqref{gwm}$ into the $(\rho,\tilde \chi)$ frame we notice we use the basic duality map
\ba
&& \tilde x^{\mu} = x^{\mu} + \frac{1}{\Lambda^{\sigma}} \partial^{\mu} \pi(x)  \\
&&  x^{\mu} = \tilde x^{\mu} - \frac{1}{\Lambda^{\sigma}} \partial^{\mu} \rho(\tilde x) \\
&& \chi(x) = \tilde \chi(\tilde x)
\ea
and defining $\Sigma_{\mu\nu} = \partial_{\mu} \partial_{\nu} \rho/\Lambda^{\sigma}$ then using the identities 
\ba
&& \int \d^dx\ \det(1+\Pi(x))=\int \d^d\tilde{x}\\
&& \int \d^dx=\int \d^d\tilde{x}\ \det{(1-\Sigma(\tilde{x}))}\\
&& (\partial\pi(x))^2=(\tilde{\partial}\rho(\tilde{x}))^2
\ea
in the action we have
\ba
S[\pi,\chi]&=&\int \d^dx \left[  -\frac{1}{2}\det(1+\Pi(x))(\partial\pi(x))^2+{\cal L}_m(\pi(x),\chi(x)) \right] \nn\\
&=&\int\d^d\tilde{x} \left[  -\frac{1}{2}(\tilde{\partial}\rho(\tilde{x}))^2+\det{(1-\Sigma(\tilde{x}))}{\cal L}_m\left(\rho(\tilde{x})-\frac{1}{2}\frac{1}{\Lambda^{\sigma}}(\tilde{\partial}\rho(\tilde{x}))^2,\tilde{\chi}(\tilde{x})\right)\right] \nn\\
&=&\int\d^dx \left[  -\frac{1}{2}(\partial\rho(x))^2+\det{(1-\Sigma(x))}{\cal L}_m\left(\rho(x)-\frac{1}{2}\frac{1}{\Lambda^{\sigma}}(\partial\rho(x))^2,\tilde{\chi}(x)\right) \right] \nn\\
&=&\tilde{S}[\rho,\tilde \chi] \ .
\ea
In the absence of matter we simply have
\be
S_{\text{Galileon}}[\pi]=\int \d^dx \left[  -\frac{1}{2}\det(1+\Pi(x))(\partial\pi(x))^2 \right] =\int\d^dx \left[ -\frac{1}{2}(\partial\rho(x))^2 \right] =S_{\text{free}}[\rho]\ .
\ee

\section{Galileon Wightman functions}

\label{sec:Wightman}

In a local field theory, the set of position space Wightman functions encodes all the information about the system. As we have argued, in theories which exhibit gravitational non-locality, the position space Wightman function may simply not exist in position space, without appropriate smearing, or at best may exist only in regions of spacetime separations (the quasi-local case). This is the price that must be paid for the exponentially growing spectral density associated with gravitational non-locality. However what may exist, and be calculable as we now show, are their Fourier transforms, i.e. momentum space representation of the Wightman functions. The momentum space Fourier transforms can be encoded in a set of spectral densities by a simple generalization of the \KL representation for the two point function. 

In this section we will provide an explicit computation of this spectral density for the case of the $(d+1)$'th Galileon which is classically equivalent to a free theory. We will provide an operational definition of the quantum theory, and show that this definition is meaningful in a basis of a complete set of momentum eigenstates. 

\subsection{Closed formula for the Duality Map}

The implicit nature of the duality map as defined in the previous section makes it unclear how to perform the duality transformation beyond perturbation theory. We shall now present a useful set of closed formulae for the duality map which can be used to give a non-perturbative definition. We shall make explicit use of these formula in what follows in defining the quantum theory. \\

To begin with let us consider how to construct the field $\pi(x)$ out of $\rho(x)$. We can define $\pi(x)$ by its Fourier transform as
\be
\pi(x) = \int \frac{\d^d k }{(2 \pi)^d} \,  \pi(k) \, e^{ik.x}
\ee
We may then determine $\pi(k)$ from the inverse Fourier transform
\be
\pi(k) = \int \d^d y \, e^{-i k.y} \pi(y) \, .
\ee
The trick is to now perform the duality transformation within this integral. According to the above map this is then simply
\be
\pi(k) =  \int \d^d \tilde y \,  \det{(1-\Sigma(\tilde{y}))}\ \, e^{-i k.( \tilde y- \Lambda^{-\sigma} \tilde \partial \rho(\tilde y)) } \left( \rho(\tilde y)-\frac{1}{2\Lambda^{\sigma}}(\tilde \partial\rho(\tilde y))^2 \right) \, .
\ee
Changing the dummy integration variable $\tilde y $ to $y$ we obtain
\be
\pi(x)= \int  \frac{\d^d k }{(2 \pi)^d} \int \d^d y  \,  e^{ik.(x-y)} \, U(\rho(y)) \, e^{\frac{i k. \partial \rho(y)}{\Lambda^{\sigma} } }  \, ,
\ee
where 
\be
U(\rho(y)) =  \det{(1-\Sigma(y))} \left( \rho(y)-\frac{1}{2\Lambda^{\sigma}}(\partial\rho( y))^2 \right)  \, .
\ee
This formula is now {\bf explicit} unlike the previous duality formula. Furthermore it is non-perturbative, there is no need to perform an expansion in inverse powers of $\Lambda$. \\

We can perform similar transformations to infer $\tilde \chi$ from $\chi$ and $\rho$. Specifically
\be
\tilde \chi(\tilde x) = \chi(x) = \chi(\tilde x - \Lambda^{-\sigma} \tilde \partial \rho(\tilde x) )  
\ee
Switching the dummy label $\tilde x \rightarrow x$ we have the simpler formula
\be
\tilde \chi( x) = \chi(x - \Lambda^{-\sigma} \partial \rho(x) ) \, , 
\ee
or in terms of the Fourier transform of $\chi(x)$
\be
\label{eq:dualitymatter2}
\tilde \chi( x) = \int  \frac{\d^d k }{(2 \pi)^d}  e^{ik.x } \, \chi_k \, e^{\frac{-i k. \partial \rho(x)}{\Lambda^{\sigma} }} \, .
\ee
We see that both of these formula contain the crucial exponential factors $e^{\frac{-i k. \partial \rho(x)}{\Lambda^{\sigma} }} $ which will be responsible for much of the quantum properties.

\subsection{Formal definition of Galileon operators ($\pi,\chi$) frame}

Our aim is to define the quantum theory of a $(d+1)$'th Galileon which is dual to a free theory. At the perturbative level this map is unambiguous and a direct perturbative computation of the S-matrix demonstrates that the $(d+1)$'th Galileon theory is indeed free. However, within the loop diagrams we will nevertheless need to, in a given quantization scheme, add an infinite number of irrelevant counterterms in order to verify this statement. These counterterms arise from the measure of the path integral and are typically power law divergent.
An analogous statement is that in attempting to define the Galileon operators, we must choose a very specific ordering of the operators to maintain the equivalence with a free theory. We shall now see that normal ordering is sufficient to achieve this and remove all UV divergences. In this way, by the simple act of normal ordering, we may give a {\it formal definition of a unitary quantum theory which is classically equivalent to the $(d+1)$'th Galileon theory}. This will be equivalent to the theory obtained perturbatively via canonical quantitzation with a specific choice of counterterms but non-perturbatively it is not canonical.  \\

The construction method proceeds as follows: If the $(d+1)$'th Galileon theory is dual to a free theory at the quantum level then it must occupy the same Hilbert space, the Fock space of a free quantum field ${\cal H}={\cal F}$. Given this, our goal is to define an operator $\hat \pi(x)$ which acts in this Hilbert space, and in the classical limit is equivalent to the classical Galileon field. Any such operator must be built out of creation and annihilation operators acting in the Fock space which are used to define the dual field $\hat \rho(x)$. In other words the operator $\hat \pi(x)$ must be built out of the operator $\hat \rho(x)$. 

At the classical level the duality map in $d$ dimensions can be expressed as
\be
\pi(x) = \int \d^d y \int \frac{\d^d k}{(2 \pi)^d} \, U(\rho(y)) \, \, e^{ i k.(x-y) + \frac{i k . \partial \rho(y)}{\Lambda^\sigma}} \, . 
\ee
We thus define at the quantum level the following operator
\be
\label{eq:operatordefinition}
\hat \pi(x) = \int \d^d y \int \frac{\d^d k}{(2 \pi)^d} \, e^{i k.(x-y)} : U(\hat \rho(y)) \, \, e^{ \frac{i k . \partial \hat \rho(y)}{\Lambda^\sigma}} : \, ,
\ee
where $::$ denotes the normal ordering with respect to $\hat \rho$'s creation and annihilation operators and
\be
\hat \rho(x) = \int \d \tilde k \,  \left[ \hat a_k e^{i k.x} + \hat a_k^{\dagger}  e^{-i k.x}  \right]\, .
\ee
 In choosing this normal ordering we are automatically discarding self-contractions, i.e. tadpole diagrams, when we come to define correlation functions. Of course, from a perturbative point of view this corresponds to an infinite number of subtractions. It is in this step that we have implicitly defined the UV theory which will contain no additional divergences, at least for those quantities that have physical meaning. This procedure is analogous to the definition of vertex operators in string theory which are also normal ordered. However, a fundamental difference is that the string vertex operators will have a fixed conformal dimension whereas these operators do not exhibit any conformal behaviour at high energies. Our definition of the UV will not be associated with any UV fixed point of an RG flow. This is reflected in the fact that the high energy behaviour strongly depends on the dimensionful scale $\Lambda$. This is the `asymptotic fragility' idea \cite{Dubovsky:2012wk}. \\

The operator definition (\ref{eq:operatordefinition}) is further justified by the fact that if we evaluate this operator between two coherent states for the $\rho$ field, then we will automatically get the classical configuration for $\pi$ which is associated with the appropriate classical configuration for $\rho$
\be
\langle \beta | \hat \pi(x) | \alpha \rangle = \pi_c(x) \, ,
\ee
where 
\be
\pi_c(x) = \int \d^d y  \int \frac{\d^d k}{(2 \pi)^d} \, U(\rho_c(y)) \, \, e^{ i k.(x-y) + \frac{i k . \partial \rho_c(y)}{\Lambda^\sigma}} \, . 
\ee
and
\be
\rho_c(x) =\langle \beta | \hat \rho(x) | \alpha \rangle = \int \d \tilde k \,  \left[ \alpha_k e^{i k.x} + \beta_k^*  e^{-i k.x}  \right]\, .
\ee
In other words this definition corresponds to working in the coherent state representation for the path integral and performing the duality transformation of the coherent state variables.   \\

We stress again that this definition of the operator is only formal, since we do not expect the position space correlators to exist without appropriate smearing. More precisely we will find it is well-defined in perturbation theory, which amounts to expanding the exponential out, but becomes ill-defined non-perturbatively without the process of smearing. The correlation functions are tempered distributions at finite order in perturbation theory, but in the non-perturbative limit they are no longer tempered. With this in mind we may define a regulated local field as follows\footnote{This is in effect the approach proposed by Taylor to consider locality for non-localizable field theories \cite{Taylor:1971gd,Blomer:1971ie}.}
\be
\label{piN}
\hat \pi_N(x) = \int \d^d y  \int \frac{\d^d k}{(2 \pi)^d} \, e^{i k.(x-y)} : U(\hat \rho(y)) \, \, \sum_{n=0}^N \frac{1}{n!} \left( \frac{i k . \partial \hat \rho(y)}{\Lambda^\sigma} \right)^n: \, .
\ee
This operator is a well-defined local field and interactions built out of this operator will satisfy the conventional axioms of local field theories, including polynomial boundedness and temperedness of the correlation functions. Our concern then is in taking the limit $N \rightarrow \infty$. We will see that this limit is ill-defined in position space without smearing, but well-defined in momentum space for the Wightman functions. \\

The non-existence of the position space Wightman functions may be traced to the fact that the vacuum correlation of two of these fields will formally take the form
\ba
\langle 0 | \hat \pi(x) \hat \pi(x') | 0 \rangle &=&\int \d^d y \int \d^d y'  \int \frac{\d^d k}{(2 \pi)^d} \, e^{i k.(x-y)} \int \frac{\d^d k}{(2 \pi)^d} \, e^{i k'.(x'-y')} \\
&& \langle 0 |  : U(\hat \rho(y)) \, \, e^{ \frac{i k . \partial \hat \rho(y)}{\Lambda^\sigma}} : : U(\hat \rho(y')) \, \, e^{ \frac{i k' . \partial \hat \rho(y')}{\Lambda^\sigma}} : | 0 \rangle \, ,
\ea
which by translation invariance of the vacuum expectation value ($k'=-k$) is
\ba
\langle 0 | \hat \pi(x) \hat \pi(x') | 0 \rangle &=&\int \d^d y  \int \frac{\d^d k}{(2 \pi)^d} \, e^{i k.(x-x'-y )}  \langle 0 |  : U(\hat \rho(y)) \, \, e^{ \frac{i k . \partial \hat \rho(y)}{\Lambda^\sigma}} : : U(\hat \rho(0)) \, \, e^{ \frac{-i k . \partial \hat \rho(0)}{\Lambda^\sigma}} : | 0 \rangle \, , \nn \\
&=&\int \d^d y   \int \frac{\d^d k}{(2 \pi)^d} \, e^{i k.(x-x'-y )} F(y,k)  \langle 0 | :e^{ \frac{i k . \partial \hat \rho(y)}{\Lambda^\sigma}} : : e^{ \frac{-i k . \partial \hat \rho(0)}{\Lambda^\sigma}} : | 0 \rangle
\ea
where the function $F(y,k)$ is a polynomial in $k$ coming from the contractions of the exponential with $U$ and two $U$s together. Since this part contains only a finite number of terms and does not affect the convergence of the integral we ignore for now its precise form.\\

Using the elementary result that 
\be
 \langle 0 | :e^{ \frac{i k . \partial \hat \rho(y)}{\Lambda^\sigma}} : : e^{ \frac{-i k . \partial \hat \rho(0)}{\Lambda^\sigma}} : | 0 \rangle = e^{- \frac{1}{2} k^{\mu} k^{\nu} \partial_{\mu} \partial_{\nu} \langle 0|\rho(y) \rho(0) | 0 \rangle/\Lambda^{2 \sigma}}
 \ee
and the expression for the $d$ dimensional Wightman function for a free massless field
\ba
\langle\hat{\rho}(x)\hat{\rho}(0)\rangle=\frac{a_d}{(d-2)(|\vec x|^2 - (x^0 - i \epsilon)^2)^{{(d-2)}/2}} \, ,
\ea
where 
\be
a_d= \frac{(d-2)}{4} \pi^{-d/2} \Gamma(d/2-1) 
\ee
we find 
\ba
\hspace{-20pt} \langle 0 | \hat \pi(x) \hat \pi(x') | 0 \rangle &=& \int \d^d y  \int \frac{\d^d k}{(2 \pi)^d} \, e^{i k.(x-x'-y)} F(y,k)  e^{- \frac{1}{2} k^{\mu} k^{\nu} \partial_{\mu} \partial_{\nu} \langle 0|\rho(y) \rho(0) | 0 \rangle/\Lambda^{2 \sigma}} \, , \\
&=& \int \d^d y  \int \frac{\d^d k}{(2 \pi)^d} \, e^{i k.(x-x'-y )} F(y,k)  \exp\left[\frac{a_d}{\Lambda^{2 \sigma}}\left(\frac{k^2}{y_-^d}-d\frac{(k\cdot y)^2}{y_-^{d+2}}\right)\right] \, ,
\ea
where by $1/y_-^{d}$ we mean
\be
\frac{1}{y_-^{d}} = \frac{1}{(|\vec y|^2 - (y^0 - i \epsilon)^2)^{{d}/2}} \, .
\ee
However this expression is meaningless as it stands since for any dimension $d>1$ regardless of whether $k$ is spacelike or timelike, there is always at least one direction in which the integrand grows exponentially and this cannot be compensated for by the polynomial behaviour of $F(y,k)$. Indeed, this expression is equally ill-defined in the Euclidean, and so we may not Wick rotate and use the Osterwalder-Schrader reconstruction, nor define this function as the boundary value in the complex time plane as usual in axiomatic formulations of the Wightman functions. \\

By contrast, the computation of $\langle 0 | \hat \pi_N(x) \hat \pi_N(x') | 0 \rangle $ is well-defined and gives rise to a specific generalized function/distribution which may be extended to the Euclidean. This suggests that we should look for quantities that are well defined in the limit $N \rightarrow \infty$.
It is conceivable that $\langle 0 | \hat \pi(x) \hat \pi(x') | 0 \rangle$ may be defined by analytic continuation, as was considered often in the context of the Efimov-Fradkin/superpropagator method. While it is possible to analytically continue this integral into a finite integral, this mathematical process will in general contradict the physical requirement of unitarity. Since our aim is to construct a manifestly unitary theory we regard such mathematical manipulations as unacceptable. Indeed many of the problems associated with this historical approach to non-renormalizable field theories can be attributed to using analytic continuation in a way which breaks unitarity. \\
  
 The clue to making progress is to note that the Fourier transform of the operator $\hat \pi(k) = \int \d^d x \, e^{-i k.x} \hat \pi(x)$ is well-defined when evaluated between two multi-particle momentum eigenstates. To see this we need to use the basic fact that for every Heisenberg operator we can write $\hat \rho(x) = e^{-i \hat P.x} \hat \rho(0) e^{ i \hat P.x}$ and similarly $\partial_{\mu}\hat \rho(x) = e^{-i \hat P.x}  \partial_{\mu} \hat \rho(0) e^{ i \hat P.x}$. 
 With this in mind we may evaluate $\hat \pi(k)$, for which 
 \be
 \hat \pi(k) = \int d^d x \, e^{-i k.x} : U(\hat \rho(x)) e^{\frac{i k. \partial \hat \rho(x)}{\Lambda^{\sigma}}}: \, ,
 \ee
 between two momentum eigenstates as follows
 \ba
 \langle P_f | \hat \pi(k) | P_i \rangle &=& \int \d^d x e^{-i k.x }  \langle P_f | : U(\hat \rho(x)) \, \, e^{ \frac{i k . \partial \hat \rho(x)}{\Lambda^\sigma}} :  | P_i \rangle \\
 &=& \int \d^d x\,  e^{-i k.x}  \langle P_f | e^{-i \hat P_f.x}: U(\hat \rho(0)) \, \, e^{ \frac{i k . \partial \hat \rho(0)}{\Lambda^\sigma}} :  e^{i \hat P_i.x} | P_i \rangle \\
  &=& \int \d^d x \, e^{-i k.x}  e^{i (P_i-P_f).x}\langle P_f | : U(\hat \rho(0)) \, \, e^{ \frac{i k . \partial \hat \rho(0)}{\Lambda^\sigma}} :   | P_i \rangle \\
   &=& (2\pi)^d \delta^d(k-(P_i-P_f))\langle P_f | : U(\hat \rho(0)) \, \, e^{ \frac{i (P_i-P_f) . \partial \hat \rho(0)}{\Lambda^\sigma}} :   | P_i \rangle \, .
 \ea
 This last expression is perfectly well-defined between any two multi-particle states. If both states contain a finite number of particles then the contribution from the exponential truncates at finite order implying that this function is polynomial bounded in $k$. \\
 
 In other words the following limit is well-defined
 \be
  \langle P_f | \hat \pi(k) | P_i \rangle  = \lim_{N \rightarrow \infty }  \langle P_f | \hat \pi_N(k) | P_i \rangle  \, .
 \ee
 Since taking the limit $N \rightarrow \infty$ is precisely the limit of resumming the perturbation series, then we may say that $  \langle P_f | \hat \pi(k) | P_i \rangle $ exists as a meaningful non-perturbative expression. 
 Furthermore if $| P_i \rangle$ and $| P_f \rangle$ correspond to states with a finite number of particles, then since the expansion of the exponential truncates, $  \langle P_f | \hat \pi(k) | P_i \rangle$ grows at most as a polynomial in $k$. Thus we may always perform the Fourier transform, to give a well-defined (in the distributional sense) expression for 
 \be
  \langle P_f | \hat \pi(x) | P_i \rangle  = \int \frac{\d^d k}{(2 \pi )^d } e^{ik.x}\langle P_f | \hat \pi(k) | P_i \rangle =    e^{i(P_i-P_f).x}  \langle P_f | : U(\hat \rho(0)) \, \, e^{ \frac{i (P_i-P_f) . \partial \hat \rho(0)}{\Lambda^\sigma}} :   | P_i \rangle  \, .
 \ee
 This is consistent with the fact that we already know a well-defined expression for $\langle \beta | \hat \pi(x) | \alpha \rangle = \pi_c(x) $ which may in turn be inferred from the double sum over complete sets of momentum eigenstates as
 \be
 \langle \beta | \hat \pi(x) | \alpha \rangle = \sum_{n,m} \langle \beta | P_n \rangle \langle P_n| \hat \pi(x) | P_m \rangle \langle P_m | \alpha \rangle \, .
 \ee
 
 \subsection{Formal definition of Galileon operators $(\rho, \tilde \chi)$ frame}
 
 \label{formal2}
 
 It is straightforward to extend the discussion to the $\rho, \tilde \chi$ frame. At lowest order in perturbations in the coupling between $\rho$ and $\tilde \chi$, we can treat $\rho$ as a free field. Then we can compute the matrix elements of $\tilde \chi$ from the matrix elements of $\chi$ using the operator analogue of the duality map (\ref{eq:dualitymatter2})
\be
\hat{\tilde \chi}( x) = \int  \frac{\d^d k }{(2 \pi)^d}  e^{ik.x } \,  \hat \chi_k \, : e^{\frac{-i k. \partial \hat \rho(x)}{\Lambda^{\sigma} }} : \, ,
\ee
which in momentum space is
\be
\hat{\tilde \chi}_k = \int \d^d x \int  \frac{\d^d k' }{(2 \pi)^d}  e^{i(k'-k) .x } \,  \hat \chi_{k'} \, : e^{\frac{-i k'. \partial \hat \rho(x)}{\Lambda^{\sigma} }} : \, .
\ee
Evaluating between two momentum eigenstates we have
\ba
\langle P_f | \hat{\tilde \chi}_k  | P_i \rangle &=& \int \d^d x \int  \frac{\d^d k' }{(2 \pi)^d}  e^{i(k'-k) .x } \langle P_f | \hat \chi_{k'} \, : e^{\frac{-i k'. \partial \hat \rho(x)}{\Lambda^{\sigma} }} : | P_i \rangle \\
&=&  \int \d^d x \int  \frac{\d^d k' }{(2 \pi)^d}  e^{i(k'-k) .x } e^{i(P_i-P_f).x} \langle P_f | \hat \chi_{k'} \, : e^{\frac{-i k'. \partial \hat \rho(0)}{\Lambda^{\sigma} }} : | P_i \rangle \\
&=&  \langle P_f | \hat \chi_{k+(P_f-P_i)} \, : e^{\frac{-i (k+P_f-P_i). \partial \hat \rho(0)}{\Lambda^{\sigma} }} : | P_i \rangle \, . 
\ea
Once again, this expression is well-defined, and when evaluated between finite particle number states is polynomial in the momenta (i.e. only a finite number of terms in the expansion of the exponential contribute).
 
 \subsection{Operator Products and Spectral representations}
  
 As we have seen there is a well-defined expression for $ \langle P_f | \hat \pi(x) | P_i \rangle$, meaning that we know all the matrix elements of the operator $\hat \pi(x)$ in the momentum basis, which may be used as an operational definition of this operator. 
The departure from standard local field theories comes from trying to define operator products. In ordinary field theories we can make sense of such operator products via the operator product expansion (OPE).  The product of two operators at points $x$ and $y$ may be expressed as an expansion in $(x-y)$ in which there are a finite number of leading singular terms. Divergences may be removed with a finite number of counterterms in the Lagrangian.
In the present case since the operator $\hat \pi(x)$ is built out of an infinite number of powers of $\hat \rho(x)$ this entails an infinite number of subtractions. Furthermore there is no conformal behaviour at high energies since there is no UV fixed point. These fields then do not have a standard Wilson OPE, a reflection of the fact that they are fundamentally non-localizable.\footnote{We note in passing that Isham, Salam and Strathdee proposed a generalization of the Wilson OPE to theories of the strictly localizable type in \cite{Isham:1973by}. } The theory never becomes conformal or loses knowledge of the dimensional scale $\Lambda$.  \\

In the language of axiomatic field theory, we are no longer dealing with operator valued tempered distributions which is an essential assumption of the standard Wightman axioms. To give meaning to the integral that formally defines the two-point function we can introduce test functions $f(x)$ and $g(x)$ and define the smeared two-point function
\ba
&& \langle 0 | \hat \pi(f) \hat \pi(g) | 0 \rangle =  \\
&& \int \d^d y    \int \frac{\d^d k}{(2 \pi)^d} e^{ik.y}F(y,k) f^*(k) g(k) \exp\left[\frac{a_d}{\Lambda^{2 \sigma}}\left(\frac{k^2}{y_-^d}-d\frac{(k\cdot y)^2}{y_-^{d+2}}\right)\right] \ \, , \nn
\ea
where $\pi(f) = \int \d^d x f(x) \hat \pi(x)$and similarly for $g$. As we have discussed, the test functions must be drawn from the appropriate \GS spaces $S_{\alpha}$.  
However, while the process of smearing is useful to give a well-defined mathematical meaning to this expression, it disguises the physics and, in particular, the unitarity of this expression. Furthermore the above integrals are difficult to perform as they stand. \\

 A more practical approach is to recognize that since $ \langle P_f | \hat \pi(x) | P_i \rangle $ is well-defined, we may define the product of two operator by inserting a complete set of momentum eigenstates  to define the vacuum correlation as
\ba
  \langle 0 | \hat \pi(x)  \pi(y) | 0 \rangle  = \sum_{n} \langle 0 | \hat \pi(x)  | P_n \rangle \langle P_n |\pi(y) | 0 \rangle  \, .
\ea
Again this expression is formal because the sum over intermediate states does not converge. However we may diagnose this lack of convergence by using the technique of spectral representations. Following the standard derivation of the \KL (K-L) spectral representation we may rewrite this as
\ba
  \langle 0 | \hat \pi(x)  \hat \pi(y) | 0 \rangle  &=& \int_0^{\infty} \d \mu \int \frac{\d^d k}{(2 \pi)^d } e^{i k.(x-y) } \rho(-k^2)  2 \pi \delta(k^2 + \mu) \theta(k^0)  \\
   &=& \int_0^{\infty} \d \mu \int \frac{\d^{d-1} k}{(2 \pi)^{d-1} 2 \omega_k(\mu)} e^{i k.(x-y) } \rho(\mu) \, .
\ea
where $\omega_k(\mu) = \sqrt{{\vec k}^2  + \mu}$ and we have defined the spectral density in the usual way by
\be
\theta(k^0) \theta(-k^2) 2 \pi \rho(-k^2) = \sum_n (2 \pi)^d \delta^{(d)}(P_n -k) | \langle P_n | \hat \pi(0) | 0 \rangle |^2 \, .
\ee
Unitarity requires that $\rho(\mu) \ge 0$ which will be manifest at any finite order in perturbation theory. In particular, since $| \langle P_n | \hat \pi(0) | 0 \rangle |$ exists and the sum over momenta is finite for a given $\mu$ then there is no problem computing the spectral density $\rho(\mu)$ as we shall do below. Thus the content of the two-point function is contained in the well-defined finite spectral density $\rho(\mu)$. This spectral density determines the Fourier transform of the correlation function
\be
\langle 0 | \hat \pi(k) \hat \pi(k') |  0 \rangle =  ( 2\pi)^d \delta^d(k+k') 2 \pi \rho(-k^2)  \, ,
\ee
which is well-defined. The spectral density will have an exponential growth which prevents us from performing the Fourier transform back to position space necessary to define the position space correlations, but we find no difficulties directly working in momentum space.  

Despite the absence of position space correlators, we may define the operators smeared by test functions from the appropriate \GS space as
\be
\langle 0 | \hat \pi(f) \hat \pi(g) | 0 \rangle = \int \frac{\d^d k}{(2 \pi)^d} f(k) g^*(k)  2 \pi \rho(-k^2) \, . 
\ee
Once again this expression is well-defined when the test functions decay fast enough at large $|k|$ to balance the exponential growth of the spectral density. This expression then encodes a class of well-defined observables in this theory. \\

The most general class of observables will be the set of $N$-point correlation functions. These may be defined in an analogous way by defining their generalized K-L spectral representations which are well-known \cite{Kallen:1958ifa,Kallen:1961wza,Kallen:1959kza} and then defining the observables either directly in momentum space, or by using smeared functions
\be
\langle 0 | \hat \pi(f_1) \hat \pi(f_2) \dots \hat \pi(f_N)| 0 \rangle 
\ee 
where the $f$'s are drawn from the same \GS space $S_{\alpha}$. \\

The need to smear the correlation functions in order to give finite values to them in position space is an indication that the theory exhibits some degree of non-locality. The precise degree to which it is non-local depends on the precise \GS space needed to define these correlators. This in turn is determined by the precise exponential growth of the spectral densities. In other words, it is the exponential growth of the spectral density which is the signature of non-locality. By contrast,  in a local field theory these test functions can be taken arbitrarily close to delta functions. 

\subsection{Direct calculation of Galileon Spectral Density: $(\pi$, $\chi)$ frame}

We now come to the computation of the spectral density. We shall do this for simplicity in 4 dimensions (for which $a_4 =1/(2 \pi^2)$), although the following argument is easily generalized to any dimensions. Our naive expression for the two point function 
\ba
\hspace{-20pt} \langle 0 | \hat \pi(x) \hat \pi(x') | 0 \rangle  &=& \int \d^4 y  \int \frac{\d^4 k}{(2 \pi)^4} \, e^{i k.(x-x'-y )} F(y,k)  \exp\left[\frac{1}{2 \pi^2 \Lambda^{6}}\left(\frac{k^2}{y_-^4}-4\frac{(k\cdot y)^2}{y_-^{6}}\right)\right] \, ,
\ea
implies that the spectral density is given by
\be
\theta(k^0) \theta(-k^2) 2 \pi \rho_{\pi}(-k^2)  = \int \d^4 y  \, e^{-i k.y } F(y,k)  \exp\left[\frac{1}{2 \pi^2 \Lambda^{6}}\left(\frac{k^2}{y_-^4}-4\frac{(k\cdot y)^2}{y_-^{6}}\right)\right] \, .
\ee
Where, by $1/y_-^{2n}$ we mean
\be
\frac{1}{y^{2n}_-} = \frac{1}{(\vec y^2 - (y^0-i \epsilon)^2)^{n}} \, .
\ee
To make sense of the integral we perform an expansion in the coupling constant $1/\Lambda^6$
\be
\theta(k^0) \theta(-k^2)2 \pi \rho_{\pi}(-k^2)  = \int \d^4 y  \, e^{-i k.y } F(y,k)  \sum_{n=0}^{\infty} \frac{1}{n!}   \left[\frac{1}{2 \pi^2 \Lambda^{6}}\left(\frac{k^2}{y_-^4}-4\frac{(k\cdot y)^2}{y_-^{6}}\right)\right]^n  \, .
\ee
The validity of this expansion will be justified after the fact by demonstrating that the spectral density is an entire function of $1/\Lambda^6$.
Neglecting the contribution from $F(y,k)$, then each term in the sum is associated with contributions from $n$ particle intermediate states. This is consistent with the definition of the \KL spectral density, each term must be positive. Now $F(y,k)$ contains a finite number of terms of which the first one is $1/(4 \pi^2) y_-^{-2}$. Let us consider this first contribution
\ba
&& \vspace{-10pt} \theta(k^0) \theta(-k^2) 2 \pi \rho_{\pi,0}(-k^2)  = \int \d^4 y  \, e^{-i k.y } \frac{1}{4 \pi^2 y_-^2}  \sum_{n=0}^{\infty} \frac{1}{n!}   \left[\frac{1}{2 \pi^2 \Lambda^{6}}\left(\frac{k^2}{y_-^4}-4\frac{(k\cdot y)^2}{y_-^{6}}\right)\right]^n  \, , \nn \\
&=& \int \d^4 y  \, e^{-i k.y } \frac{1}{4 \pi^2 y_-^2}  \sum_{n=0}^{\infty} \frac{1}{n!}   \left[\frac{1}{2 \pi^2 \Lambda^{6}}\left(-\frac{(-k^2)}{y_-^4}+4\frac{(-ik\cdot y)^2}{y_-^{6}}\right)\right]^n  \, , \\
&=& \int \d^4 y  \, e^{-i k.y } \frac{1}{4 \pi^2}  \sum_{n=0}^{\infty} \sum_{r=0}^{n} \frac{(-1)^r 4^{n-r}  }{r! (n-r)! (2 \pi^2 \Lambda^{6})^n}    (-k^2)^r (-ik.y)^{2 n-2 r} \frac{1}{y_-^{6n-2r+2}}  \, .
\ea
Now by a simple extension of the result in Appendix~\ref{app1} 
\be
\int \d^4 y  \, e^{-i \alpha k.y } \frac{1}{(4 \pi^2)^N y_-^{2N}} = \theta(k^0) \theta(-k^2) \Omega_N(-k^2) \alpha^{2(N-2)}  \, ,
\ee
we may by repeated differentiation with respect to $\alpha$ followed by setting $\alpha=1$ infer that
\be
\int \d^4 y  \, e^{-i k.y } \frac{(-i k.y)^{2m}}{(4 \pi^2)^N y_-^{2N}} = \theta(k^0) \theta(-k^2) \Omega_N(-k^2)  \frac{(2N-4)!}{(2N-4-2m)!} \, ,
\ee
and so the spectral density is given by the double sum (using $N= 3 n -r+1$),
\ba 
&&   2 \pi \rho_{\pi,0}(\mu ) = \frac{1}{4 \pi^2}  \sum_{n=0}^{\infty} \sum_{r=0}^{n} \frac{\mu^{3n-1}}{\Lambda^{6n}} \frac{(-1)^r 4^{n-r}  }{r! (n-r)! (2 \pi^2 )^n}     \frac{(4 \pi^2)^{3n-r+1} (6n-2r-2)!}{(16 \pi^2)^{3n-r} (4n-2)! (3n-r-1)! (3n-r)!} \, . \nn
\ea
At large $n$ we may approximate 
\ba
{}_2F_{1}\left(-3n,-n,\frac{3}{2}-3n,\frac{1}{4}\right)\approx a\  _2F_{1}\left(-3n,-n,-3n,\frac{1}{4}\right)\approx a\left(\frac{3}{4}\right)^n
\ea
where $a\approx 0.85$.  We can therefore say
\ba
2 \pi \rho_{\pi,0}(\mu ) &=&2 \pi \delta(\mu) +a\sum_{n=1}^{\infty} \frac{\mu^{3n-1}}{\Lambda^{6n}}  \frac{2^{n-2} \pi^{-1/2-2n}  (3 (2n-1)/2)! \,  {}}{n! (3n)! (4n-2)!}\left(\frac{3}{4}\right)^n \nn \\
&=&2 \pi \delta(\mu)+\frac{3a\mu^2}{128\pi^2\Lambda^6}\ _PF_Q\left(\frac{5}{6},\frac{7}{6};\frac{3}{4},\frac{5}{4},\frac{4}{3},\frac{5}{3},2,2;\frac{3\mu^3}{512\pi^2\Lambda^6}\right) \, ,
\ea
which at large $\mu$ is approximated as
\ba
2 \pi \rho_{\pi,0}(\mu ) &=&2 \pi \delta(\mu)+\frac{a}{3\pi\mu}\sqrt{\frac{2}{15}}e^{\left(\frac{5\cdot 3^{1/5}}{2^{9/5}\pi^{2/5}}\right)\left(\frac{\mu^3}{\Lambda^6}\right)^{1/5}}\ .
\ea
The full contribution to $\rho$ coming from the additional terms in $F$ can be evaluated using the same method with a qualitatively similar result. In particular the full $\rho(\mu)$ is manifestly positive guaranteeing unitarity, and has order of growth $\alpha = 3/5$. This same method may then be easily generalized to any dimension to give $\alpha = (d+2)/(2(d+1))$.\\

As we have seen, the spectral density is an entire function of exponential growth. One way to understand the exponential growth is to use the fact that at large $\mu$ the sum will be dominated by large $n$, and so may be safely approximated by an integral. The growth of the term $\mu^{3n}$ is compensated by the fall off of the factorial coefficients in such a way that there is a peak value of $n$ which determines the leading contribution to the sum. Performing a saddle-point approximation the peak value of $n$ is found to be
\be
n \sim \left( \frac{\sqrt{\mu}}{\Lambda}\right)^{6/5} \, .
\ee
However, the sum over $n$ corresponds to the sum over $n=N$-particle intermediate states in the \KL spectral density. We thus see that the resonant contribution to the spectral density of the Galileon operator comes principally from $N$-particle intermediate states (generalizing to $d$ dimensions) with
\be
N \sim \left( \frac{E}{\Lambda}\right)^{(d+1)/(d+2)} \sim E \, r_*(E) \, . 
\ee
with $E \sim \sqrt{\mu}$. The typical energy of a quanta is $E/N = 1/r_*(E)$, i.e the frequency associated with the Vainshtein/classicalization radius. \\

Although this is a free theory, and there will be no scattering, assuming this property to hold true for the Wightman functions of an interacting theory, and given the relation between the scattering amplitudes and these functions, it is reasonable to suspect that the scattering amplitudes for scattering of two hard quanta with energy $E$ will be similarly dominated by $N$-particle out states with $N \sim E r_*(E)$. This is precisely the conjecture of the classicalization scenario \cite{Dvali:2010jz,Dvali:2010ns,Dvali:2011nj,Dvali:2011th}\cite{Alberte:2012is}. There the intermediate states are viewed as bound states of $N$ soft quanta which ultimately decay into free quanta. These bound states may be viewed at large energies as approximately coherent states, i.e. solutions of the semi-classical equations of motion. This is precisely what we have above, except the states are not truly bound since we are dealing with a free theory.  \\

To reiterate the central point, we have confirmed the exponential growth of the spectral density, which implies that there is a fundamental bound on the locality of the theory: local effective field theory is recovered only for 
\be
|x| \gg r_*(E) = \Lambda^{-1} \left( \frac{E}{\Lambda}\right)^{\frac{1}{(d+1)}} \, .
\ee
Nevertheless, unitarity and Lorentz invariance are maintained up to arbitrarily high energies.

\subsection{Direct calculation of Galileon Spectral Density: $(\rho$, $\tilde \chi)$ frame}

In the dual $(\rho$, $\tilde \chi)$ frame, the dual Galileon field $\rho$ is simply a free field with a standard two-point function. On the other hand, the matter field $\tilde \chi$ satisfies a non-trivial equation which is dependent on $\rho$. The non-localizability does not disappear in this frame, but is contained entirely in the matter field spectral densities. To compute this we use the formal operator definition given in Sec.~\ref{formal2} from which we infer
\be
\langle 0 | \hat{\tilde \chi}_k  \hat{\tilde \chi}_q | 0 \rangle = \int \d^d x \int \d^d y \int \frac{\d^d k'}{(2 \pi)^d}  \int \frac{\d^d q'}{(2 \pi)^d}  \, e^{i(k'-k).x}e^{i(q'-q).y} \langle 0 | \hat \chi_{k'} \hat \chi_{q'} |0 \rangle \langle 0 |  :e^{\frac{-i k'. \partial \hat \rho(x)}{\Lambda^{\sigma}}} :     :e^{\frac{-i q'. \partial \hat \rho(y)}{\Lambda^{\sigma}}} :   | 0 \rangle \, .
\ee 
Using
\be
 \langle 0 | \hat \chi_k \hat \chi_q |0 \rangle  = \theta(k^0) (2 \pi)^d \delta^d(k+q)  2 \pi \delta(k^2)
\ee
then
\be
 \langle 0 | \hat{\tilde \chi}_k  \hat{\tilde \chi}_q | 0 \rangle = \int \d^d x \int \d^d y \int \frac{\d^d k'}{(2 \pi)^d}    \, e^{i(k'-k).x}e^{i(-k'-q).y} \theta({k^0}')  2 \pi \delta({k'}^2) \langle 0 |  :e^{\frac{-i k'. \partial \hat \rho(x)}{\Lambda^{\sigma}}} :     :e^{\frac{i k'. \partial \hat \rho(y)}{\Lambda^{\sigma}}} :   | 0 \rangle \,.
\ee
Then defining
\be
\langle 0 | \hat{\tilde \chi}_k  \hat{\tilde \chi}_q | 0 \rangle = (2 \pi)^d \delta^d(k+q) 2 \pi \rho_{\chi}(k) \, ,
\ee
we have
\ba
& 2 \pi \rho_{\chi}(k) &= \int \d^d y  \int \frac{\d^d k'}{(2 \pi)^d}   \, e^{i(k'- k).y}\theta({k^0}')  2 \pi  \delta({k'}^2) \exp\left[\frac{a_d}{\Lambda^{2 \sigma}}\left(\frac{{k'}^2}{y_-^d}-d\frac{(k'\cdot y)^2}{y_-^{d+2}}\right)\right]  \, \\   
&=&  \int \d^d y  \int \frac{\d^d k'}{(2 \pi)^d}   \, e^{i(k'- k).y}\theta({k^0}')  2 \pi  \delta({k'}^2) \exp\left[-\frac{d a_d}{\Lambda^{2 \sigma}} \frac{(k'\cdot y)^2}{y_-^{d+2}} \right] \, , \\
&=&   \sum_{n=0}^{\infty} \int \frac{\d^d k'}{(2 \pi)^d}   \, \frac{1}{n!}  \theta({k^0}')  2 \pi  \delta({k'}^2) \left(\frac{d a_d}{\Lambda^{2 \sigma}}  \right)^n   \frac{\partial^{2n}}{\partial \alpha^{2n}} \int \d^d y  \, e^{i( \alpha k'- k).y} \frac{1}{y_-^{n (d+2)}}  |_{\alpha=1}    \ .\nn\\
\ea 
Performing the integral over $k'$ we have
\ba
 \hspace{-20pt}  &&  2 \pi \rho_{\chi}(k) =  \, \frac{1}{n!} \left(\frac{d a_d}{\Lambda^{2 \sigma}}  \right)^n    \frac{a_d}{(d-2)}\frac{\partial^{2n}}{\partial \alpha^{2n}} \int \d^d y  \, \alpha^{-(d-2)} e^{-i k.y} \frac{1}{y_-^{n (d+2) +(d-2)}}   |_{\alpha=1}  \, ,  \\
 &=&  \sum_{n=0}^{\infty}    \, \frac{1}{n!} \left(\frac{d a_d}{\Lambda^{2 \sigma}}  \right)^n    \frac{a_d}{(d-2)} \frac{\Gamma(3-d)}{\Gamma(3-d+2n)} \int \d^d y  \, e^{-i k.y} \frac{1}{y_-^{n (d+2) +(d-2)}}   |_{\alpha=1}  \, ,  \\
 &=&   \sum_{n=0}^{\infty}    \, \frac{1}{n!} \left(\frac{d a_d}{\Lambda^{2 \sigma}}  \right)^n    \frac{a_d}{(d-2)} \frac{\Gamma(3-d)}{\Gamma(3-d+2n)} (4 \pi^2)^{(n (d+2) +(d-2))/2} \Omega_{(n (d+2) +(d-2))/2}(-k^2) \ .\nn\\ 
\ea
The spectral density is then
\be
2 \pi \rho_{\chi}(\mu) = 2 \pi \delta(\mu) +  \sum_{n=1}^{\infty}    \, \frac{1}{n!} \left(\frac{d a_d}{\Lambda^{2 \sigma}}  \right)^n    \frac{a_d}{(d-2)} \frac{\Gamma(d+2 n-2)}{\Gamma(d-2)} (4 \pi^2)^{(n (d+2) +(d-2))/2} \Omega_{(n (d+2) +(d-2))/2}(\mu) \, .
\ee
In particular in four dimensions we have
\be
2 \pi \rho_{\chi}(\mu) = 2 \pi \delta(\mu) +  \sum_{n=1}^{\infty}    \,  \left(\frac{2 }{ \pi^2 \Lambda^{6}}  \right)^n    \frac{1}{4 \pi^2} \frac{(2n+1 )!}{n !} (4 \pi^2)^{3 n+1} \Omega_{3 n+1}(\mu) \, .
\ee
This result is derived from a different point of view in Appendix \ref{Density2}. Explicitly this is
\ba
2 \pi \rho_{\chi}(\mu) &=& 2 \pi \delta(\mu) +  \sum_{n=1}^{\infty}    \,  \left(\frac{2 }{ \pi^2 \Lambda^{6}}  \right)^n    \frac{1}{4 \pi^2} \frac{(2n+1 )!}{n !} (4 \pi^2)^{3 n+1} \frac{1}{(16 \pi^2)^{3n} (3n-1)! (3n)!} \mu^{3n-1} \, , \nn  \\
&=& 2 \pi  \delta(\mu) + \frac{\mu^2}{64 \pi^2 \Lambda^6} {}_PF_Q \({\frac{5}{2};\frac{4}{3},\frac{4}{3},\frac{5}{3},5,3,2 ;} \frac{\mu^3}{5832 \Lambda^6 \pi^2} \) \, ,
\ea
which at large $\mu$ behaves as
\be
2 \pi \rho_{\chi}(\mu) \approx 2 \pi  \delta(\mu) + e^{\frac{5 \mu^{3/5} }{3^{6/5} 2^{3/5} \pi^{2/5}\Lambda^{6/5}}} \frac{2^{1/10}}{3^{9/5} \sqrt{5} \pi^{11/10}  \mu^{1/10} \Lambda^{18/10}} \, ,
\ee
once again confirming the predicted growth $\alpha = 3/5$. This demonstrates that regardless of which duality frame we use, the field theory will remain non-localizable with the same order of growth when there are more than one interacting species.

\subsection{From Wightman to Feynman to Schwinger}

\label{propagators}

The calculation of the S-matrix requires knowledge of the time ordered correlation functions (Feynman) or equivalently the generalized retarded products. Once the Wightman functions are known then the time ordered correlators determine the retarded products uniquely (and vice versa). However both require an off-shell extension not needed in specifying the Wightman functions. The existence of well-defined Wightman functions in momentum space is not sufficient to uniquely specify the Feynman or retarded correlators. That is because their usual definition requires a Lorentz invariant notion of time-ordering. However, in the case of a non-localizable field theory the position space correlators do not exist and so there is no sense in which we can time order them to define the time-ordered or retarded products. \\

There is another way to see this problem, which also gives us a clue to the solution. Consider an ordinary local field theory for which the Wightman function spectral density growth is bounded polynomially by $\rho \le C \mu^N$ for integer $N$. A naive definition of the Feynman propagator in momentum space would be given by
\be
G_F(k) = \int_0^{\infty} \d \mu \, \rho(\mu) \frac{-i}{k^2 + \mu - i \epsilon}  \, .
\ee
This is a dispersion relation which shows that the Feynman propagator is an analytic function of $z=-k^2$ in the whole complex $z$ plane, modulo a branch cut along the real $z \ge 0$ axis (see Fig.~\ref{fig:contour}).
\be
G_F(z) = \int_0^{\infty} \d \mu \, \rho(\mu) \frac{i}{z-\mu }  \, .
\ee
with the physical propagator defined as the limit to the real axis from the upper half $z$ plane. \\

\begin{figure}  
\begin{center}  
\includegraphics[height=3in,width=3in,angle=0]{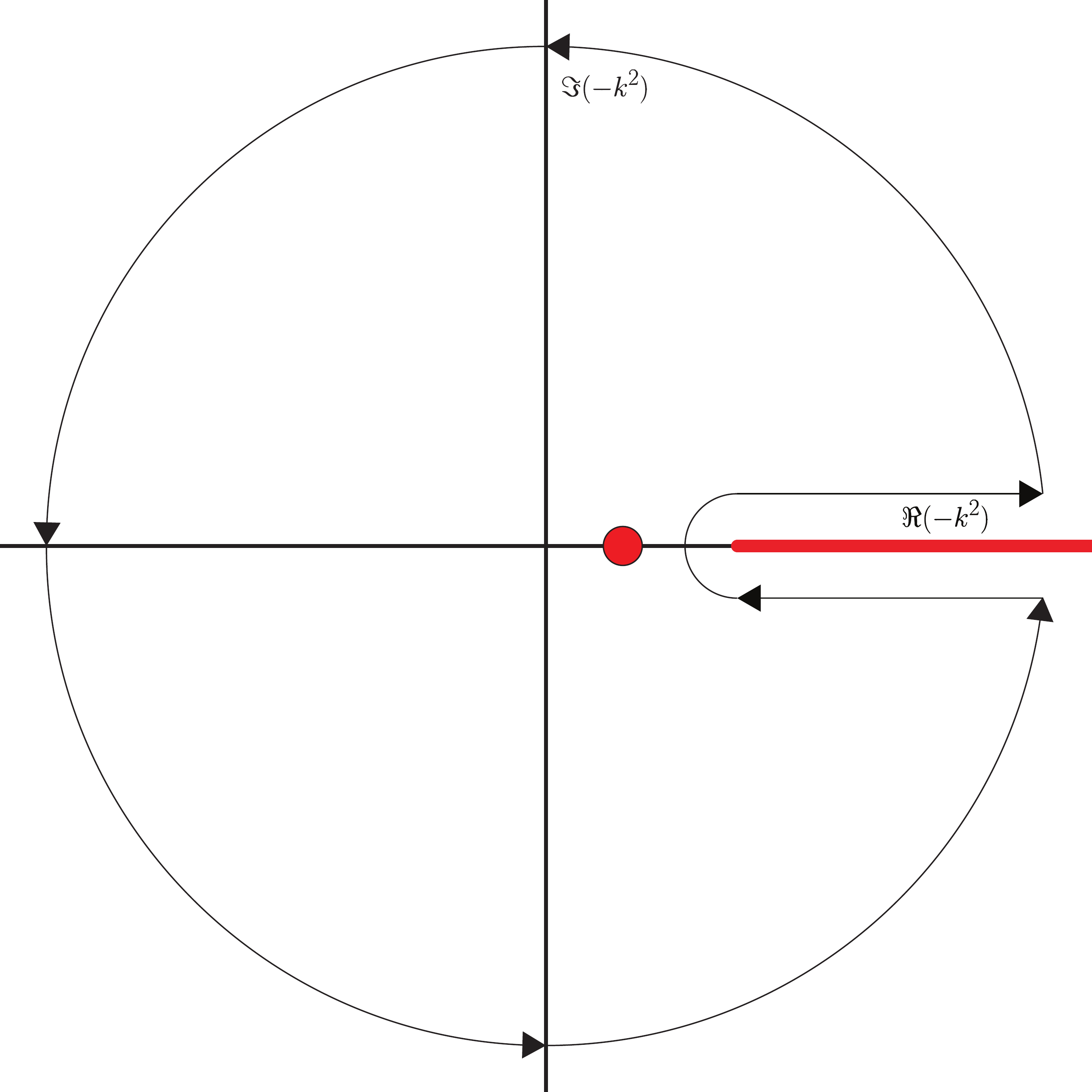}  
\caption{\small \sl Contour used to define Dispersion relation for Feynman propagator. The branch cut and pole are shown in red.\label{fig:contour}}  
\end{center}  
\end{figure}  

However this expression is ill-defined as it diverges at large $\mu$ due to the polynomial growth of the spectral density. This divergence shows up in the imaginary part only, the real part, being determined by the pole at $\mu = - k^2$, is finite. As is usual for dispersion relations the resolution is to perform a finite number of subtractions at some reference point $k^2=\mu_0$
\ba
&& G_F(k) =  -i \int_0^{\infty} \d \mu \, \rho(\mu) \left[  \frac{1}{k^2 + \mu - i \epsilon} - \sum_{n=0}^{N-1} (-1)^n \frac{(k^2 - \mu_0)^n}{(\mu+ \mu_0)^{n+1}} + \sum_{n=0}^{N-1} (-1)^n \frac{(k^2 - \mu_0)^n}{(\mu+ \mu_0)^{n+1}} \right] \, ,  \nn \\
&=& i \sum_{n=0}^{N-1} c_n(\mu_0) (k^2-\mu_0)^n + \int_0^{\infty}   \d \mu \, \rho(\mu)  \left[  \frac{-i}{k^2 + \mu - i \epsilon} - \sum_{n=0}^{N-1} (-1)^n \frac{(k^2 - \mu_0)^n}{(\mu+ \mu_0)^{n+1}} \right] 
\ea
where the coefficients $c_n(\mu_0)$ are formally defined by
\be
c_n(\mu_0) = - \int_0^{\infty} \d \mu  (-1)^n  \rho(\mu) \frac{(k^2 - \mu_0)^n}{(\mu+ \mu_0)^{n+1}} \, .
\ee
The divergence of the original expression is now confined to these finite number of coefficients and they may consequently be renormalized by subtracting infinite counterterms which amounts to allowing the $c_n$ to be arbitrary finite coefficients. 

In position space these terms correspond to derivatives of delta functions, so-called contact terms, as they are all localized at the coincidence or `contact' point $x=y$. They are thus all associated with local counterterms needed to subtract UV divergences. The implication is that even in a local field theory whose spectral density grows only polynomially, the time ordered correlation function for two fields is defined only up to counterterms which renormalize the contact terms even once the Wightman function has been determined. \\

The freedom of choice of these renormalized contact terms is the same ambiguity that resides in defining a time-ordered product. For instance, the time-ordered product between two fields $\hat A$ and $\hat B$ can be expressed as
\be
T \hat A(t) \hat B(t') = \theta(t-t')  \hat A(t) \hat B(t') + \theta(t'-t)  \hat B(t') \hat A(t)  + \sum_n \hat C_n(t)  \frac{\partial^{2 n} }{\partial t^{2 n}} \delta(t-t')  \, .
\ee
Since all the contact terms, $\hat C_n$, vanish away from $t-t'=0$, any expression of this form is an acceptable definition of a time-ordered product. An explicit example of the ambiguity in defining time ordered products is the difference between the normal $T$ product used in canonical Hamiltonian perturbation theory and the $T^*$ product used in the covariant Lagrangian perturbation theory. The equivalence of the Hamiltonian and Lagrangian definitions of perturbation theory rests precisely on the existence of those nonzero contact terms which arise in the different definitions of time ordering (this is sometimes known as Matthews theorem \cite{PhysRev.76.684.2}, a more general discussion can be found in the textbook of Bogoliubov and Shirkov). \\

An equivalent way to express this ambiguity is to introduce a polynomial {\bf indicator/indicatrix} function $g(-k^2) = \sum_n d_n (-k^2)^n $ which has the same polynomial growth at high $k$ as the spectral density and has no zeros for positive real arguments $d_n \ge 0$. Then we may define the Feynman propagator via
\be
G_F(k) = g(-k^2) \int_0^{\infty}   \d \mu \, \frac{\rho(\mu)}{g(\mu)} \frac{-i}{k^2 + \mu - i \epsilon}  + i  \, u(-k^2)\, . 
\ee
where $u(-k^2)$ is a real polynomial in $k^2$.
To see the equivalence of this with the previous expression it is sufficient to note that
\be
 \frac{1}{k^2 + \mu - i \epsilon} - \sum_{n=0}^{N-1} (-1)^n \frac{(k^2 - \mu_0)^n}{(\mu+ \mu_0)^{n+1}}  = g(-k^2) \frac{1}{g(\mu) (k^2 +\mu - i \epsilon)} \, ,
\ee
where $g(\mu) = (\mu  + \mu_0)^N$. The polynomial $u(-k^2)$ accounts for the contact terms. More generally it is sufficient to take $g(\mu)$ to be any $N$-th order polynomial whose zeros do not lie on the positive real axis. Reversing, the argument we see that for a generic polynomial $N$-th order polynomial $g(\mu)$ we can write
\be
\frac{1}{k^2 + \mu - i \epsilon} = \frac{g(-k^2)}{g(\mu) (k^2+\mu-i \epsilon)} + \sum_{n=0}^{N-1}\frac{1}{(n+1)!} \frac{g^{(n+1)}(\mu)}{g(\mu)} (-k^2-\mu)^n \, . 
\ee
Modifying the form of $g(\mu)$ just modifies the contact terms and so any real polynomial $g(\mu)$ is an acceptable choice.
\\

It is straightforward to see that regardless of the precise form of $g(\mu)$, as long as $g(\mu)$ is real the on-shell part is
\be
Re[G_F(k)] = \pi \rho(-k^2) \, .
\ee
The off-shell part affects only the contact terms. We may see this easily by passing to position space for which the Feynman propagator will take the form
\be
G_F(x) = g( \Box) \left( \theta(x^0) W_g(x)  + \theta(-x^0) W^*_g(x)\right) + i u(\Box) \delta^4(x)
\ee
where we have defined the regulated Wightman function
\be
W_g(x) = \int \frac{\d^4 k}{(2 \pi)^4} \theta(-k^2) \theta(k^0) \frac{\rho(-k^2)}{g(-k^2)} e^{ik.x} \, .
\ee
Due to the indicator function, $W_g$ is well-defined in position space. Since it is Lorentz invariant we may time order it in the standard way. Then the polynomial differential operator $g(\Box)$ acts on this time ordered Wightman function giving two sets of terms. The action of $g(\Box)$ on $W_g$ will just give back the original Wightman function, whereas the action of $g(\Box)$ on the $\theta(x^0)$ functions will give contact terms. Since $g(\Box)$ is a polynomial there are a finite number of such terms and so we find
\be
G_F(x) = \left( \theta(x^0) W(x)  + \theta(-x^0) W^*(x)\right) + \text{finite number of contact terms} \, .
\ee

\subsubsection*{{Generalization to non-localizable field theories}}

In passing from a standard local field theory to a strictly localizable or non-localizable field theory, we pass from needing a finite number of contact terms (counterterms) to an infinite number. Mathematically this corresponds to replacing the polynomial $g(-k^2)$ with an entire function that satisfies the same requirements that has the same order of growth as $\rho(\mu)$ and has no zeros for $-k^2$ real and positive. Then we may define the universal form of the momentum space Feynman propagator as
\be
G_F(k) = g(-k^2) \int_0^{\infty}   \d \mu \, \frac{\rho(\mu)}{g(\mu)} \frac{-i}{k^2 + \mu - i \epsilon} + i u(-k^2)  \, , 
\ee
where $u(-k^2)$ is now an entire function. Once more this definition is unique on-shell, and all the ambiguity is contained in the off-shell part. 
It is naturally extended to the complex plane $z=-k^2$ via the dispersion relation
\be
G_F(z) = g(z) \int_0^{\infty}   \d \mu \, \frac{\rho(\mu)}{g(\mu)} \frac{i}{z - \mu } + i u(z)  \, .
\ee
In the case of non-localizable field theories this expression cannot be directly Fourier transformed without appropriate smearing with test-functions. However, in momentum space it is well-defined.\\

The analogous definition of the retarded propagator is now uniquely specified in terms of the Wightman function and Feynman propagator to be
\be
G_R(k) = G_F(k) -W(k)  = g(-k^2) \int_0^{\infty}   \d \mu \, \frac{\rho(\mu)}{g(\mu)} \frac{-i}{\vec k^2 - (k^0 + i \epsilon)^2 + \mu } + i u(-k^2) \, . 
\ee
Since the function $g(-k^2)$ is entire both the Feynman and retarded propagators have the same analyticity properties as for a standard local field theory. However, since they are not polynomial bounded we may not give the standard finite number of subtractions dispersion relations for them. Since it is apparent that different choices of $g(-k^2)$ will give rise to different $u(-k^2)$, we may set $u(-k^2)$ to zero, understanding that this extra freedom is already built into $g(-k^2)$. \\

This procedure may be generalized to give time-ordered correlation functions for any number of fields \cite{Pfaffelhuber:1969ih,Pfaffelhuber:1972ny}. Consider a N-point Wightman function which may be expressed in momentum space in the form
\be
\langle 0 | \hat \phi(x_1) \dots \hat \phi(x_N) |0 \rangle = \int \frac{\d^4 k_1}{(2 \pi)^4} \dots \int \frac{\d^4 k_N}{(2 \pi)^4} (2 \pi)^4 \,  \delta^{(4)}\( \sum_{n=1}^N k_n \) \tilde W(k_1 \dots k_N) \, . 
\ee
We introduce a symmetric N-point indicator function $g(k_1, \dots k_N)$ which is an entire function in all its arguments, has no zeros in the physical region, and has the same growth at large arguments as $ \tilde W(k_1 \dots k_N)$, then the time-ordered correlation functions in position space are given formally by
\be
G_F(x_1 , \dots , x_N) = g( - i \partial_1 , \dots -i \partial_N) \left[ \sum_{\text{permutations}}  \theta(x_1-x_2) \dots \theta(x_{N-1}-x_N) \, W_g(x_1 , \dots , x_N)  \right]
\ee
and 
\be
W_g(x_1 , \dots,  x_N) =  \int \d^4 x_1 \dots \int \d^4 x_N \, e^{i \sum_n k_n.x_n}\delta^{(4)}\( \sum_{n=1}^N k_n \) \frac{\tilde W(k_1 \dots , k_N) }{g(k_1, \dots , k_N)} \, .
\ee
To these we add entire function contributions which vanish on-shell. As usual these formal expressions become well-defined when appropriately smeared with test-functions. Here the sum over permutations is the standard time-ordering operation.
The anti-time ordered correlation functions may be defined analogously and are given by the complex conjugates $G_F^*(x_1 , \dots , x_N) $. 
The generalized retarded products which are usually formally defined as
\be
\label{eq:retarded1}
\hat {\cal R}(x,x_1,\dots, x_n) = (-i)^n \sum_{\rm Permutations} \theta(x-x_1) \dots \theta(x_{n-1}-x_n) [ \dots [[\hat \phi(x), \hat \phi(x_1) ] , \hat \phi(x_2) ] \dots \hat \phi(x_n)] \, ,
\ee
are determined unambiguously once the time ordered products have been specified
\be
\hat {\cal R}(x,x_1,\dots, x_n)  = i^n \sum_{k=0}^{n} \frac{(-1)^k}{k!(n-k)!} G^*_F(x_k, \dots , x_n)  G_F(x, x_1 , \dots , x_k) \, .
\ee

\subsubsection*{Schwinger functions}

Having constructed the time-ordered correlation functions, and using the fact that the indicator functions are entire and hence well-defined under Wick rotation, it is now straightforward to Wick rotate to give the Euclidean correlation functions (Schwinger functions)
\be
G_E(x_1 , \dots x_N) = g( - i \partial_1 , \dots -i \partial_N) \left[ G_{E,g} (x_1 , \dots x_N) \right] \, ,
\ee
where $G_{E,g} (x_1 , \dots x_N)  $ is the Wick rotated version of 
\be
\left[ \sum_{\text{permutations}}  \theta(x_1-x_2) \dots \theta(x_{N-1}-x_N) \, W_g(x_1 , \dots x_N)  \right] \rightarrow^{\text{Wick rotation}} \rightarrow G_{E,g} (x_1 , \dots x_N) \,.
\ee
In particular, in the case of the two-point function, we have in momentum space
\be
G_E(k) = g(-k^2) \int_0^{\infty}   \d \mu \, \frac{\rho(\mu)}{g(\mu)} \frac{1}{k^2 + \mu } - u(-k^2)\, . 
\ee
This is the obvious Wick rotation of the Feynman propagator in momentum space. It is not, however, the analytic continuation of the Wightman function since the Wightman function contains no reference to $g(-k^2)$. This is an important difference with standard local field theories, the latter having the property that the Wightman function at spacelike separations is the Euclidean correlation function. We also see that it is impossible to uniquely specify the Schwinger functions without reference to the indicator. This is a reflection of the fact that Euclidean theories are always off-shell, and are consequently always sensitive to the ambiguity of the off-shell extension. In path integral language this is related to the inherent ambiguity in specifying the measure of the path integral. \\

The most striking consequence of this distinction between the Schwinger function and the spacelike Wightman function is that for the case of a non-localizable field theory, even though the spacelike Wightman function may not exist in position space without appropriate smoothing, the Euclidean correlator can exist since it is determined by the behaviour of $g(-k^2)$ for $k^2>0$, i.e. for negative arguments. Even though $g(-k^2)$ grows exponentially for large timelike $k$, for certain theories it may decrease exponentially or as a power for large spacelike $k$.  
An example of a non-localizable spectral density with this property is $\rho(\mu) = e^{\alpha \mu}$ for which we may choose $g(-k^2) = e^{-\alpha k^2}$. This renders the UV behavior of the Euclidean theory much better than a standard local field theory. This fact is the basis of Efimov's non-local field theory program (see e.g. \cite{efimov1967non,Alebastrov:1973vw})  which utilizes this improved UV behavior to regulate loops in a perturbative expansion. Our perspective is somewhat different though since we imagine that this behavior comes from the nonperturbative properties of a classically local field theory and it is not clear that there is any sense in which we may find a perturbative expansion in which the Euclidean propagator is replaced by an expression of the above form. More generally, there is no expectation that $g(-k^2)$ decreases exponentially at spacelike $k^2$. It may, for example, simply oscillate.

\section{Superluminality, Macro-causality}

\subsection{Causality and the Jost-Lehmann-Dyson representation}

One of the most commonly used arguments against the possibility of UV completing Galileon theories and massive gravity theories in a Lorentz invariant manner is the apparent existence of superluminal solutions in the low energy effective field theory. Indeed the superluminal solutions that arise in massive gravity are already present in its Galileon decoupling limit (for a discussion on this connection see the review \cite{deRham:2014zqa}), and it is sufficient to analyze the decoupling limit theory to understand these. \\

These arguments can be stated in slightly different but entirely equivalent  ways
\begin{itemize}

\item The LEEFT admits classical solutions whose perturbations have superluminal group and/or phase velocities.

\item A characteristic analysis of the classical equations of motion implies that there are initial data for which the Cauchy problem becomes ill-defined - the `Velo-Zwanziger problem'. \cite{Velo:1969bt} (For a discussion of this for massive gravity see \cite{Izumi:2013poa,Deser:2013eua}).

\item There exist solutions for which the momentum-velocity relation for the field may not be inverted. As such there is a conflict between the desire to canonically quantize the theory using equal time commutation relations and Lorentz invariance - the `Johnson-Sudarshan problem'. \cite{Johnson:1960vt}

\end{itemize}

\noindent As a consequence, these results are often argued to be fatal to any attempt to UV complete such theories, at least in a Lorentz invariant way. It is argued that:
\begin{itemize}

\item Lorentz invariance is fundamentally broken in these theories.

\item The existence of superluminal propagation implies the formation of Closed Time-like curves and hence the violation of causality.

\end{itemize}

Although all of these statements are correct classically, their importance in the quantum theory is more subtle. To begin with, neither the low energy phase nor group velocity has anything to do with causality, rather as is well known this is determined by the front velocity which is the high momenta limit of the phase velocity $v_F = \lim_{k \rightarrow 0} v_P(k)$. The characteristic analysis correctly determines the classical front velocity, however since this velocity is determined by the high frequency limit, in theories built on the Vainshtein mechanism the field is highly strongly coupled there, and the tree level calculation cannot be trusted (we shall see this explicitly below). The fact that the momentum-velocity relation cannot be inverted is not a problem semi-classically since those solutions for which it cannot be inverted will not be described by the LEEFT. Quantum mechanically, this observation is consistent with our earlier arguments (Sec.~\ref{canonical}) that Vainshtein type theories should not be canonically quantized and, as we have emphasized throughout, there is no obstruction to quantizing in a Lorentz invariant manner.  Finally, the classical formation of closed time-like curves is never in the regime of validity of the LEEFT theory \cite{Burrage:2011cr}. \\

A concrete realization of the apparent superluminal propagation can already be seen in the case of the special Galileon which is dual to a free theory. In any dimension, the dual Galileon theory exhibits classical solutions which exhibit superluminal propagation (for details on these arguments see \cite{deRham:2013hsa,deRham:2014lqa,Creminelli:2014zxa,Kampf:2014rka}). In our current language this is the statement that if we evaluate the two point function of the operator $\hat \pi(x)$ in a coherent state $| \alpha \rangle$, and maintain only the contributions that arise at tree-level, then there will exist some coherent state $| \alpha \rangle $ for which the tree-level two point function has a commutator which is non-zero outside of the Lorentz invariant light cone. In fact, the simplest example of this is in the $(\rho$, $\tilde \chi)$ dual frame. Evaluating the two-point function for $\tilde \chi $ in a coherent state for $\rho$ we obtain
\be
\langle \alpha | \hat {\tilde \chi}(x) \hat {\tilde \chi}(y) | \alpha \rangle =  \int \d \tilde k  \, \exp \[ ik.\left( x-y - \frac{\partial \rho_c(x)}{\Lambda^{\sigma}} + \frac{\partial \rho_c(y)}{\Lambda^{\sigma}} \right) \] \, ,
 \ee
 where
 \be
 \langle \alpha | \hat \rho(x) | \alpha \rangle = \rho_c(x) \, 
 \ee
 defines the background field in the $(\rho$, $\tilde \chi)$ dual frame.
This clearly propagates in a lightcone determined by
\be
\( x-y - \frac{\partial \rho_c(x)}{\Lambda^{\sigma}} + \frac{\partial \rho_c(y)}{\Lambda^{\sigma}} \right)^2  \le 0 \, ,
\ee
which can easily be chosen to lie outside the usual lightcone $(x-y)^2 \le 0$ for some choice of $\rho_c(x)$.
This example is discussed in more detail in \cite{deRham:2014lqa}. However, the above is only a tree-level calculation, and the all-loop (orders in $\hbar$) order answer 
\ba
&& \langle \alpha | \hat {\tilde \chi}(x) \hat {\tilde \chi}(y) | \alpha \rangle =  \\
&& \nn \int \d \tilde k  \, \exp \[ ik.\left( x-y - \frac{\partial \rho_c(x)}{\Lambda^{\sigma}} + \frac{\partial \rho_c(y)}{\Lambda^{\sigma}} \right) \]\exp\left[\hbar \frac{a_d}{\Lambda^{2 \sigma}}\left(-d\frac{(k\cdot (x-y))^2}{(x-y)_-^{d+2}}\right)\right]  \, ,
 \ea
is only meaningful after smearing with test functions. The additional exponential which arise from loop corrections (we have included a factor of $\hbar$ to make clear its dependence) completely modifies the high $k$ behaviour of the correlator, and this is relevant since causality is determined by the front velocity which is the high $k$ limit. The above expression shows clearly that the tree-level approximation cannot be trusted in calculating this. \\

To analyze this two-point function, as in the vacuum case, it is helpful to work with a spectral density which is finite. Since we are no longer in vacuum, the relevant spectral density is the one provided by the Jost-Lehmann-Dyson (JLD) representation \cite{jost1957integral,Dyson:1997gw}.
The relevant physical question is: {\bf Does the UV completion of Galileon theories allow for macroscopic superluminal propagation}? Here, by macroscopic, we mean at distances larger than the locality bound. This distinction is important since in a trivial sense superluminality occurs in the region forbidden by the locality bound since their is no notion of locality there, but this superluminality is fundamentally harmless. Provided that the theory respects a Lorentz invariant notion of macroscopic causality, then there will be no real superluminal propagation.  The tree-level calculation on the other hand implies a superluminal propagation which would be seen at arbitrary large distances.
\\

A meaningful definition of macroscopic causality is as follows. We define a pair of test functions $f_{x_0}(x),g_{x_0}(x) $ from the space $S_{\alpha}$ which are localized at the point $x=x_0$ to the maximum extent allowed by the locality bound of the theory. We define the smeared field 
\be
\hat \phi(f_{x_0}) = \int \d^d x f_{x_0}(x) \hat \phi(x) \, .
\ee
Then macro-causality requires that 
\be
 \[\hat \phi(f_{x_0}), \hat \phi(g_{y_0}) \] \rightarrow 0  \,, \quad (x_0-y_0)^2 \rightarrow + \infty
\ee
for two such localized test functions $f,g$, i.e. the vanishing of the smeared commutator at large space-like separations. Furthermore this vanishing should be exponentially rapid.
We further require this for all allowed pairs of test functions from the space $S_{\alpha}$, including those test functions obtained by a Lorentz boost from the original one. In this sense, this definition of macro-causality is Lorentz invariant. This condition alone is sufficient to forbid superluminal propagation since any superluminal propagation would inevitable lead to an enlargening of the causal cone at large distances, leading to a contribution which falls off only as a power law.  \\

Although superluminality is usually diagnosed around classical backgrounds, i.e. in coherent states, it is simpler at the quantum level to work with momentum eigenstates. A momentum eigenstate with a finite number of particles will not exhibit any superluminality, and so we must still work with states with infinite number of particles as in the case of coherent states. This is easily achieved by projecting the coherent state $|\alpha \rangle$ onto its momentum eigenstate as follows
\be
|\alpha, P \rangle = \int \d^d y \, e^{-i P.y} | \alpha'_y \rangle \, ,
\ee
where $\alpha'_y$ is the translated coherent state whose vev is $\phi_c(x+y)$ if $\alpha$ has vev $\phi_c(x)$, so that $|\alpha, P \rangle $ defined a momentum eigenstate containing an infinite number of particles.
More precisely, if the original coherent state takes the normalized form
\be
|\alpha \rangle = e^{-\frac{1}{2} \int \d \tilde k  \, | \alpha(k)|^2}\exp \left[ \int \d \tilde k \,  \alpha(k) \hat a_k^{\dagger}\right] |0 \rangle \, ,
\ee
then
\be
|\alpha , P\rangle = \int \d^d y \, e^{-\frac{1}{2} \int \d \tilde k \, | \alpha(k)|^2} e^{-i P.y}\exp \left[ \int \d \tilde k \, \alpha(k) e^{ik.y} \hat a_k^{\dagger}\right] |0 \rangle \, .
\ee
The original coherent state is obtained as a superposition of such states
\be
|\alpha \rangle = \int \frac{\d^d P}{(2 \pi)^d}|\alpha, P \rangle \, .
\ee
Given this, to address causality it is sufficient to analyze the commutator between two such coherent momentum eigenstates. Denoting these states by the further shorthand $|P_i \rangle $ and $|P_f \rangle $ then we consider 
\be
\langle P_f | \[\hat \phi(x) , \hat \phi(y) \] |P_i \rangle  = e^{i (P_i-P_f) \frac{1}{2} (x+y)} \langle P_f | \[\hat \phi(z/2) , \hat \phi(-z/2) \] |P_i \rangle \, ,
\ee
where $z =  (x-y) $. \\

In a strictly localizable field theory we would require that $\langle P_f | \[\hat \phi(z/2) , \hat \phi(-z/2) \] |P_i \rangle $ vanishes outside the lightcone $z^2<0$ for all $|P_i \rangle $ and $|P_f \rangle$. This would be sufficient to ensure the absence of superluminal propagation and micro-causality. 
In a non-localizable theory we demand the weaker condition of macro-causality. A sufficient condition for Lorentz invariant macro-causality is that
\be
\label{macrocausal1}
\langle P_f | \[\hat \phi(z/2) , \hat \phi(-z/2) \] |P_i \rangle = g(\Box_z) F(z,P_i,P_f)  \, ,
\ee
where $ F(z,P_i,P_f) $ vanishes outside the lightcone $z^2<0$ and $g(\mu)=h(\mu)^2$ is an entire function of the desired order of growth built out of the square of an entire function $h(\mu)$. 
To see why this is sufficient consider the smeared commutator
\ba
\langle P_f | \[\hat \phi(f_{x_0}) , \hat \phi(g_{y_0}) \] |P_i \rangle &=& \int \d^d x \int \d^d y f_{x_0}(x) g_{y_0}(y) e^{i (P_i-P_f) \frac{1}{2} (x+y)} \langle P_f | \[\hat \phi(z/2) , \hat \phi(-z/2) \] |P_i \rangle \nn   \\
&=&  \int \d^d \bar x \int \d^d z f_{x_0}(\bar x+z/2) g_{y_0}(\bar z-z/2) e^{i (P_i-P_f) \bar x} h(\Box_z)h(\Box_z) F(z,P_i,P_f) \, , \nn  \\
&=&  \int \d^d \bar x \int \d^d z \tilde f_{x_0}(\bar x+z/2) \tilde g_{y_0}(\bar z-z/2) e^{i (P_i-P_f) \bar x} F(z,P_i,P_f) \, ,
\ea
with $\bar x = (x+y)/2$ and we have defined the new test functions
\ba
\tilde f_{x_0}(\bar x+z/2) = h(\Box_z) f_{x_0}(\bar x+z/2) \, , \\
\tilde g_{x_0}(\bar x-z/2) = h(\Box_z) g_{x_0}(\bar x-z/2)  \, .
\ea
If the original test functions are drawn from the space $S_{\alpha}$, then the new test functions will remain localized at $x_0$ and $y_0$ respectively (i.e. the growth of the Fourier transform of $h(k^2)$ is compensated by the fall of of the Fourier transform of the test function). \\

Given this as $(x_0-y_0)^2 =L^2 \rightarrow + \infty$, the product of test functions $\tilde f_{x_0}(\bar x+z/2) \tilde g_{y_0}(\bar z-z/2) $ will drop off exponentially away from $z^2 =L^2 \rightarrow  + \infty$. However, by our assumptions $F(z,P_i,P_f)=0$ for $z^2=L^2>0$. Consequently, the only contribution to the integral comes from the tail of the overlap of the test functions which tends to zero exponentially as $(x_0-y_0)^2 =L^2 \rightarrow + \infty$, thus fulfilling the requirements of macro-causality. \\

Thus to demonstrate Lorentz invariant macro-causality for a given field $\hat \phi(x)$, it is sufficient to demonstrate that the following quantity vanishes outside the lightcone:
\be
F(z,P_i,P_f)  = \int \frac{\d^d k}{(2\pi)^d} \, e^{ik.z} \frac{1}{g(k)} \int \frac{\d^d y}{(2\pi)^d}\, e^{-ik.y} \langle P_f | \[\hat \phi(y/2) , \hat \phi(-y/2) \] |P_i \rangle   =0 , \quad z^2<0\, .
\ee
To prove that this is the case, let us consider the $(\pi,\chi)$ frame, and consider the local field $\pi_N(x)$ defined in Eq.~(\ref{piN}). Since this field is local, and its correlation functions are tempered distributions, then its commutator vanishes outside the lightcone
\be
[\hat \pi_N(x) , \hat \pi_N(y)] =0 \,,  \quad (x-y)^2 <0\, .
\ee
Given this, it is subject to all the requirements for there to exist a JLD representation \cite{jost1957integral,Dyson:1997gw}.
We may thus write
\be
\langle P_f | \[\hat \pi_N(y/2) , \hat \pi_N(-y/2) \] |P_i \rangle = \int_0^{\infty} \d \mu  \int \frac{\d^d k}{(2 \pi)^d} \int \d^d u \, e^{ik.y} D_N(u,\mu,P_i,P_f) \, 2 \pi \epsilon(k^0-u^0) \delta( (k-u)^2+\mu) \, ,
\ee
where $D_N(u,\mu,P_i,P_f) $ is the JLD spectral representation which depends on the specific momentum eigenstates and is the analogue of the \KL spectral density when not in vacuum. Since $\pi_N$ is a tempered field $D_N(u,\mu,P_i,P_f)$ grows at most as a polynomial in $\mu$. This is sufficient to ensure the convergence of the integral since
\ba
\langle P_f | \[\hat \pi_N(y/2) , \hat \pi_N(-y/2) \] |P_i \rangle &=&   \int  \frac{\d^d k}{(2\pi)^d}\int  \frac{\d^d u}{(2\pi)^d} \, e^{ik.y} D_N(u,-(k-u)^2 ,P_i,P_f) \, 2\pi \epsilon(k^0-u^0)  \, , \nn \\
&=&   \int \d^d k \int \d^d u \, e^{ik.y}  e^{i u.y } D_N(u,-k^2 ,P_i,P_f) \, 2 \pi \epsilon(k^0)  \, .
\ea
for which the $k$ integral is convergent. Hence for this field we have
\be
F_N(z,P_i,P_f) = \int_0^{\infty} \d \mu  \int  \frac{\d^d k}{(2\pi)^d} \int \frac{\d^d u}{(2\pi)^d}\, e^{ik.z}  \frac{D_N(u,\mu ,P_i,P_f)}{g(-k^2)} \frac{g(-(k-u)^2)}{g(\mu)} \, 2 \pi \epsilon(k^0-u^0)\delta( (k-u)^2+\mu)  \, .
\ee
Now this can be rearranged in the form
\be
F_N(z,P_i,P_f) = \int_0^{\infty} \d \mu  \int  \frac{\d^d u}{(2\pi)^d} \, e^{iu.z}  \frac{D_N(u,\mu ,P_i,P_f)}{g(\mu)} \frac{g(-(-i \partial_z)^2)}{g(-(-i \partial_z+u)^2)} \, \Delta_{\mu}(z)  \, ,
\ee
where $ \Delta_{\mu}(z) = \int  \frac{\d^d k}{(2\pi)^d} e^{ik.z}  2 \pi \epsilon(k^0)\delta( k^2+\mu)   $ is the free-field of mass $\sqrt{\mu}$ commutator which automatically vanishes for $z^2<0$. \\

We may now take the limit $N \rightarrow \infty$, since although 
\be
D(u,\mu ,P_i,P_f) = \lim_{N \rightarrow \infty} D_N(u,\mu,P_i,P_f) \, ,
\ee
will have an exponential growth in this limit $D(u,\mu ,P_i,P_f) \sim e^{\sigma \mu^{\alpha}}$ this will be compensated by the indicator function $g(\mu)$. As we have already discussed, any expression which remains finite in the limit $N \rightarrow \infty$ will give the appropriate definition for the Galileon field $\hat \pi(x)$. Thus we require that the following expression is finite and vanishes outside the lightcone
\be
F(z,P_i,P_f) = \int_0^{\infty} \d \mu  \int  \frac{\d^d u}{(2\pi)^d} \, e^{iu.z}  \frac{D(u,\mu ,P_i,P_f)}{g(\mu)} \frac{g(-(-i \partial_z)^2)}{g(-(-i \partial_z+u)^2)} \, \Delta_{\mu}(z)  \, .
\ee
First let us consider the part
\be
\frac{g(-(-i \partial_z)^2)}{g(-(-i \partial_z+u)^2)} \, \Delta_{\mu}(z)  =   \int  \frac{\d^d k}{(2\pi)^d} \frac{g(-k^2)}{g(-(k+u)^2)} 2 \pi \epsilon(k^0) \delta( k^2+\mu)  e^{ik.z} \, .
\ee
Since $g(-k^2)$ is an entire function, then the ratio $\frac{g(-k^2)}{g(-(k+u)^2)} $ is an analytic function over the whole complex plane modulo poles at $\mu_i$. 
We may thus write 
\be
\frac{g(-k^2)}{g(-(k+u)^2)}  = h(k,u) +\sum_i \frac{R_i(k,u)}{-(k+u)^2 - \mu_i } \, .
\ee
where $h(k,u)$ is an entire function of $k$ and $R_i(k,u)$ are the residue terms at the poles which are also entire functions of $k$. Here we have assumed that all of the poles are simple poles for simplicitly (we can always view multiple poles as limits of combinations of simple poles). \\

In any contour integral, the contributions from the poles are harmless since their residues will give rise to contributions to $F(z,P_i,P_f) $ that vanish when acted on by $g(-\Box_z)$. This is just a reflection of the fact that the only reason these poles exist is because we have chosen to extract the operator $g(-\Box_z)$ outside of the integral, but the inverse of this operator is ambiguous due to the poles.

To make this more precise we note that we can write
\be
g(\Box) F(z,P_i,P_f)  = g(\Box)  F_1(z,P_i,P_f) + \sum_i g'_i(\Box) F_2^i (z,P_i,P_f)
\ee
where  $g'_i(\mu) = g(\mu)/(\mu-\mu_i)$ is an entire function with a given zero removed, and
\ba
&& F_1(z,P_i,P_f) = \int_0^{\infty} \d \mu  \int  \frac{\d^d u}{(2\pi)^d} \, e^{iu.z}  \frac{D(u,\mu ,P_i,P_f)}{g(\mu)} h(- i \partial_z,u) \, \Delta_{\mu}(z)  \,  , \\
&& F_2^i(z,P_i,P_f)= \int_0^{\infty} \d \mu  \int  \frac{\d^d u}{(2\pi)^d} \, e^{iu.z}  \frac{D(u,\mu ,P_i,P_f)}{g(\mu)} R_i(-i \partial_z,u) \, \Delta_{\mu}(z) \, .
\ea
In the case of higher order poles of multiplicity $N$ a similar approach may be taken in which $g'_i(\mu) = g(\mu)/(\mu-\mu_i)^N$ and $R_i(k,u) $ is replaced by the appropriate higher order pole residue. We may now refine our statement of macro-causality that independently we must have $F_1(z,P_i,P_f) $ and $F^i_2(z,P_i,P_f) $ vanish for $z^2>0$. 
\\

The key observation is that, if $g(\mu) $ is an entire function of $\mu$ of order $\alpha$, i.e. $g(\mu) \sim e^{\sigma \mu^{\alpha}}$ then both $h(k,\mu)$ and $R_i(k, \mu)$ are entire functions of $k$ of order $2 \alpha - 1$. To see this, it is sufficient to look at the fixed $u$, $|k| \rightarrow \infty$ asymptotics of 
\be
\frac{g(-k^2)}{g(-(k+u)^2)}  \sim e^{\sigma (-k^2)^{\alpha} - \sigma ( -(k+u)^2 )^{\alpha}} \sim e^{2 \sigma \alpha (-k^2 )^{\alpha-1 } k.u } \le e^{2 \sigma \alpha |k|^{2\alpha-1 } |u| } \, .
\ee
Thus, if $g(-k^2)$ has the exponential growth of a non-localizable field with $1/2<\alpha<1$ then the ratio $\frac{g(-k^2)}{g(-(k+u)^2)}$ has the exponential growth of a strictly localizable field. For instance, in the case of Galileons in four dimensions, $\alpha = 3/5$ and so 
\be
\frac{g(-k^2)}{g(-(k+u)^2)}  \sim e^{\frac{6}{5} \sigma  (-k^2 )^{-2/5 } k.u } \le e^{\frac{6}{5} \sigma |k|^{1/5} |u|} \, ,
\ee
and hence both $h(k,u)$ and $R_i(k,u)$ are independently bounded by $e^{\frac{6}{5} \sigma |k|^{1/5} |u|}$. 
However, we know any entire function of $-i \partial $ acting on a function which vanishes outside the lightcone and has a growth less than a linear exponential in $k$, will still vanish outside the lightcone because that is how we defined strictly localizable fields in the first place!  \cite{Jaffe:1967nb,Jaffe:1966an,meiman1964causality}.\\

To see this more precisely, consider the action of $h(- i \partial , u) $ on the retarded propagator
\be
G_{\rm ret, \mu}(z) = \theta(n.z) \Delta_{\mu}(z) \, , 
\ee
where $n_{\mu}$ is a future pointing timelike vector. As long as we can show that $h(- i \partial , u) G_{\rm ret, \mu}(z) $ vanishes for $n.z<0$, for any future time-like vector $n$, then this proves the vanishing of $h(- i \partial , u) G_{\rm ret, \mu}(z) $ outside the whole lightcone since this is the only way we have a Lorentz invariant (i.e. independent of $n$) notion of time ordering. The most dangerous term then comes from when all the time derivatives act on the $\theta(n.z)$ function since this gives the contribution most likely to extend outside the lightcone. Thus
\be
h(- i \partial , u)  G_{\rm ret, \mu}(z) = (h( - i \partial, u ) \theta(n.z) ) \Delta_{\mu}(z)  + \dots , 
\ee
However, viewed as a function of the time-like direction $n.k$ alone, we know that $h(k,u)$ is an entire function which grows at most as $e^{\frac{6}{5} \sigma |n.k|^{1/5} |u|}$ which being strictly localizable, is insufficient to give support for $n.z<0$ \cite{Jaffe:1967nb,Jaffe:1966an,meiman1964causality}. Hence, since $h(- i \partial , u)  G_{\rm ret, \mu}(z)=0$ for $n.z<0$ for all future time-like $n$, then $h(- i \partial , u)  G_{\rm ret, \mu}(z)=0$ for $z^2>0$ and by extension  $h(- i \partial , u) \Delta_{\mu}(z) = h(- i \partial , u)  \left( G_{\rm ret, \mu}(z)-G_{\rm ret, \mu}(-z) \right) =0$ for $z^2>0$. 
Identically the same argument applies to $R_i(- i \partial , u)  \Delta_{\mu}(z)=0$. \\

To finally show that $F_1(z,P_i,P_f)$ and $F_2^i(z,P_i,P_f)$ vanish for $z^2>0$, it is sufficient to ensure that the remaining integrals converge. The $\mu$ integral clearly converges since our assumption is that the indicator function $g(\mu)$ is chosen to have a growth to compensate the exponential growth of $D(u,\mu ,P_i,P_f)$. Thus, the only remaining concern is for the convergence of the $u$ integral. Previously we considered the action of $h(- i \partial , u) \Delta_{\mu}(z)$ at fixed $u$. However, it is possible that $h(k, u)$ could grow exponentially in $u$ as $h(k , u) \sim e^{\sigma |u|^{2 \alpha}}$ at fixed $k$. At the very least we have the following bounds
\ba
&& h(- i \partial , u) \Delta_{\mu}(z) \le B(z) e^{\sigma |u|^{2 \alpha}}  \, , \\
&& R_i(- i \partial , u) \Delta_{\mu}(z) \le B_i(z) e^{\sigma |u|^{2 \alpha}} \, .
\ea
Thus a sufficient (but perhaps not necessary) condition for the convergence of the remaining integrals is that
\be
D(u,\mu ,P_i,P_f)e^{\sigma |u|^{2 \alpha}} \, ,
\ee
is polynomially bounded in $u$, or at least is bounded by a linear exponential growth in $u$:
\be
D(u,\mu ,P_i,P_f)e^{\sigma |u|^{2 \alpha}} \le C e^{a |u|}\, .
\ee
In fact we may now withdraw from our use of momentum eigenstates and consider the commutator between coherent states $|\alpha \rangle $ and $| \beta \rangle $ given formally as
\be
\langle \beta  | [ \hat \pi(x), \hat \pi(y) ]  | \alpha \rangle =  \int_0^{\infty} \d \mu  \int \d^d k \int \d^d u \, e^{ik.z} \tilde D(u,\mu,\alpha,\beta,\bar x) \, 2 \pi \epsilon(k^0-u^0) \delta( (k-u)^2+\mu) \, ,
\ee
where 
\be
\tilde D(u,\mu,\alpha,\beta,\bar x)  = \int \frac{\d^d P_f}{(2 \pi)^d} \int \frac{\d^d P_i}{(2 \pi)^d} e^{i(P_i-P_f) \bar x }D(u,\mu,(\beta,P_f),(\alpha,P_i))  \, ,
\ee
and $D(u,\mu,(\beta,P_f),(\alpha,P_i)) $ is the JLD spectral function associated with $| \alpha, P_i \rangle $ and $| \beta, P_f \rangle$. Hence for a given initial and final coherent state, $|\alpha \rangle $ and $ | \beta \rangle$, to exhibit no superluminalities, a sufficient (although possibly not necessary) condition is that 
\be
\tilde D(u,\mu,\alpha,\beta,\bar x) e^{\sigma |u|^{2 \alpha}} \le C e^{a |u|} \, ,
\ee
is bounded by a linear exponential in $u$.  \\

This is a restriction on the state itself since $u$ essentially encodes the momentum of the background coherent fields. To see this we note that we can write
\be
\langle \beta  | [ \hat \pi(x), \hat \pi(y) ]  | \alpha \rangle =  \int_0^{\infty} \d \mu \,  \hat D(\mu,\alpha,\beta,x,y) \, \int \frac{\d^d k}{(2 \pi)^d} \, e^{ik.z}   2 \pi \epsilon(k^0) \delta( k^2+\mu) \, ,
\ee
where
\be
 \hat D(\mu,\alpha,\beta,x,y)  = \int \frac{\d^d u}{(2 \pi)^d}  \, e^{iu.z} \tilde D(u,\mu,\alpha,\beta,\bar x) \, .
\ee
From this representation it is clear that $ \hat D(\mu,\alpha,\beta,x,y) $ is a local version of the \KL spectral density, whose $x$ and $y$ dependence will arise only because we are not evaluating this expression in vacuum, but rather between two coherent states. Because of this we do not expect $\hat D(\mu,\alpha,\beta,x,y) $ to vary more in space than the background coherent state itself. In other words if $\langle \alpha | \hat \pi (x) | \beta \rangle $ varies over a typical length scale $L_0$, then we expect the same to be true for $\hat D(\mu,\alpha,\beta,x,y) $ which in turn implies that the support of $ \tilde D(u,\mu,\alpha,\beta,\bar x)$ will be for momenta with $u \sim 1/L_0$.  \\

Now in the present case, the spatial dependence of $ \hat D(\mu,\alpha,\beta,x,y)  $ will be determined by $\langle \beta | \hat \pi (x) | \alpha \rangle $  which we know to be a solution of the classical $(d+1)$'th order Galileon equation of motion. Alternatively we may use the duality to infer $\langle \beta | \hat \pi (x) | \alpha \rangle $ from $\langle \beta | \hat \rho (x) | \alpha \rangle $. Consider a coherent state of total energy $E$. Although the Fourier transform of $\langle \beta | \hat \rho (x) | \alpha \rangle $ may fall off only polynomially, we expect that the Fourier transform of $\langle \beta | \hat \pi (x) | \alpha \rangle $ to fall off exponentially for $|k| \gg 1/r_*(E)$. The reason is that the classical equation becomes strongly interacting at distances $|x|< r_*(E)$ which will suppress the momentum support for $|k| \gg 1/r_*(E)$. Indeed following the classicalization arguments, we expect a state of energy $E$ to be build out of $N$ soft quanta with typical energy/momenta $|k| \sim E/N$ with $N = E r_*(E)$. This behaviour will simply follow from the classical equations of motion. 
This in turn implies that the support of $ \tilde D(u,\mu,\alpha,\beta,\bar x)$ is localized on modes $u$ for which $|u| \sim E/N \sim 1/r_*(E)$. As long as $E \gg \Lambda$ then this support will be for $u \ll \Lambda$ and so we may reasonably expect that $ \tilde D(u,\mu,\alpha,\beta,\bar x)$ decays exponentially for $u \gg \Lambda$. From the duality map 
\be
\langle \beta | \hat \pi (x) | \alpha \rangle =  \int \d^d y  \int \frac{\d^d k}{(2 \pi)^d} \, U(\langle \beta | \hat \rho (y) | \alpha \rangle ) \, \, e^{ i k.(x-y) + \frac{i k . \partial \langle \beta | \hat \rho (y) | \alpha \rangle }{\Lambda^\sigma}} 
\ee
we may reasonably estimate using a saddle-point approximation that this exponential decay will be at least as fast as $e^{-\sigma |u|^{(d+2)/(2 (d+1)}}$ which is sufficient for convergence of all the integrals. We have not proven that all states satisfy this property, it nevertheless seems reasonable that generic coherent states fall off in this manner. 
\\

We regard this as yet another manifestation of the locality bound. Those classical background states that are not built out of high energy quanta localized in a region less that the associated locality bound will give rise to macroscopic causality. We have not (yet) rigorously excluded the existence of states which violate boundedness of $\tilde D(u,\mu,\alpha,\beta,\bar x) e^{\sigma |u|^{2 \alpha}} $, and hence potentially give superluminal propagation, however such states attempt to localize quanta more than allowed by the locality bound. To put this more pragmatically, any semi-classical solution of the Galileon LEEFT that is built out of modes only with $k<\Lambda$ should satisfy the condition that $\tilde D(u,\mu,\alpha,\beta,\bar x) e^{\sigma |u|^{2 \alpha}} $ is polynomially bounded. This is true even for configurations of total energy $E\gg \Lambda$ since the power will effectively be cutoff in the strong coupling region $ |k|\gg 1/r_*(E) $ for which $1/r_*(E) < \Lambda$.
We will leave to future work a more precise exposition of this condition and its necessity. For now, we simply note that it is sufficient, and that there are an infinite class of non-trivial background solutions for which no macroscopic superluminality occurs in the fully quantum theory, even though the tree-level calculation would imply that there was. Any superluminality that does occur would appear to be intimately connected with the violation of locality due to the locality bound.

\subsubsection*{Extension to Retarded products}

The previous argument showed that quite generally the commutator of two fields vanishes at space-like separations. However causality requires an even sharper statement that we can define a retarded product of operators that has support only in the future light-cone. Our macroscopic version of this statement is that
\ba
\langle P_f | R \phi(f_{x_0}) \hat \phi(g_{y_0}) | P_i \rangle \rightarrow 0  \,, &&  \quad (x_0-y_0)^2 \rightarrow + \infty  \, , \nn\\
&& \text{ and } (x_0-y_0)^2 \rightarrow - \infty \text{ with } (x^0-y^0)<0  \, ,
\ea
where $R$ denotes the retarded product whose definition requires subtractions. The naive unsubtracted definition of the Fourier transform of the retarded product in terms of the JLD spectral density is
\be
\int {\d^d y}  e^{-ik.y} \langle P_f | R \hat \phi(y/2) \hat \phi(-y/2)  |P_i \rangle = \int_0^{\infty} \d \mu   \int \frac{\d^d u}{(2 \pi)^d} \, D(u,\mu,P_i,P_f) \, \frac{-i}{(\vec k - \vec u)^2 -(k^0-u^0+i \epsilon)^2 + \mu} \, ,
\ee
and so the subtracted form is
\ba
&&\int {\d^d y}  e^{-ik.y} \langle P_f | R \hat \phi(y/2) \hat \phi(-y/2)  |P_i \rangle  \nn \\
&& =  \int_0^{\infty} \d \mu   \int \frac{\d^d u}{(2 \pi)^d} \,  g(-(k-u)^2) \frac{D(u,\mu,P_i,P_f)}{g(\mu)} \, \frac{-i}{(\vec k - \vec u)^2 -(k^0-u^0+i \epsilon)^2 + \mu} \, , \\
&& = g(-k^2) \int_0^{\infty} \d \mu   \int \frac{\d^d u}{(2 \pi)^d} \,  \frac{g(-(k-u)^2)}{g(-k^2)} \frac{D(u,\mu,P_i,P_f)}{g(\mu)} \, \frac{-i}{(\vec k - \vec u)^2 -(k^0-u^0+i \epsilon)^2 + \mu} \, , \nn
\ea
up to the addition of entire subtraction functions. 

The argument proceeds as before: in order to satisfy macro-causality it is sufficient to have
\be
\langle P_f | R \hat \phi(y/2) \hat \phi(-y/2)  |P_i \rangle  = g(\Box_y) F_R (y,P_i,P_f)
\ee
where $F_R (y,P_i,P_f)$ has support only in the future lightcone.  Up to possible subtraction terms we have
\ba
&& F_R (y,P_i,P_f) = \\
&&  \int_0^{\infty} \d \mu   \int \frac{\d^d u}{(2 \pi)^d}  \int \frac{\d^d k}{(2 \pi)^d}  e^{ik.y} \,  \frac{g(-(k-u)^2)}{g(-k^2)} \frac{D(u,\mu,P_i,P_f)}{g(\mu)} \, \frac{-i}{(\vec k - \vec u)^2 -(k^0-u^0+i \epsilon)^2 + \mu}  \, . \nn
\ea
Following the previous reasoning, this is a linear superposition of free-space retarded propagators which will automatically vanish outside the future light-cone provided that the integrals converge and that $ \frac{g(-(k-u)^2)}{g(-k^2)} $ has the order of growth of a strictly-localizable field, which again requires only $\alpha <1$. The convergence of the $u$ integral is ensured since $ \frac{D(u,\mu,P_i,P_f)}{g(\mu)}$ can be made polynomially bounded, and so the only doubt is the convergence of the $u$ integral which will be satisfied under the previously stated conditions.

\subsection{Causality and S-matrix Analyticity for Non-localizable theories}

\label{sec:appA}

A related statement of causality is that the S-matrix should be an analytic function of energy, modulo branch cuts associated with thresholds and poles associated with physical particles. We shall give a more thorough discussion of the properties of the scattering amplitude in \cite{KeltnerTolley2}, for now we give a simple example using the language of non-relativistic scattering that demonstrates our point of view that it is analyticity alone, and not polynomial boundedness, that is sufficient to preserve the notion of causality in a non-local theory. More precisely, we argue that analyticity alone implies macro-causality and only together with polynomial boundedness does it imply micro-causality. The latter condition is safely violated in a non-localizable theory, but the former can be maintained. We will give a more thorough discussion of these issues elsewhere \cite{KeltnerTolley2}. \\

To keep things as simple as possible let us consider a quantum mechanical S-matrix between two specific states of energy $E$, $\hat S_{fi}(E)$, which is assumed to be analytic in the upper half-plane, modulo the required poles and branch cuts, but non-polynomially bounded. Assume the existence of an entire function $g(E)= \sum_n c_n E^n$ which has the same order of growth at high energies as $\hat S_{fi}(E)$. 
Then, given our analyticity assumption, we may define a subtracted dispersion relation for the S-matrix element of the form
\be
\hat S_{fi}(E) = u_{fi}(E) + g(E) \int_{-\infty}^{\infty} \frac{ \d E'}{2 \pi} \frac{1}{g(E')}\frac{\rho_{fi}(E')}{E'-E-i \epsilon} \, ,
\ee
where $u_{fi}(E)$ is an entire function of $E$. $u_{fi}(E)$ may grow no faster than $g(E) \times \text{polynomial}$ so that the Fourier transform of $\tilde u_{fi}(E)=u_{fi}(E)/g(E)$ is a tempered distribution. \\

In the case of a local field theory it would be sufficient to take $g(E)$ as a polynomial, then associated with this scattering matrix we could define a retarded operator
\ba
\hat G_R(t)  &=& \int_{-\infty}^{\infty} \frac{ \d E}{2 \pi}e^{-i E t}  \hat S_{fi}(E) \, ,  \\
  &=&   \int_{-\infty}^{\infty} \frac{ \d E}{2 \pi}e^{-i E t}    u_{fi}(E) + \int_{-\infty}^{\infty} \frac{ \d E}{2 \pi}e^{-i E t}    g(E) \int_{-\infty}^{\infty} \frac{ \d E'}{2 \pi}\frac{1}{g(E')}\frac{\rho_{fi}(E')}{E'-E- i \epsilon}   \, .    
\ea
The first term, being the Fourier transform of a polynomial, can be expressed in terms of derivatives of delta functions. The second term can be organized into derivatives of $\theta$ functions. We then have
\ba
\label{eqGR1}
\hat G_R(t)    &=&   u_{fi}\[i \frac{\partial}{\partial t}\] \delta(t) -i g\[i \frac{\partial}{\partial t}\] H_{fi}(t)    \, ,
\ea
where
\be
H_{fi}(t) =  \theta(t) \int_{-\infty}^{\infty} \frac{ \d E'}{2 \pi}\, e^{-i E' t}  \frac{1}{g(E')} \rho_{fi}(E') \, .
\ee
$H_{fi}(t)$ manifestly vanishes for $t<0$. Thus, again in a local field theory, $G_R$ is built out of a finite number of derivatives of delta functions, which vanish away from $t=0$ and a contribution which vanishes for $t<0$ 
\be
\hat G_R(t)  =  v_{fi}\[i \frac{\partial}{\partial t}\] \delta(t) -i \theta(t) \int_{-\infty}^{\infty} \frac{ \d E'}{2 \pi} \, e^{-i E' t}   \rho_{fi}(E')     \, .    
\ee
where $v_{fi}$ includes the contributions from the second term of (\ref{eqGR1}) where at least one of the time derivatives acts on the $\theta(t)$ inside $H_{fi}(t) $. \\

Thus we see that in a local theory, analyticity and polynomial boundedness would guarantee $G_R(t)=0$ for $t<0$. This is the usual statement of micro-causality as seen through the analyticity and polynomial boundedness of the S-matrix. This will continue to be true for strictly localizable field theories for which all of the above Fourier transforms are well-defined since the usual exponential oscillations from $e^{-iEt}$ are sufficient to ensure convergence. Consequently we anticipate that the micro-causality condition holds for all strictly localizable theories $0 \le \alpha<1/2$. This has been proven for fixed physical momentum transfer in relativistic theories in \cite{Epstein:1969bg}. \\

In the case of a non-localizable field theory the micro-causality condition $G_R(t)=0$ for $t<0$  can no longer hold due to non-polynomial boundedness of the S-matrix which is captured by the fact that the growth of $g(E) \sim e^{\sigma |E|^{2 \alpha}}$ for some complex $E$ grows too fast to ensure convergence of the contour integrals when $\alpha>1/2$. However, as we have seen, this also means this expression for $G_R(t)$ does not exist, (at least not as a tempered distribution). This is quite a different situation than a genuinely acausal theory for which $G_R(t)$ does exist but has support for $t<0$. \\

However, what does exist in the non-localizable case is the causal response from any incoming state which is sufficiently delocalized in time, i.e. it is drawn from the same \GS space $S_{\alpha}$ as the test functions needed to define the operators. The formal response from that initial state is given by
\ba
\hat \phi_{\rm out}(t) &=&     \int_{-\infty}^{\infty} \d t' \hat G_R(t-t') \phi_{\rm in}(t')  \, .
\ea
or in Fourier space
\be
\hat \phi_{\rm out}(E)=  u_{fi}(E) \phi_{\rm in}(E) + g(E) \int_{-\infty}^{\infty} \frac{ \d E'}{2 \pi} \frac{1}{g(E')}\frac{\rho_{fi}(E')}{E'-E-i \epsilon} \phi_{\rm in}(E) \, .
\ee
We may rewrite this as 
\be
\hat \phi_{\rm out}(E)=  \tilde u_{fi}(E) \tilde \phi_{\rm in}(E) +  \int_{-\infty}^{\infty} \frac{ \d E'}{2 \pi} \frac{1}{g(E')}\frac{\rho_{fi}(E')}{E'-E-i \epsilon} \tilde \phi_{\rm in}(E) \, .
\ee
where $\tilde \phi_{\rm in}(E) = g(E) \phi_{\rm in}(E)$ and $\tilde u_{fi}(E) = u_{fi}(E)/g(E)$,  so that
\be
\hat \phi_{\rm out}(t) =     \int_{-\infty}^{\infty} \d t' \hat G_R'(t-t') \tilde \phi_{\rm in}(t')  \, .
\ee
Here $G_R'(t)$ is given by
\be
\hat G_R'(t)   =   \int_{-\infty}^{\infty} \frac{ \d E}{2 \pi}e^{-i E t}    \tilde u_{fi}(E) -i \theta(t)  \int_{-\infty}^{\infty} \frac{ \d E'}{2 \pi}\frac{1}{g(E')} \rho_{fi}(E') e^{-i E' t}   \, .    
\ee
Unlike $G_R(t)$, $G'_R(t)$ exists and vanishes for $t<0$. This is the S-matrix equivalent of the causality condition imposed in Eq.~(\ref{macrocausal1}). Micro-causality is only violated in the sense that $\tilde J(t)$ has support in the past of $\phi_{\rm in}(t)$, but only to the extent consistent with the locality bound. \\

In this sense analyticity alone in the upper half-plane is enough to ensure macro-causality. We regard this is as sufficient to save us from the na\"ive perils of closed timelike curves. Any closed timelike curves formed will be localized in the region where the locality bound forbids us from giving a well defined notion for time. Nevertheless macroscopically a well defined notion of time and time ordering emerges. This is consistent with the perspective of Steinmann that analyticity in momentum/energy space is a sufficient notion for causality in a non-localizable theory \cite{Steinmann:1970cm}.
This discussion should be contrasted with the case where analyticity is violated. For instance, imagine $\hat S_{fi}(E)$ has a simple pole in the upper half-plane at $E = E_R + i E_I$. Then $G_R$ will pick up an exponentially decaying contribution from the residue $|G_R| \sim e^{- E_I t}$. However this contribution blows up exponentially in the past destroying any notion of causality. A similar result would follow from the existence of branch cut (a branch cut may be viewed as a continuum of an infinite number of poles).

\subsection{An example of a non-localizable theory: Asymptotic Fragility of Fluctuating Infinitely Long Strings}

A simple example of a field theory which is consistent with our above reasoning is provided by the `asymptotic fragility' for fluctuations in the effective field theory of infinitely long strings \cite{Dubovsky:2012wk} and, in particular, the model S-matrix considered in \cite{Cooper:2013ffa} which in our language is an explicit example of a S-matrix for a non-localizable field theory with $\alpha=1$. \\

Consider the gauge fixed Nambu-Goto action for a string propagating in $D$ dimensions in terms of its $D-2$ transverse coordinates
\be
\label{NG}
S = - l_s^2 \int \d^2 \sigma \sqrt{- Det[\eta_{ab} - l_s^2 \partial_a X^i \partial_b X^i]} \, .
\ee
In the case $D=24$ this is of course the critical bosonic string. From the perspective of a two-dimensional field theory, this is, like the Galileon models, a non-renormalizable field theory for which in perturbative calculations we expect an infinite number of counterterms to be included all at the scale $l_s$. Thus from the standard EFT logic we may only expect this theory to be meaningful for $E \ll l_s^{-1}$. \\

Despite this, it is argued in \cite{Dubovsky:2012wk} that the S-matrix for scattering of worldsheet perturbations on an infinitely long string takes the remarkably simple exact form
\ba
&& A(s)=e^{2 i \delta(s) } = e^{i s l_s^2/4} \, , \quad  {\rm Im}(s) >0 \, ,\\ 
&& A(s) = e^{2 i \delta(s) } = e^{-i s l_s^2/4} \, , \quad {\rm Im}(s) <0 \, ,
\ea
where $s$ is the usual Mandelstam variable. This S-matrix, which is defined by a single complex phase because we are scattering in two dimensions, is manifestly Lorentz invariant, hermitian analytic $A^*(s) = A(s^*)$, and crossing symmetric $A(-s) = A(s)$. It thus defines a UV completion valid at arbitrarily high energies $E\gg l_s^{-1}$, and it can be shown at low energies to match the expectations based on the LEEFT (\ref{NG}). This example already has many similarities with our proposal for Galileons. In this case, the strong coupling scale $\Lambda$ is $1/l_s$ and the theory contains no UV fixed point, rather its high energy behavior is determined by $1/l_s$. The authors of \cite{Dubovsky:2012wk} refer to this as `asymptotic fragility' to distinguish this behaviour from the UV fixed point of `asymptotic safety'. \\

This high energy behaviour is precisely what is expected based on semi-classical scaling arguments. In this case the energy scales as 
\be
E \sim \int \d \sigma \,  l_s^2 \sqrt{- Det[\eta_{ab} - l_s^2 \partial_a X^i \partial_b X^i]}  + \dots
\ee
and defining dimensionless coordinates $\sigma = r_*(E) \hat \sigma$, and dimensionless fields $X^i = l_s^{-1} r_*(E) \hat X^i$ then we find
\be
E \sim l_s^2 r_*(E) \, ,
\ee
and so the Vainshtein/classicalization radius for this system is 
\be
r_*(E) \sim l_s^{-2} E \, ,
\ee
which grows with energy. The scattering amplitude is determined by the statement that scattering is purely elastic with the phase shift determined by the classical action $\delta \sim S \sim E r_*(E) \sim s l_s^2$ with $s=E^2$. Since the scattering is elastic there are no `classicalons' in this picture, nevertheless despite the perturbative non-renormalizability of the LEEFT, the high energy scattering amplitude is unitarized in a manner consistent with semi-classical arguments along the lines discussed in Sec.~\ref{sec:unitarizationclassical}. \\

Since the S-matrix is analytic, we may express it in terms of a dispersion relation. In this case, the S-matrix is polynomially bounded, since as $Im(s) \rightarrow \pm \infty$ , $e^{2 i \delta(s) } \rightarrow e^{- |{\rm Im} (s)|  l_s^2/4} $. In particular we can express the S-matrix in terms of a dispersion relation with a single subtraction
\be
A(s) = e^{2i \delta(s)} = 1 -  \frac{s}{\pi} \int_{-\infty}^{\infty} \d \mu \,  \frac{\sin(\mu l_s^2/4)}{\mu (s- \mu)} \, .
\ee
which defines the amplitude over the whole complex $s$ plane, consistently with the previous definition. \\

However as noted in \cite{Cooper:2013ffa}, since the S-matrix is an entire function of $l_s^2$, there appears to be no logical obstruction to replacing $l_s^2$ with $-l_s^2$ for which the Nambu-Goto action would become 
\be
\label{NG}
S =  l_s^2 \int \d^2 \sigma \sqrt{- Det[\eta_{ab} + l_s^2 \partial_a X^i \partial_b X^i]} \, .
\ee
This action is the analogue of the wrong sign anti-DBI model considered in Sec.~\ref{antiDBI} for which it is known that superluminal propagation appears to arise in the LEEFT. At the same time, this is the sign associated with a functioning Vainshtein mechanism. The S-matrix, which is now trivially 
\ba
&& e^{2 i \delta(s) } = e^{-i s l_s^2/4} \, , \quad  {\rm Im}(s) >0 \, ,\\ 
&& e^{2 i \delta(s) } = e^{i s l_s^2/4} \, , \quad {\rm Im}(s) <0 \, .
\ea
satisfies all of the desirable requirements of unitarity, crossing symmetry and analyticity, despite the apparent pathologies of the LEEFT. Once again the high energy behaviour is consistent with semi-classical expectations. It is precisely an analogous situation we are arguing is possible for Galileons with the distinction being that there we generally expect the scattering to be inelastic. \\

We can easily see that from our language, this is an example of a non-localizable field theory. Indeed it is sufficient to note that in this case, the S-matrix is not polynomially bounded, since now as $Im(s) \rightarrow \pm \infty$ , $e^{2 i \delta(s) } \rightarrow e^{+ |{\rm Im} (s)| l_s^2/4} $. Thus, we may still express the S-matrix in terms of a dispersion relation, but as in the case of all non-localizable field theories, we must use an entire indicator function $g(\mu)$ to account for the infinite number of subtractions. This entire function must have the same growth at large arguments over the whole complex $s$ plane as the amplitude itself. There are an infinity of such choices, but one consistent choice is\footnote{Note that although we use the same notation, the indicator function $g(\mu)$ that controls the behavior of the S-matrix is in general not the same as that for the Feynman propagator.} 
\be
g(\mu)  = \mu ( 1+\frac{1}{2}\sin \left( \mu l_s^2/4\right)) \, .
\ee
This choice has no zeros on the real axis, and has the same exponential growth as ${\rm Im }(s) \rightarrow \pm \infty$ as the S-matrix. The resulting dispersion relation is
\be
A(s) = e^{2i \delta(s)} = v(s) + \frac{s}{\pi} \( 1+\frac{1}{2}\sin \left( s l_s^2/4\right) \)  \int_{-\infty}^{\infty} \d \mu \,  \frac{\sin(\mu l_s^2/4)}{\mu \( 1+\frac{1}{2}\sin \left( \mu l_s^2/4\right) \) (s- \mu)} \, ,
\ee
for a real entire function $v(s)$ with the same order of growth. We see then that a simple change in the sign of $l_s^2$ has a profound change of the dispersion relation due to the violation of polynomial boundedness. If there exists a field theory description of this system, the fields must necessarily be non-localizable with an order of growth $\alpha=1$. \\

Following the discussion in the previous section, this model can still satisfy macro-locality. Reverting to non-relativistic language, this example corresponds to taking $g(E)=E^2 ( 1+\frac{1}{2}\sin \left( E^2 l_s^2/4\right)) $. Since we are dealing with non-localizable fields of order $\alpha=1$, the incoming state localized at $t=t_0$ should be taken to be at least as localized as a Gaussian.
\be
\tilde \phi_{\rm in}(t) =\frac{1}{ 2 L \sqrt{\pi} } e^{-\frac{(t-t_0)^2}{4 L^2}} \, .
\ee
with $L = a l_s$ and $a \sim {\cal O}(1)$. We may view this as a smeared delta function.
The response from this initial state is causal in terms of the modified initial state $\tilde \phi_{\rm in}$
\be
\hat \phi_{\rm out}(t) =     \int_{-\infty}^{\infty} \d t' \hat G_R'(t-t')  \tilde \phi_{\rm in}(t')  \, .
\ee
Here the modified initial state is
\ba
&& \tilde \phi_{\rm in}(t) = \int_{-\infty}^{\infty} \frac{ \d E}{2 \pi}  E^2 ( 1+\frac{1}{2}\sin \left( E^2 l_s^2/4\right))  e^{-i E(t- t_0)} e^{-L^2 E^2} \\
&=& \frac{\sqrt{\pi}}{4}e^{-\frac{(t-t_0)^2}{4 L^2}} \frac{(2 L^2-(t-t_0)^2)}{L^5}  - 2 \sqrt{\pi} {\rm Im} \left( e^{-\frac{(t-t_0)^2}{4 L^2-i l_s^2}} \frac{4L^2-i l_s^2-2(t-t_0)^2}{(4L^2-i l_s^2)^{5/2})}\right)  \, .
\ea
Since $\hat G_R'(t)=0$ for $t<0$ it is clear that $\hat \phi(t)$ will decay exponentially for $t<t_0$ as 
 \be
 \hat \phi_{\rm out}(t) \sim \, e^{- B (t-t_0)^2} , \quad \quad \text{for} \quad t<t_0 \, ,
 \ee
 where
 \be
 B = {\rm Min} \(\frac{1}{4L^2} ,  \frac{8 L^2 }{16 L^4 + l_s^4} \) \, .
 \ee
Since the spread in energy of the incoming state is $1/L$, the region in which locality/causality should hold is given by the bound
\be
(t-t_0) \gg r_*(E) = E l_s^2 \sim l_s^2/L
\ee
and so we have in this region
 \ba
 \hat \phi_{\rm out}(t)  \ll \exp{\( -{\rm Min} \(\frac{l_s^4}{4L^4} ,  \frac{8 l_s^4 }{16 L^4 + l_s^4} \) \)} , \quad \quad \text{for} \quad t<t_0  \, \quad, 
 \ea
and regardless of the value of $L\le l_s$, the acausal contribution to $\hat \phi_{\rm out}(t)$ will be exponentially small outside the region subject to the locality bound. This is macro-causality without micro-causality! Despite the apparent superluminal propagation of the wrong sign Nambu-Goto action, Lorentz invariance and analyticity will ensure that there can be no macroscopic superluminal communication.

\section{Summary}

Using an explicit form for the Galileon duality transformation, we have defined a non-perturbative, unitary, Lorentz invariant and UV finite quantization of a $(d+1)$'th order Galileon model that is dual to a free theory. We shall confirm elsewhere that the S-matrix is trivial \cite{KeltnerTolley2} , establishing quantum equivalence to a free theory. By an explicit calculation of the Wightman functions of this theory, in two distinct duality frames, we have confirmed that the \KL spectral densitys grow exponentially as $\rho \sim e^{E r_*(E)}$ where $r_*(E)$ is the Vainshtein radius. This occurs because the resonant contribution to the spectral densities are dominated by $N$-particle states with $N \sim E r_*(E)$.
This exponential growth implies that this Galileon model is a Jaffe type non-localizable field theory with growth index $\alpha= (d+1)/(2(d+2))$. For such theories micro-locality and micro-causality are weakened to macro-locality/causality for distances larger than the locality bound
\be
|x| \ge r_*(E) \, ,
\ee
where $E$ is the typical scattering energy. This is a manifestation of the Vainshtein effect at the quantum level, and we conjecture that similar properties hold true for all Vainshtein type theories.  \\

Although most calculations were done using the Galileon which is dual to a free field, we have provided evidence that suggest all Galileon models are non-localizable field theories. The consequences of non-localizability for the S-matrix will be that it is no longer expected to be polynomially bounded, and hence Galileon scattering does not have to respect the Froissart bound. The consequence of this will be discussed elsewhere  \cite{KeltnerTolley2}. However, we have argued, following Steinmann \cite{Steinmann:1970cm}, that there is no obstruction to maintaining analyticity of the retarded products, and hence analyticity of the S-matrix, despite the failure of polynomial boundedness. Furthermore, analyticity of the scattering amplitude alone is sufficient to ensure macro-causality \cite{Steinmann:1970cm}.  \\

Although we have explicity constructed a quantization of the dual to a free theory, the UV completion of an interacting Galileon remains conjectured. However, our arguments support the view that the UV completion cannot be a local field theory, i.e. that of a non-Wilsonian completion, consistent with the arguments of the `classicalization' proposal \cite{Dvali:2010jz,Dvali:2010ns,Dvali:2011nj,Dvali:2011th,Dvali:2014ila} . In particular, we note that our UV description of the special Galileon contains no new degrees of freedom, despite becoming strongly coupled at the scale $\Lambda$. In line with the `asymptotic fragility' proposal \cite{Dubovsky:2012wk,Cooper:2013ffa}, our picture of the UV theory does not contain any UV fixed point or conformal behaviour. Rather the scale $\Lambda$ determines the exponential growth properties of the Wightman functions and scattering amplitudes. We have argued that the superluminal group and phase velocities found in the LEEFT are generically absent in the UV completion. Macro-causality will enforce luminality by modifying the high energy properties of the retarded propagator, which are what actually determine causal propagation speeds through the front velocity. \\

If our conjecture is correct, then this resolves many of the technical issues with IR completions of Galileons such as Massive Gravity/Multi-Gravity theories \cite{deRham:2010kj,Hassan:2011zd,Hinterbichler:2012cn}. We have argued that the non-locality that arises in the Galileon models is of the same nature as we would expect in any gravitational theory, and this fits well with the role of the Galileon as the helicity-zero mode of the graviton in these theories. For example, when two high energy particles scatter in GR at impact parameters $b<R_S(E)$ then a black hole is expected to form. The production of this extended semi-classical object is tied to the mild gravitational non-locality of GR as proposed by Giddings and Lippert \cite{Giddings:2001pt,Giddings:2004ud}.  When the same happens in Massive Gravity, as well as the Schwarzshild radius, there is the Vainshtein radius which describes the profile of the helicity-zero mode. The usual gravitational non-locality of GR is now pushed out to the larger Vainshtein/classicalization radius $r_*(E)$. In taking the Galileon decoupling limit \cite{deRham:2010ik}, although the usual gravity switches off, the gravitational non-locality of the helicity-zero mode remains and explains why the UV completion of Galileons must be non-localizable. This does not undermine the classical picture of the Vainshtein mechanism, of relevance to cosmological phenomenology, just as black holes are well-described classically. However it has profound implications for quantum fluctuations and the scattering of high energy Galileon quanta.

\acknowledgments

AJT is supported by a Department of Energy Early Career Award DE-SC0010600.  AJT would like to thank the Perimeter Institute for Theoretical Physics for hospitality while part of this work was being completed. We would like to thank Andrew Matas, Nick Ondo, Shuang-Yong Zhou, Kurt Hinterbichler, and Austin Joyce for useful comments on the manuscript and especially Claudia de Rham and Matteo Fasiello for discussions and comments. 
\appendix

\section{$N$-particle phase space density}

\label{app1}

The $n$-particle phase space density in 4 dimensions defined as 
\be
\Omega_n(-k^2) = \left[ \prod_{i=1}^n \int \d \tilde k_i \right] \, (2 \pi)^4 \delta^{(d)}(k - \sum_{i=1}^n k_i) \, ,
\ee
can be computed as follows
\ba
\Omega_n(\mu) &=& \int \d^4 x  \left[ \prod_{i=1}^n \int \d \tilde k_i \right] \,  e^{-i(k - \sum_{i=1}^n k_i).x} \, , \\
&=& \int \d^4 x  \, e^{-i k.x} \frac{1}{(4 \pi^2)^n}  \frac{1}{(\vec x^2 - (x^0-i \epsilon)^2)^n}  \,.
\ea
Rather than computing this integral directly we can use the indirect relation for the Euclidean correlation functions
\ba
 \frac{1}{(4 \pi^2)^n}  \frac{1}{x^{2n} } & =& \int_0^{\infty} \d \mu \int \frac{\d^4 k}{(2 \pi)^4} e^{ik.x}  \frac{\Omega_n(\mu)}{k^2+\mu} \, , \\
&=&  \int_0^{\infty} \d s \int_0^{\infty} \d \mu \int \frac{\d^4 k}{(2 \pi)^4} e^{ik.x}  \Omega_N(\mu) e^{-s (k^2+\mu)} \, , \\
&=&  \int_0^{\infty} \d s \int_0^{\infty} \d \mu  \frac{1}{2^4 \pi^2} \frac{1}{s^2}e^{-\frac{1}{4s} x^2}  \Omega_n(\mu) e^{-s \mu}  \, .
\ea
Since we know dimensionally $\Omega_n(\mu) = a_n \mu^{n-2}$ then performing the $\mu$ integral
\ba
\frac{1}{(4 \pi^2)^n}  \frac{1}{x^{2n} } & =&a_n (n-2)! \frac{1}{2^4 \pi^2} \int_0^{\infty} \d s   \, s^{-1-n} e^{-\frac{1}{4s} x^2}  \, ,\\
& =&a_n (n-2)! (n-1)! \frac{4^{n}}{2^4 \pi^2} \frac{1}{x^{2n}}  \, ,
\ea
whence
\be
\Omega_n = \frac{1}{ (16 \pi^2)^{n-1} (n-2)! (n-1)!}\mu^{n-2} \, .
\ee

\section{Spectral density as a Double Borel Transform}

\label{doubleBorel}

In field theory, perturbative expansions are nearly always asymptotic and so must be resumed by some means, for example by Borel resummation. Our technique of defining Wightman functions through their spectral densities can be viewed as a double Borel transform.

To see this we note that the Euclidean propagator can be written as
\ba
G_E(x) &=& \int_0^{\infty} \d \mu \int \frac{\d^4 k}{(2 \pi)^4} \rho(\mu) e^{ik.x} \frac{1}{k^2 + \mu}  \, , \\
&=&  \int_0^{\infty} \d \mu  \int_0^{\infty} \d s \int \frac{\d^4 k}{(2 \pi)^4} \rho(\mu) e^{ik.x} e^{-s( k^2 + \mu)} \, \\
&=& \int_0^{\infty} \d \mu  \int_0^{\infty} \d s \, \frac{1}{16\pi^2} \frac{1}{s^2}e^{-\frac{1}{4s} x^2}  \rho(\mu) e^{-s \mu} \, .
\ea
We see that the Euclidean propagator is given by a double Laplace transform of the spectral density. This means that the convergence properties of $\Omega(\mu)$ are always better than $G_E(x)$. Specifically, if $G_E(x)$ is given by the series
\be
G_E(x) = \frac{1}{4 \pi^2 x^2 }+ \sum_{n=2}^{\infty }\frac{d_n}{(4 \pi^2)^n x^{2 n}}
\ee
then performing the two inverse Laplace transforms we find
\be
\rho(\mu) =  \delta(\mu) + \sum_{n=2}^{\infty }\frac{d_n \mu^{n-2}}{(16 \pi^2)^{n-1} (n-2)! (n-1)!} \, .
\ee
This is characteristic of a double Borel transform and has a vastly improved convergence from the additional $1/( (n-2)! (n-1)!)$ at large $n$.
In particular, if we define an operator via
\be
\hat O(x) = \phi(x) + \sum_{n=2} c_n : \phi(x)^n : \, ,
\ee
then from Wick's theorem $d_n = n! \, c_n^2$. The spectral density will converge provided only that $c_n $ grows no faster than $\sqrt{n!}$.

\section{Spectral density calculation in $(\rho, \tilde \chi)$ frame.}

\label{Density2}

A slightly different approach to the calculation of the spectral density is via a Euclidean calculation. The naive definition of the Euclidean two point correlation function for the matter field $\tilde{\chi}$ in $d$ dimensions can be written as 
\ba
\tp=\int\frac{d^dk}{(2\pi)^d}\ \frac{1}{k^2}e^{ik\cdot x}\exp\left[\frac{a_d}{  \Lambda^{2 \sigma}}\left(\frac{k^2}{x^d}-d\frac{(k\cdot x)^2}{x^{d+2}}\right)\right]\ .
\ea
where $x=\tilde{x}_1-\tilde{x}_2$  and $a_d= \frac{(d-2)}{4} \pi^{-d/2} \Gamma(d/2-1) $.  Transforming into spherical coordinates,
\ba
\tp=\frac{V_{d-2}}{(2\pi)^d}\sum_{n=0}^{\infty}\frac{1}{n!}\left(\frac{a_d}{\Lambda^{2 \sigma} x^d}\right)^n\int_0^{\infty}dk\ k^{d+2n-3}\int_{-1}^1dw\ (1-w^2)^{\frac{1}{2}(d-3)}(1-dw^2)^ne^{ikxw} \, ,\nonumber
\ea
where $V_{d-2}$ is the volume integral over a $(d-2)$ sphere and $w=\cos(\theta)$.  Taking advantage of binomial expansions we may write
\ba
(1-dw^2)^n=\sum_{s=0}^n\frac{n!(-d)^s}{s!(n-s)!}w^{2s} \, ,
\ea
and expanding $\exp(ikxw)$ we have
\ba
\tp=\frac{V_{d-2}}{(2\pi)^d}\sum_{n=0}^{\infty}\sum_{s=0}^n\frac{(-d)^s}{s!(n-s)!}\left(\frac{a_d}{ \Lambda^{2 \sigma} x^d}\right)^n\int_0^\infty dk\ k^{d+2n-3}\nonumber\\
\times\sum_{m=0}^{\infty}\frac{(ikx)^m}{m!}\int_{-1}^1dw\ w^{2s+m}(1-w^2)^{\frac{1}{2}(d-3)}\ .
\ea
Doing the $w$ integral and then the $m$ sum gives us
\ba\label{k}
\tp=\frac{V_{d-2}\sqrt{\pi}}{(2\pi)^d}\Gamma\left[\frac{1}{2}(d-1)\right]\sum_{n=0}^{\infty}\sum_{s=0}^n\frac{(-d)^s}{s!(n-s)!}\Gamma\left[s+\frac{1}{2}\right]\left(\frac{a_d}{ \Lambda^{2 \sigma} x^d}\right)^n\nonumber\\
\int_0^\infty dk\ k^{d+2n-3} \ _{P}\tilde{F}_Q\left(s+\frac{1}{2};\frac{1}{2},s+\frac{d}{2};-\frac{1}{4}k^2x^2\right) \, ,
\ea
where $_{P}\tilde{F}_Q$ is the regularized hypergeometric function.  

\subsubsection*{Four dimensions}

Focusing on the $d=4$ case  and doing the $k$ integral we have 
\ba
\tp=\frac{4\pi}{(2\pi)^4}\sum_{n=0}^{\infty}\sum_{s=0}^n\frac{(-4)^s}{s!(n-s)!}\left(\frac{1}{ 2 \pi^2 \Lambda^6 x^6}\right)^n\left(-\frac{\sqrt{\pi}\cos(n\pi)\Gamma[2n+2]\Gamma\left[s-n-\frac{1}{2}\right]}{2x^{2(n+1)}\Gamma[s+n-1]}\right)\nonumber\ .
\ea
Finally, doing the $s$ sum leaves us with
\ba
\tp=\frac{4\pi^2}{(2\pi)^4}\frac{1}{x^2}\sum_{n=0}^\infty\frac{4^n(2n+1)!}{n!}\left(\frac{1}{ 2 \pi^2 \Lambda^6 x^6}\right)^n\ .
\ea
This series is asymptotic, as we would expect, since the real space two-point function should not exist even in the Euclidean. However term by term we may infer the contribution to the associated spectral density using the results of Appendix \ref{app1} to give
\ba
\rho(\mu) &=& \delta(\mu) +\frac{4\pi^2}{(2\pi)^4}  \sum_{n=1}^{\infty}  \frac{1}{(2 \pi^2\Lambda^6)^n}  \frac{4^n(2n+1)!}{n!} (4 \pi^2)^{3n+1} \Omega_{3n+1}(\mu)  \, .
\ea
This agrees with the result used in the text.

\bibliography{references}

\end{document}